\documentclass[11pt]{article}

\usepackage{fullpage}
\usepackage{setspace}
\usepackage{parskip}
\usepackage{titlesec}
\usepackage[section]{placeins}
\usepackage{xcolor}
\usepackage{breakcites}
\usepackage{lineno}

\usepackage{times}

\PassOptionsToPackage{hyphens}{url}
\usepackage[colorlinks = true,
            linkcolor = blue,
            urlcolor  = blue,
            citecolor = blue,
            anchorcolor = blue]{hyperref}
\usepackage{etoolbox}

\usepackage{natbib}

\renewenvironment{abstract}
  {{\bfseries\noindent{\abstractname}\par\nobreak}\footnotesize}
  {\bigskip}

  {\begin{tabular}{|p{13cm}}}
  {\end{tabular}}

\titlespacing{\section}{0pt}{*3}{*1}
\titlespacing{\subsection}{0pt}{*2}{*0.5}
\titlespacing{\subsubsection}{0pt}{*1.5}{0pt}

\usepackage{authblk}

\usepackage{graphicx}
\usepackage[space]{grffile}
\usepackage{latexsym}
\usepackage{textcomp}
\usepackage{longtable}
\usepackage{tabulary}
\usepackage{booktabs,array,multirow}
\usepackage{amsfonts,amsmath,amssymb}
\providecommand\citet{\cite}
\providecommand\citep{\cite}
\providecommand\citealt{\cite}
\newif\iflatexml\latexmlfalse

\AtBeginDocument{\DeclareGraphicsExtensions{.pdf,.PDF,.eps,.EPS,.png,.PNG,.tif,.TIF,.jpg,.JPG,.jpeg,.JPEG}}

\usepackage[utf8]{inputenc}
\usepackage[english]{babel}

\usepackage{float}

\usepackage[margin=0.85in]{geometry}

\usepackage{cleveref}[2012/02/15]%
\crefformat{footnote}{#2\footnotemark[#1]#3}

\usepackage{subfig}
\usepackage[export]{adjustbox}

\usepackage{pifont}%
\newcommand{\cmark}{\ding{51}}%
\newcommand{\xmark}{\ding{55}}%

\begin{document}

\title{Time Series Analysis via Network Science: Concepts and Algorithms\footnote{Accepted authors manuscript (AAM) published in Data Mining and Knowledge Discovery, 10.1002/widm.1404. Final source webpage: https://doi.org/10.1002/widm.1404 CONTACT Vanessa Freitas Silva. Email: vanessa.silva@dcc.fc.up.pt }}

\author[1]{Vanessa Freitas Silva}%
\author[2]{Maria Eduarda Silva}%
\author[1]{Pedro Ribeiro}%
\author[1]{Fernando Silva}%

\affil[1]{CRACS and INESC-TEC, DCC, Faculdade de Ciências, Universidade do Porto}%
\affil[2]{CIDMA, Faculdade de Economia, Universidade do Porto}%

\vspace{-1em}

  \date{}

\begingroup
\let\center\flushleft
\let\endcenter\endflushleft
\maketitle
\endgroup

\selectlanguage{english}
\begin{abstract}
There is nowadays a constant flux of data being generated and collected in all types of real world systems. These data sets are often  indexed by time, space or both requiring appropriate approaches to analyze the data. In univariate settings, time series analysis is a mature and solid field.  
However, in multivariate contexts, time series analysis still presents many limitations. In order to address these issues, the last decade has brought  approaches  based on network science. These methods involve transforming an initial time series data set into one or more networks, which can be analyzed in depth to provide insight into the original time series. This review provides a comprehensive overview of existing mapping methods for transforming time series into networks for a wide audience of researchers and practitioners in machine learning, data mining and time series. Our main contribution is a structured review of existing methodologies, identifying their main characteristics and their differences. We describe the main conceptual approaches, provide authoritative references and give insight into their advantages and limitations in a unified notation and language. 
We first describe the case of univariate time series, which can be mapped to single layer networks, and we divide the current mappings based on the underlying concept: visibility, transition and proximity. We then proceed with multivariate time series discussing both single layer and multiple layer approaches. Although still very recent, this research area has much potential and with this survey we intend to pave the way for future research on the topic.

\end{abstract}%

\par\null

\section{Introduction}
The recent developments in technology are leading to the generation and collection of huge volumes of data associated with various real world systems.
These data sets typically exhibit an inherent temporal dimension that is crucial to their understanding.
The analysis of a single collection of observations indexed in time in a univariate setting is a mature and solid field, usually referred to as time series analysis~\citep{Sumway2017}, which mostly involves statistical linear analysis. 
Given that many observed data sets exhibit non-Gaussian and non-linear characteristics, in the last decades several statistical non-Gaussian and non-linear models have been proposed in the literature~\citep{Stoffernonlinear}. 
These models have often been developed tailored to model data from specific areas and under several assumptions that hinder their wide application~\citep{wei2018multivariate}. 
In fact, in many settings both univariate (e.g. high frequency signals, non-stationarity) and high dimensional (e.g. multivariate, spatio-temporal, panels of time series)  time series analysis still presents many limitations. 
An alternative approach to time series data analysis has been developed under the framework of dynamical systems theory. This approach, denominated nonlinear time-series analysis~\citep{bradley2015nonlinear} can be overly effective when the data model is based o deterministic dynamics in some state space. 

\newpage
Network science is nowadays a well established field of science that studies a wide range of systems in nature and society, by representing the existing interactions through complex network (or graph) structures~\citep{Barabasi2016}. 
There exists a vast set of topological graph measurements available~\citep{Costa2007}, a well-defined set of problems and a large track record of successful application of complex network methodologies to different fields~\citep{vespignani2018twenty}. 

In the last decade, several approaches for time series analysis based on network science methodologies were proposed, leveraging the large body of research in network analysis and providing new insight and novel angles on which to understand the structure of time series~\citep{zou2018complex}. 
This new framework for time series analysis inevitably involves mapping the time series into networks. 
This survey aims precisely to provide a comprehensive overview of the existing mapping methods for this purpose. 
Our main contribution is to produce a general and structured review of existing mapping algorithms, highlighting their similarities and differences, the data characteristics they capture and the main references and results. 
We propose here a conceptual division of the methods at the level of the time series dimensionality and concomitant choice of the resulting  network structure. From our point of view, this high-level conceptual division encompasses all existing methods in the literature, for both univariate and multivariate frameworks. 

\begin{figure}[hbt!]
	\centering 
	\includegraphics[width=0.95\textwidth]{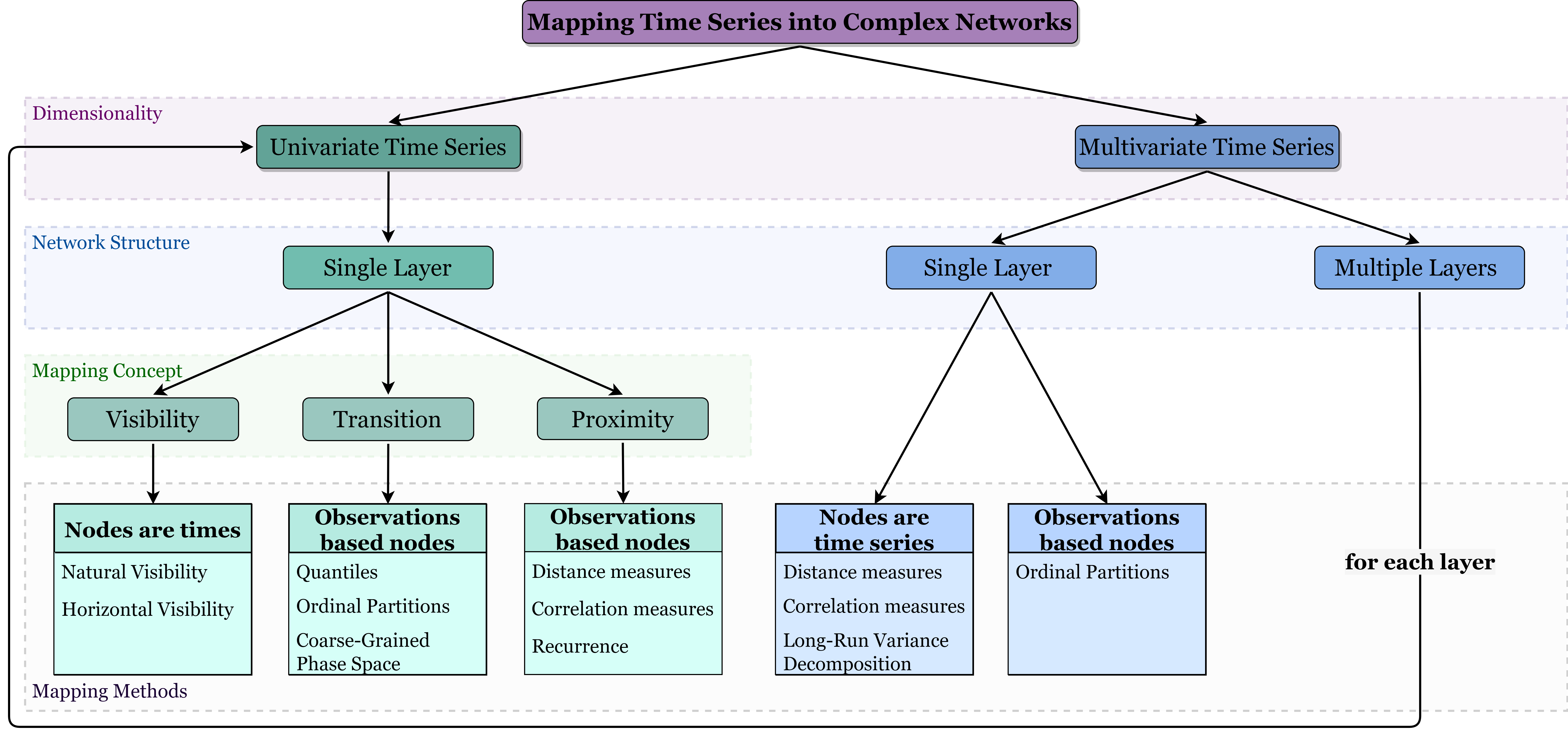}

	\caption[Figure 1]{Overview of our survey. Taxonomy of algorithms for mapping time series into complex networks based on the dimensionality of time series, resulting network structure, mapping concept, and main mapping methods.}
	\label{figure:1}
\end{figure}

The proposed high-level conceptual division is depicted in the diagram of Figure~\ref{figure:1}, and corresponds to the organization of our paper. We first divide the mappings into two large groups, depending on the source data set being a univariate or a multivariate time series. The first group results in single layer networks, while multivariate series can be mapped both in single layer and in multiple layer networks (such as multiplex networks). 
\vspace{-1.5mm}

The set of univariate series mappings proposed in the literature may be divided into three main types depending on the underlying concept: visibility, transition and proximity. 
Visibility mapping concepts establish connections (edges) between the time stamps of the series (as nodes) using visibility lines (with or without restrictions) between observations. 
For transition mappings, the methods are based on the transition probabilities between states (or partitions) defined by dividing the time space in a set of temporal states (or dividing the series support/observations into partitions) that will be the nodes of the network. 
Finally, proximity mappings establish connections using measures of distance or similarity between time points (or states) become network nodes. 
Regarding multivariate time series, there are methods leading to a single layer network, where each series (or patterns based on observations from multivariate series) is represented by a node in the network and the connections are established based on relationships between the series (or patterns). 
We can also find mappings leading to multiple layer network structures, where each layer represents  a time series and, in fact, each layer can be mapped based on one of the univariate time series mapping methods, separately. Research on multivariate mappings is however still in its infancy. 

We aim at reaching a wide community of researchers and practitioners who may benefit from these methods in machine learning, data mining and time series analysis tasks. 
With this purpose in mind, the mappings are described  from an algorithmic point of view geared towards software implementations. 
We provide quick reference tables summarizing   the algorithms, properties of the mappings (univariate and multivariate) and of  the resulting  networks. Each mapping algorithm  is depicted using a toy time series (the same for all methods for comparison purposes) to facilitate its understanding. 
In comparison to the survey by~\cite{zou2018complex}, whose focus is on non-linear time series analysis by means of complex network methods, we add more recent aspects of the literature as well as approaches stemming  from the statistical literature. 

Given the multidisciplinary of time series, as well as complex networks, our approach in this review can impact several fields and provide new insights and structured information to practitioners of several areas.

\par\null

\section{Preliminaries}

{\label{sec:concepts}}

This section presents the background concepts and terminology on time series and network science and establishes the necessary notation for the remainder of the document.

\subsection{Time Series}

{\label{sec:ts}}

Formally, a time series $\boldsymbol{Y}=(Y_1, \ldots, Y_T)$ is a finite realization of a \textit{stochastic process} which is a sequence of random variables $\{Y_t\}_t$ indexed by time $t$, or of a \textit{dynamic deterministic process} that can be fully represented by an explicit mathematical relation. In this work we consider $t$ discrete and varying in the integers $t=0,\pm1, \pm 2, \ldots.$
The main characteristic of a time series is the serial dependence between the observations which restricts the applicability of many conventional statistical methods developed under the assumption of independent and identically distributed (i.i.d.) observations. 
Time series analysis refers to the collection of procedures developed to systematically solve the statistical problems posed by the serial correlations. Thus, its main purpose is to develop mathematical models that provide plausible descriptions of data characteristics with a view to forecasting and simulation~\citep{Box2015}.

A complete description of a time series $Y_{1}, \ldots, Y_{T}$ is provided by the joint distribution function $F(c_1, \ldots, c_T) = P(Y_1 \leq c_1, \ldots, Y_T \leq c_T)$. This joint distribution can be obtained easily  only in the particular case of jointly Gaussian random variables and, even so, under the assumption of some regularity of the time series over time.  
Thus, time series analysis relies on a set of concepts and measures that capture the essential feature of the data, the serial correlation. 
The main statistical tools in time series analysis are the following. 
The \textit{mean function} defined as $\mu_t = \mbox{\rm E}(Y_t)$. 
The \textit{autocovariance function}  defined as the covariance between  the variables at times $t$ and $s$: 
\begin{equation}
	\gamma(s, t) = \mbox{\rm cov}(Y_s, Y_t) = \mbox{\rm E}[(Y_s-\mu_s)(Y_t-\mu_t)],
\end{equation}
for all $t$ and $s.$ 
We should note that $\gamma(s,t) = \gamma(t,s)$ for all-time points $s$ and $t$ and $\gamma(t, t) = \mbox{\rm E}[(Y_t-\mu_t)^2] = \mbox{\rm var}(Y_t).$ The autocovariance measures the linear dependence between variables at different times. 
The \textit{autocorrelation function} (ACF) which measures the linear predictability of the series at time $t$ using only the variable at time $s,$ $Y_s,$ is defined as: 
\begin{equation}
	\rho(s,t) = \mbox{\rm corr}(Y_t, Y_s) =  \frac{\gamma(s,t)}{\sqrt{\gamma(s,s)\gamma(t,t)}},
\end{equation}
$-1 \leq \rho(s,t) \leq 1$ with $\rho(s,t)=1$ indicating perfect positive correlation and $\rho(s,t)=-1$ indicating perfect negative correlation. 
Hence, the ACF provides a  rough measure of the ability to forecast the series at time $t$ from the value at time $s$~\citep{Sumway2017}.

The notion of regularity referred to above is defined using the concept of \textit{stationarity}.
Technically, we may distinguish strict (or strong) stationarity from weak (or wide, or second order) stationarity. 
A time series is said to be \textit{strictly stationary} if  the probabilistic behavior of every collection of variables $\{Y_1, Y_2, \ldots, Y_t \}$ is identical to that of the time shifted set $\{Y_{1 + h}, Y_{2 + h}, \ldots, Y_{t + h} \}$. 
However, this condition is too strong for most applications, and it is difficult to assess it for a single data set. 
Rather than imposing conditions on all possible distributions of a time series, weak stationarity imposes conditions only on the first two moments of the series. 
Thus, a  time series is said to be \textit{weakly stationary} if it is a finite variance process, with a constant mean function and does not depend on time $t$, $\mbox{\rm E}(Y_t)=\mu$, and the autocovariance function depends only on the lag, $h=s-t$, $\gamma(s,t)= \gamma(s-t,0)=\gamma(h)$.  

In summary,  a time series is stationary when its statistical characteristics (mean and variance) are constant over time, that is, data fluctuate around a constant mean with the variance of the fluctuations remaining essentially the same over time. 
Moreover, stationarity implies that correlation between observations depends only on the time lag between the observations. 
Stationarity is essential in the analysis of time series since it  enables the meaningful computation of statistics such as the mean, variance and ACF from the data.

The simplest time series process is the \textit{white noise} (or purely random) process, henceforth denoted by $\epsilon_t.$ It is a sequence of i.i.d. random variables with mean $0$ and constant variance $\sigma_{\epsilon}^2.$ A particular case is the \textit{Gaussian white noise}, where $\epsilon_t$ are independent normal random variables. 

Univariate time series relate to measurements of a given variable over time, $Y_t$, such as temperature measurements. However, it is increasingly common to observe data from several variables, $Y_{i,t},$ $i = 1, 2, \ldots, m$, that may be correlated. For example, in air pollution studies  atmospheric variables such as temperature and air humidity, as well as pollutants, such as particulate matter and NO$_2$ hourly concentrations are observed and need to be analyzed jointly and over time. We represent these multivariate time series by a vector, $\boldsymbol{Y}_t = [Y_{1,t}, Y_{2,t}, \ldots, Y_{m,t}]^{'}$ where $'$ represents the transpose of a matrix/vector and  $Y_{i,t}$ is the $i$-th  component time series (a random variable for each $i$ and $t$).

Multivariate time series present not only serial dependence within each component series $Y_{i,t}$ but also interdependence between the different components  $Y_{i,t}$ and $Y_{j,s}$ when $i \neq j$, regardless of whether the times $s$ and $t$ are the same or not. 
Although the theory of univariate time series extends in a natural way to the multivariate case new concepts arise. The $m$-dimensional time series process, $\boldsymbol{Y}_t$, is a stationary process if each of its component series is a univariate stationary process and the covariance matrix function depends only on the lag. Thus, $\mbox{\rm{E}}\left(\boldsymbol{Y}_t \right) = \boldsymbol{\mu},$ is now a $m\time 1$ vector of constants, and the autocovariance  function, now a matrix function, is defined for lag $h$ as the $m\times m$ matrix  $\Gamma_{\boldsymbol{Y}}(h) = \mbox{\rm{E}}\left [ \left(\boldsymbol{Y}_t - \boldsymbol{\mu}\right)\left(\boldsymbol{Y}_{t+h} - \boldsymbol{\mu}\right)^{'}\right]$, with elements $\gamma_{i,j}(h)= \mbox{\rm{cov}} (Y_{i,t},Y_{j,t+h})$. Note that $\Gamma_{\boldsymbol{Y}}(h)=\Gamma_{\boldsymbol{Y}}^{'}(-h)$ and the element $(i,j)$ of matrix $\Gamma_{\boldsymbol{Y}}(0)$ is the contemporaneous correlation between $Y_{i,t}$ and $Y_{j,t}$ (see~\citealp{wei2018multivariate} for more details). 
Similarly, we can define the correlation matrix at lag $h$ with elements $\rho_{i,j}(h) = \mbox{\rm corr}(Y_{i,t}, Y_{j,t+h}).$  

The simplest example of an $m$-dimensional vector process is the \textit{vector white noise process}, $\{\boldsymbol{a}_t\},$  with mean vector $0$ and covariance matrix function $\boldsymbol{\Sigma}$, where $\boldsymbol{\Sigma}$ is an $m \times m$ symmetric positive definite matrix. Although the components of the white noise process are serially uncorrelated $\mbox{\rm corr}(a_{i,t}, a_{i,s})=0,$ for $t \neq s$, they may be contemporaneously correlated, $\mbox{\rm corr}(a_{i,t}, a_{j,t}) \neq 0$. The most common vector white noise process is the Gaussian white noise process for which $\boldsymbol{a}_t$ follows a multivariate normal distribution~\citep{wei2018multivariate}.

The analysis of multivariate time series requires tools, methods and models for processing information contained in multiple measurements that have both temporal and cross-sectional dependence which must be necessarily different from standard statistical theory and methods based on a random sample that assume independence. Linear and non-linear models for multivariate time series have often been developed  tailored to model data from specific areas and under several assumptions. In recent years, procedures to analyze high dimensional time series have been developed but  many issues remain open~\citep{wei2018multivariate}.

\subsection{Network Science} 

Many real world systems can be seen as complex systems with a set of elements that interact with each other and which exhibit emergent collective properties~\citep{Costa2011}. 
\textit{Graphs} provide a powerful and general abstraction for this kind of scenarios, in which the elements are represented by \textit{nodes} and the interactions by \textit{edges}. Such graphs are often described as \textit{complex networks}, since they typically exhibit non-trivial topological properties, due to the characteristics of the underlying complex system, which are neither random nor purely regular~\citep{Albert2002}.

\subsubsection{Graph Terminology and Concepts}

\begin{figure}[hbt!]
	\centering 
	\subfloat[Directed graph]{
		\includegraphics[scale=0.15,keepaspectratio,valign=m]{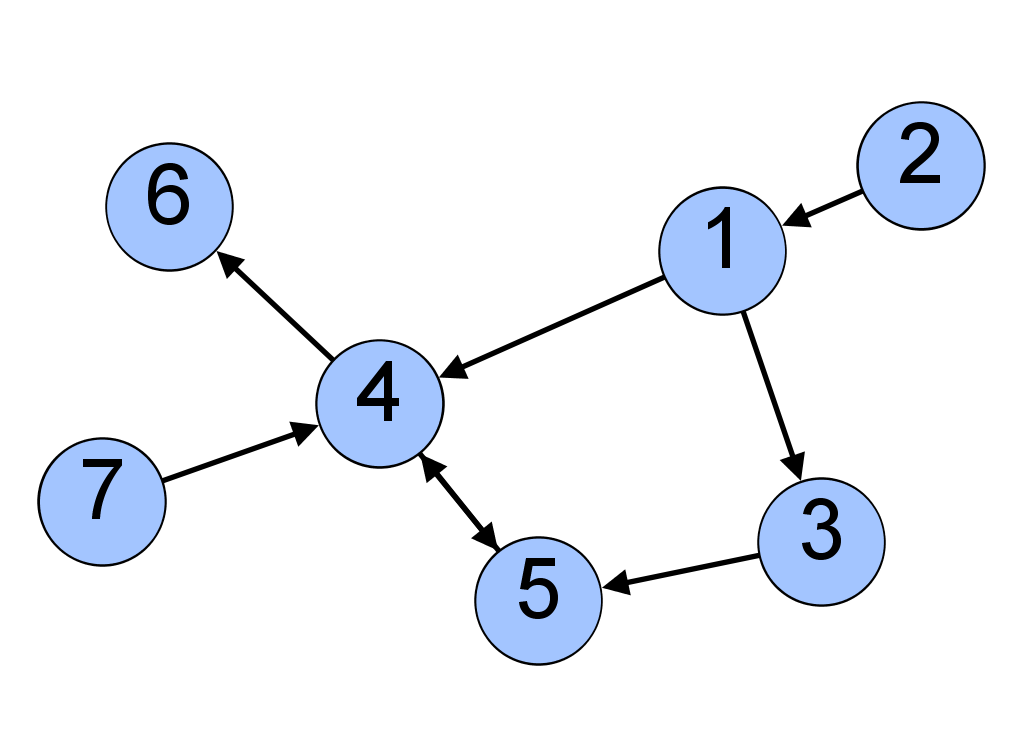}
	}
	\quad
	\quad
	\quad
	\subfloat[Undirected weighted graph]{
		\includegraphics[scale=0.15,keepaspectratio,valign=m]{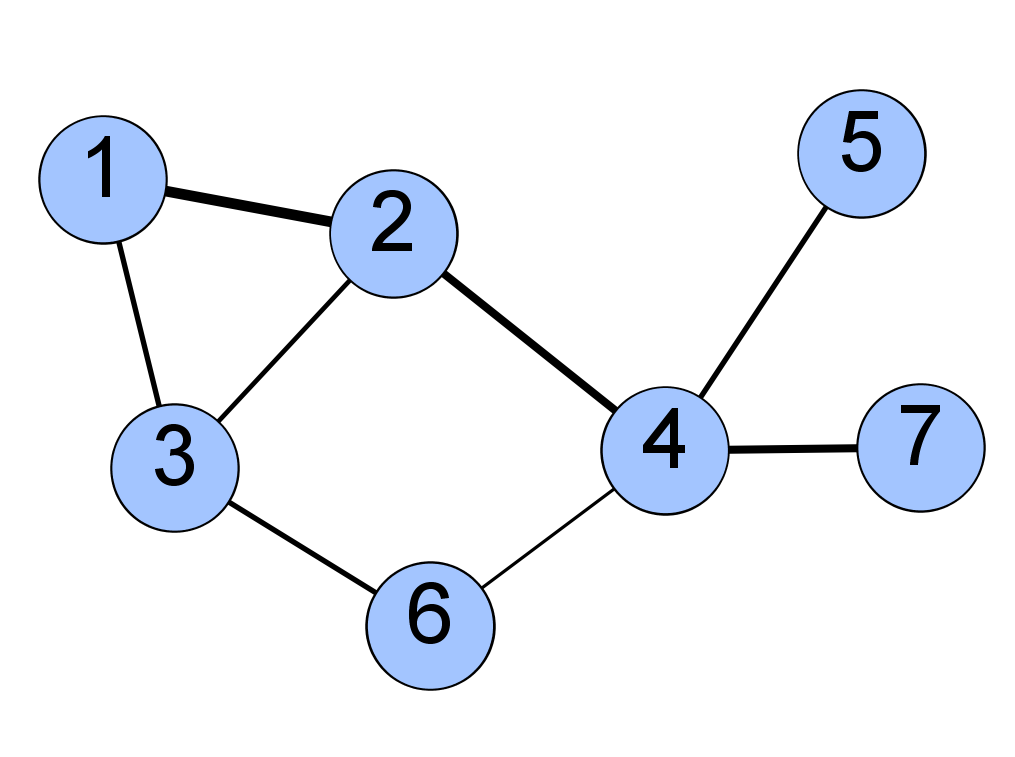}
	}
	\caption[Figure 2]{A representation of (a) a simple directed graph and (b) a simple undirected weighted graph.}
	\label{figure:2}
\end{figure}

A \textit{graph}, $G$, is an ordered pair $(V(G), E(G))$, where $V(G)$ represents the set of \textit{nodes} (or vertices) and $E(G)$ the set of \textit{edges} (or links) between pairs of elements of $V(G)$. The number of nodes, also known as the \textit{size} of the graph, is written as $|V(G)|$ and the number of edges as $|E(G)|.$ 
Two nodes $v_i$ and $v_j$ are \textit{neighbors} (or \textit{adjacent}) if they are linked, that is, if $(v_i,v_j) \in E(G).$ We can distinguish between \textit{directed} edges, which connect a source node to a target node, and \textit{undirected} edges, when there is no such concept of orientation. In the first case the graph is called \textit{directed} or \textit{digraph}.
A graph is \textit{weighted}, if there is a weight (or cost) $w_{i,j}$ associated the edge $(v_i, v_j).$ 
It is classified as \textit{simple} if it does not contain multiple edges (two or more edges connecting the same pair of nodes) and it does not contain self-loops (an edge connecting a node to itself). 
Figure~\ref{figure:2} illustrates the concepts here defined and depicts a simple directed graph and a simple undirected weighted graph, respectively.
\vspace{-1.8mm}

A graph $G$ is usually represented mathematically as an \textit{adjacency matrix} denoted as $\boldsymbol{A}$, and $A_{i,j}$ is $1$ (or $w_{i,j}$) when $(v_i,v_j) \in E(G)$ and is $0$ otherwise. In the context of this paper, all nodes are labeled with consecutive integer numbers starting from  $1$ to $|V(G)|$.

Most topological measures/metrics are related to the concepts of \textit{paths} and graph \textit{connectivity}. 
A \textit{path} is a sequence of nodes in which each consecutive pair of nodes in the sequence is connected by an edge. It may also be useful to think of the path as the sequence of edges that connect those nodes. 
Two nodes are \textit{connected} if there is a path between them and are \textit{disconnected} if no such path exists. A set of nodes is called a connected component if all its node pairs are connected. 

The analysis of complex networks has made enormous progress in the last decade, leading to the development of statistical properties of certain types of complex systems that are related and codified in their topology~\citep{Albert2002,newman2003structure}. 
There exists therefore a vast set of topological metrics that are able to characterize a network, each reflecting some particular features of the system under analysis. 
A more detailed list of the most important topological metrics can be found in the literature~\citep{Albert2002,Costa2007,Barabasi2016}.

\subsubsection{Multilayer Networks}\label{susbsubsec:mnet}
Most real world complex systems are not isolated and interact with other systems. Furthermore, a single entity may belong to several subsystems, establishing both internal (in the same system) and external (with other systems) connections. 
The latter type of connectivity can be interpreted as an interaction between levels (or layers), leading to the concept of \textit{multilayer networks}. 

\begin{figure}[hbt!]
	\centering 
	\subfloat[Multilayer Network $(s=2)$]{
		\includegraphics[scale=0.46,keepaspectratio,valign=m]{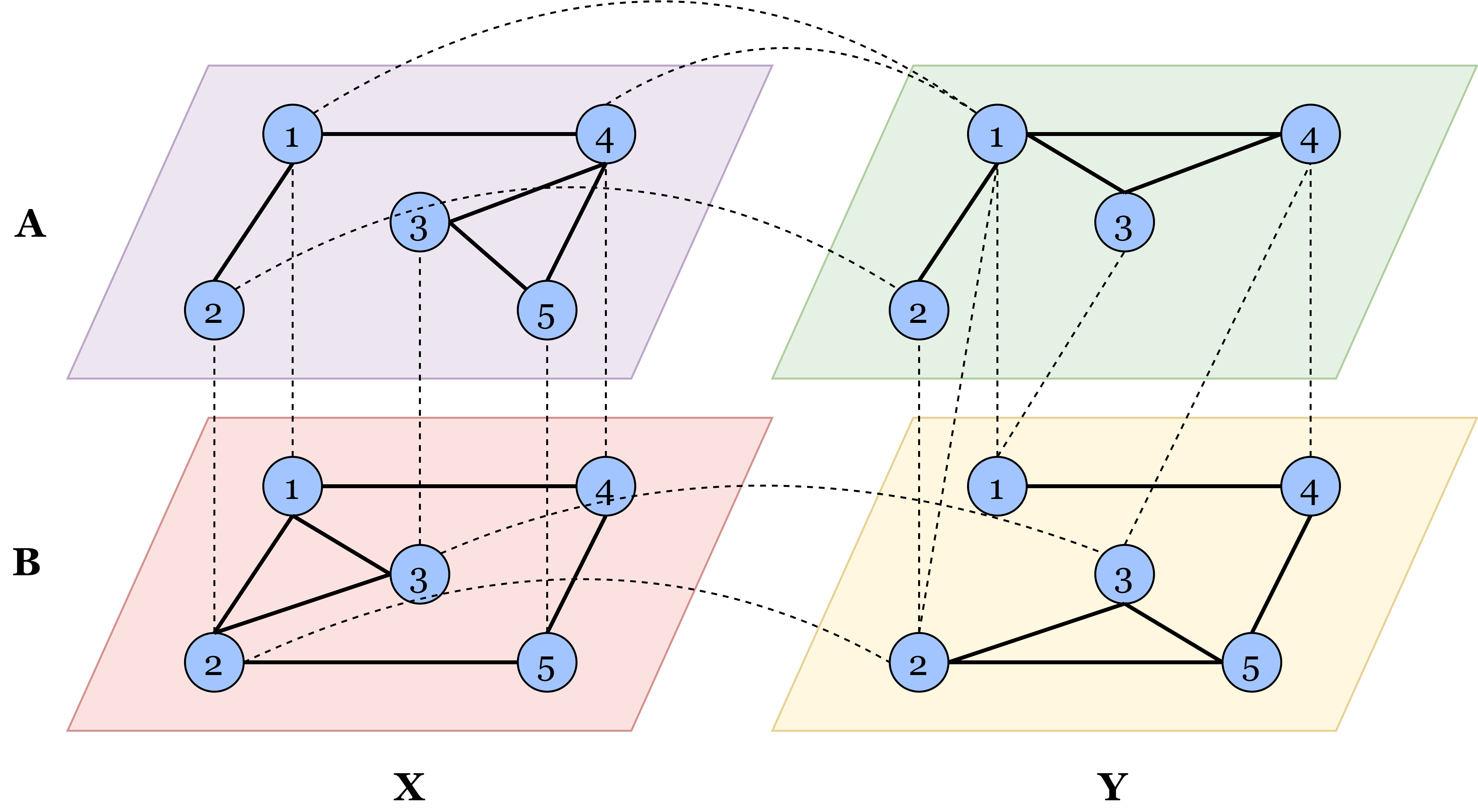}
		\label{figure:3a}
	}
	\quad
	\quad
	\quad
	\subfloat[Multiplex Network $(s=1)$]{
		\includegraphics[scale=0.46,keepaspectratio,valign=m]{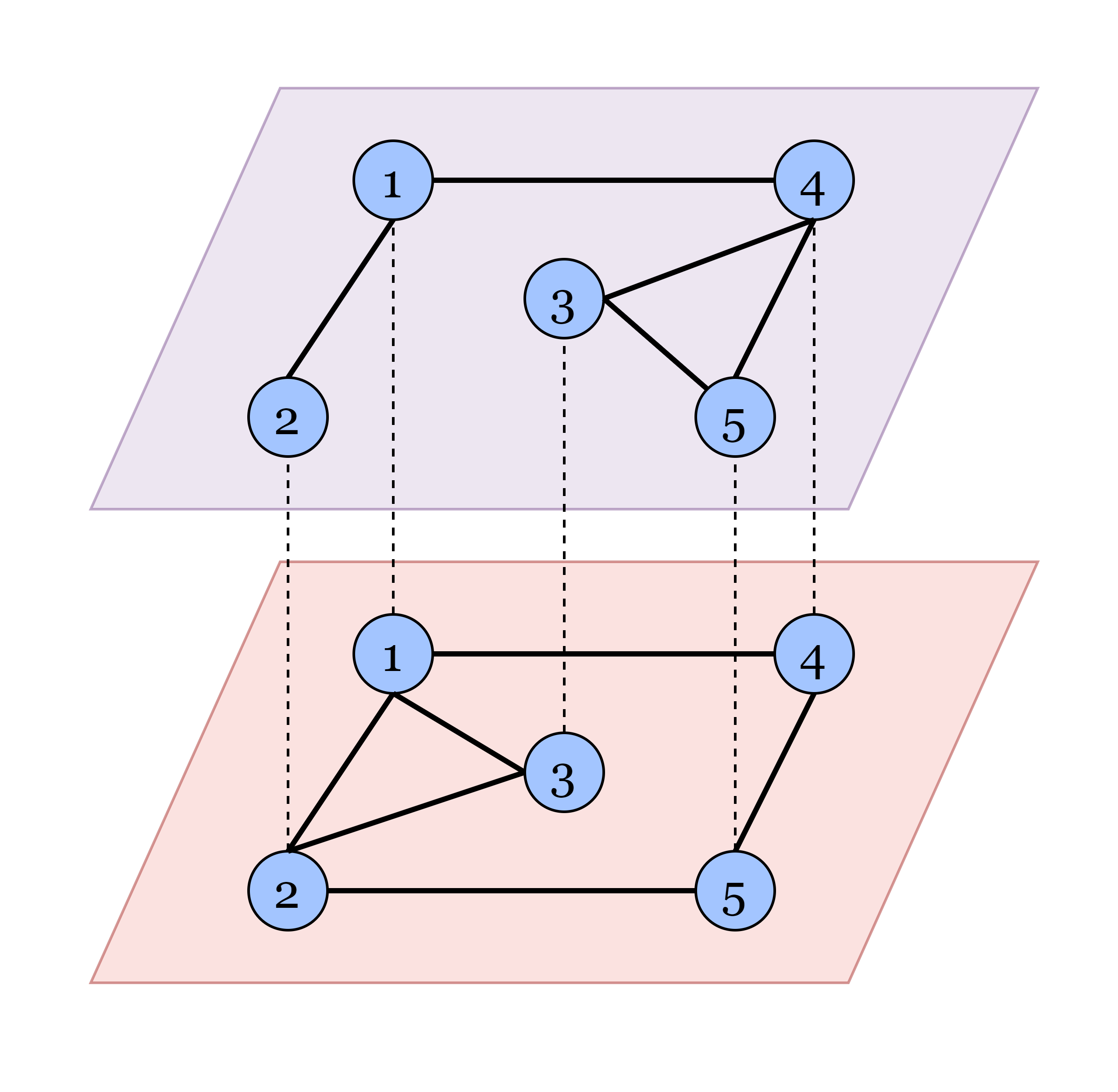}
		\label{figure:3b}
	}
	\caption[Figure 3]{An example of a general (a) multilayer network with five nodes $V= \{0,1,2,3,4\}$ and two aspects, which have the corresponding elementary-layers sets $L_1 = \{A,B\}$ and $L_2 = \{X,Y\}$; (b) multiplex network with five nodes $V= \{0,1,2,3,4\}$ and one aspect. In both figures, solid lines represent intra-layer edges and dashed lines represent inter-layer edges.}
\end{figure}

A multilayer network is defined as a quadruplet $M = (V_M, E_M, V, \boldsymbol{L})$ where 
$\boldsymbol{L} = \{L_a\}^s_{a=1}$ are sets of elementary layers (that is, there is one set of elementary layers $L_a$ for each \textit{aspect} $a$, $s$ is the number of \textit{aspects}), $V$ is a set of nodes of $M,$ $V_M \subseteq V \times L_1 \times \ldots \times L_s$ is a set of the node-layer combinations in which a node is present in the corresponding layer,
and $E_M \subseteq V_M \times V_M$ is a set of edges that contains the pairs of possible combinations of nodes and elementary layers~\citep{kivela2014multilayer}. 
A general simple representation of a multilayer network of two aspects can be seen in Figure~\ref{figure:3a}. 
Two particular cases of multilayer networks are the \textit{monoplex network}, when $s=0$ and the multilayer reduces to a single-layer graph; and the \textit{multiplex network}, when $M$ is a sequence of $l$ graphs, $\{G_\alpha\}_{\alpha=1}^l = \{(V_\alpha, E_\alpha)\}_{\alpha=1}^l$, where, usually, the node sets are the same across the different layers and the connections between nodes of different layers is only to its counterparts~\citep{boccaletti2014structure}; an example representation of a multiplex network can be seen in Figure~\ref{figure:3b}. 
\vspace{-1.5mm}

An unweighted multilayer network with $s$ aspects and the same number of nodes in each layer has an \textit{associated adjacency tensor} $\mathcal{A}$ of order $2(s+1)$. Analogous to the adjacency matrix, each directed edge in $E_M$ is associated with a $1$ (or a weight $w$) entry of $\mathcal{A}$ (or two entries in undirected edges case) and the other entries are $0.$ If the network does not have the same number of nodes in each layer, we can consider empty nodes to complete~\citep{porter2018multilayer}. 
Another representation is flattening $\mathcal{A}$ into a \textit{supra-adjacency matrix} $\boldsymbol{A}$, where intra-layer edges are associated with entries on the diagonal blocks and inter-layer edges are associated with matrix entries on the off-diagonal blocks. 

The analysis of multilayer networks is still a recent and not very mature topic, with most of the research being applied to networks with a single aspect (i.e., $s = 1$), and in the context of multiplex networks~\citep{kivela2014multilayer,porter2018multilayer}.

\par\null

\section{Mapping Univariate Time Series into Complex Networks}

In this section, we review the state of the art of univariate time series mappings into the network domain. As previously represented  in Figure~\ref{figure:1}, from a high-level perspective, and as in previous papers~\citep{donner2011recurrence,zou2018complex}, we divide the existing approaches into three different main classes, according to the underlying mapping concept. An overview of the main characteristics of the methods here described is given in Table~\ref{table:1}. In the following sections we will give an in-depth description of how these approaches emerged, how they work, what variants exist, and what are the core results.

\begin{table}[!ht]
    \centering
    \footnotesize
    \renewcommand{\arraystretch}{0.6}
    \begin{tabular}{|c|l|c|c|c|c|c|c|}
        \cline{2-8}
         \multicolumn{1}{l|}{} & \begin{tabular}{@{}l@{}}\textbf{Network Type} \\ \scriptsize(original reference)\end{tabular} & 
         \rotatebox[origin=c]{90}{\textbf{Node}} & \rotatebox[origin=c]{90}{\textbf{Edge}} & \rotatebox[origin=c]{90}{\textbf{Directed}} & \rotatebox[origin=c]{90}{\textbf{Weighted}} & \rotatebox[origin=c]{90}{\textbf{Param-free}} & \rotatebox[origin=c]{90}{\textbf{Pre proc}} \\
        \hline
        
        \parbox[c]{2mm}{\multirow{16}{*}{\rotatebox[origin=c]{90}{\textbf{Visibility}}}}
            \rule{0pt}{11pt} & Natural Visibility Network & $t$ & \multirow{2}{*}{NV} & \multirow{2}{*}{\xmark} & \multirow{2}{*}{\xmark} & \multirow{2}{*}{\cmark} & \multirow{2}{*}{\xmark} \\
            & \tiny~\citep{Lacasa2008} & & & & & &  \\
            
            \rule{0pt}{9pt} & Horizontal Visibility Network & $t$ & \multirow{2}{*}{HV} & \multirow{2}{*}{\xmark} & \multirow{2}{*}{\xmark} & \multirow{2}{*}{\cmark} & \multirow{2}{*}{\xmark} \\
            & \tiny~\citep{Luque2009} & & & & & &  \\
            
            \rule{0pt}{9pt} & Direct Horizontal Visibility Network & $t$ & \multirow{2}{*}{HV} & \multirow{2}{*}{\cmark} & \multirow{2}{*}{\xmark} & \multirow{2}{*}{\cmark} & \multirow{2}{*}{\xmark} \\
            & \tiny~\citep{lacasa2012time} & & & & & &  \\
            
            \rule{0pt}{9pt} & Limited Penetrable Natural Visibility Network & $t$ & \multirow{2}{*}{NV} & \multirow{2}{*}{\xmark} & \multirow{2}{*}{\xmark} & \multirow{2}{*}{\xmark} & \multirow{2}{*}{\xmark} \\
            & \tiny~\citep{ning2012limited} & & & & & &  \\
            
            \rule{0pt}{9pt} & Parametric Natural Visibility Network & $t$ & \multirow{2}{*}{NV} & \multirow{2}{*}{\cmark} & \multirow{2}{*}{\cmark} & \multirow{2}{*}{\xmark} & \multirow{2}{*}{\xmark} \\
            & \tiny~\citep{Bezsudnov2014} & & & & & &  \\
            
            \rule{0pt}{9pt} & Difference Visibility Network & $t$ & \multirow{2}{*}{V} & \multirow{2}{*}{\xmark} & \multirow{2}{*}{\xmark} & \multirow{2}{*}{\cmark} & \multirow{2}{*}{\xmark} \\
            & \tiny~\citep{zhu2014analysis} & & & & & &  \\
            
            \rule{0pt}{9pt} & Weighted Visibility Network & $t$ & \multirow{2}{*}{NV} & \multirow{2}{*}{\cmark} & \multirow{2}{*}{\cmark} & \multirow{2}{*}{\cmark} & \multirow{2}{*}{\xmark} \\
            & \tiny~\citep{Supriya2016} & & & & & &  \\
            
            \rule{0pt}{9pt} & Limited Penetrable Horizontal Visibility Network & $t$ & \multirow{2}{*}{HV} & \multirow{2}{*}{\xmark} & \multirow{2}{*}{\xmark} & \multirow{2}{*}{\xmark} & \multirow{2}{*}{\xmark} \\
            & \tiny~\citep{gao2016multiscale} & & & & & &  \\
         
        \hline
        
        \parbox[c]{2mm}{\multirow{8}{*}{\rotatebox[origin=c]{90}{\textbf{Transition}}}}
            \rule{0pt}{11pt} & Coarse-Grained Phase Space Network & \multirow{2}{*}{$\vec{z}_i$} & \multirow{2}{*}{TP} & \multirow{2}{*}{\cmark} & \multirow{2}{*}{\cmark} & \multirow{2}{*}{\xmark} & \multirow{2}{*}{PS} \\
            & \tiny~\citep{Gao2009} & & & & & &  \\
           
            \rule{0pt}{9pt} & Quantile Network & \multirow{2}{*}{$q_i$} & \multirow{2}{*}{TP} & \multirow{2}{*}{\cmark} & \multirow{2}{*}{\cmark} & \multirow{2}{*}{\xmark} & \multirow{2}{*}{DS} \\
            & \tiny~\citep{Campanharo2011} & & & & & & \\
           
            \rule{0pt}{9pt} & Ordinal Partitions Network & \multirow{2}{*}{$\pi_i$} & \multirow{2}{*}{TP} & \multirow{2}{*}{\cmark} & \multirow{2}{*}{\cmark} & \multirow{2}{*}{\xmark} & \multirow{2}{*}{AR} \\
            & \tiny~\citep{small2013complex} & & & & & &  \\
            
            \rule{0pt}{9pt} & Visibility Graphlets Network & \multirow{2}{*}{$G_i$} & \multirow{2}{*}{TP} & \multirow{2}{*}{\cmark} & \multirow{2}{*}{\cmark} & \multirow{2}{*}{\cmark} & \multirow{2}{*}{PS;VG} \\
            & \tiny~\citep{stephen2015visibility} & & & & & &  \\
        
        \hline
        
        \parbox[c]{2mm}{\multirow{10}{*}{\rotatebox[origin=c]{90}{\textbf{Proximity}}}}
            \rule{0pt}{11pt} & Cycle Network & \multirow{2}{*}{$c_i$} & \multirow{2}{*}{CM} & \multirow{2}{*}{\xmark} & \multirow{2}{*}{\xmark} & \multirow{2}{*}{\xmark} & \multirow{2}{*}{DC} \\
            & \tiny~\citep{zhang2006detecting} & & & & & &  \\
            
            \rule{0pt}{9pt} & Correlation Network & \multirow{2}{*}{$\vec{z}_i$} & \multirow{2}{*}{CM}  & \multirow{2}{*}{\xmark} & \multirow{2}{*}{\cmark} & \multirow{2}{*}{\xmark} & \multirow{2}{*}{PS} \\
            & \tiny~\citep{yang2008complex} & & & & & &  \\

            \rule{0pt}{9pt} & $\kappa$-Nearest Neighbor Network & \multirow{2}{*}{$\vec{z}_i$} & \multirow{2}{*}{DM} & \multirow{2}{*}{\cmark} & \multirow{2}{*}{\xmark} & \multirow{2}{*}{\xmark} & \multirow{2}{*}{PS} \\
            & \tiny~\citep{small2009transforming} & & & & & &  \\
            
            \rule{0pt}{9pt} & Adaptive Nearest Neighbor Network & \multirow{2}{*}{$\vec{z}_i$} & \multirow{2}{*}{DM} & \multirow{2}{*}{\xmark} & \multirow{2}{*}{\xmark} & \multirow{2}{*}{\xmark} & \multirow{2}{*}{PS} \\
            & \tiny~\citep{Xu2008,small2009transforming,donner2011recurrence} & & & & & &  \\
            
            \rule{0pt}{9pt} & $\epsilon$-Recurrence Network & \multirow{2}{*}{$\vec{z}_i$} & \multirow{2}{*}{DM} & \multirow{2}{*}{\xmark} & \multirow{2}{*}{\xmark} & \multirow{2}{*}{\xmark} & \multirow{2}{*}{PS} \\
            & \tiny~\citep{Donner2010} & & & & & &  \\
         
        \hline
        
    \end{tabular}
    \caption{Comparison of (univariate) time series mappings based on the properties of the corresponding algorithms and of the resulting networks. Notation: NV - natural visibility, HV - horizontal visibility, V - natural and horizontal visibility, TP - transition probability, CM - correlation measures, DM - distance measures, PS - phase space reconstruction, DS - division of the support, AR - amplitude rank, VG - directed visibility graph, DC - division into cycles.}
    \label{table:1}
\end{table}

\subsection{Visibility Networks}

{\label{section.visnet}}

Lacasa and co-authors~\citep{Lacasa2008} proposed visibility  mappings from (univariate) time series to complex networks,  based on traditional visibility algorithms from computational geometry~\citep{ghosh2007visibility}. 
The visibility networks have a geometric criterion associated with the natural ordering of the time series. 
Given its ease of interpretation and implementation, this mapping quickly attracted the research community and several results as well as variants of these methods began to emerge. 
Visibility-based algorithms incorporate, as we shall see, global and local topological characteristics of the  time series in the graph characteristics, are easy to implement,  computationally fast, and, except for the limited and parametric versions,  are  parameter-free. 

Throughout this section we present the different types of visibility algorithms proposed in the literature, which are essentially based on the natural visibility~\citep{Lacasa2008} and horizontal visibility~\citep{Luque2009} algorithms.

\subsubsection{Natural Visibility Graphs}

{\label{nvg_alg}}

\begin{figure}[hbt!]
	\centering 
	\subfloat[Toy time series]{
		\includegraphics[scale = 0.6,keepaspectratio,valign=m]{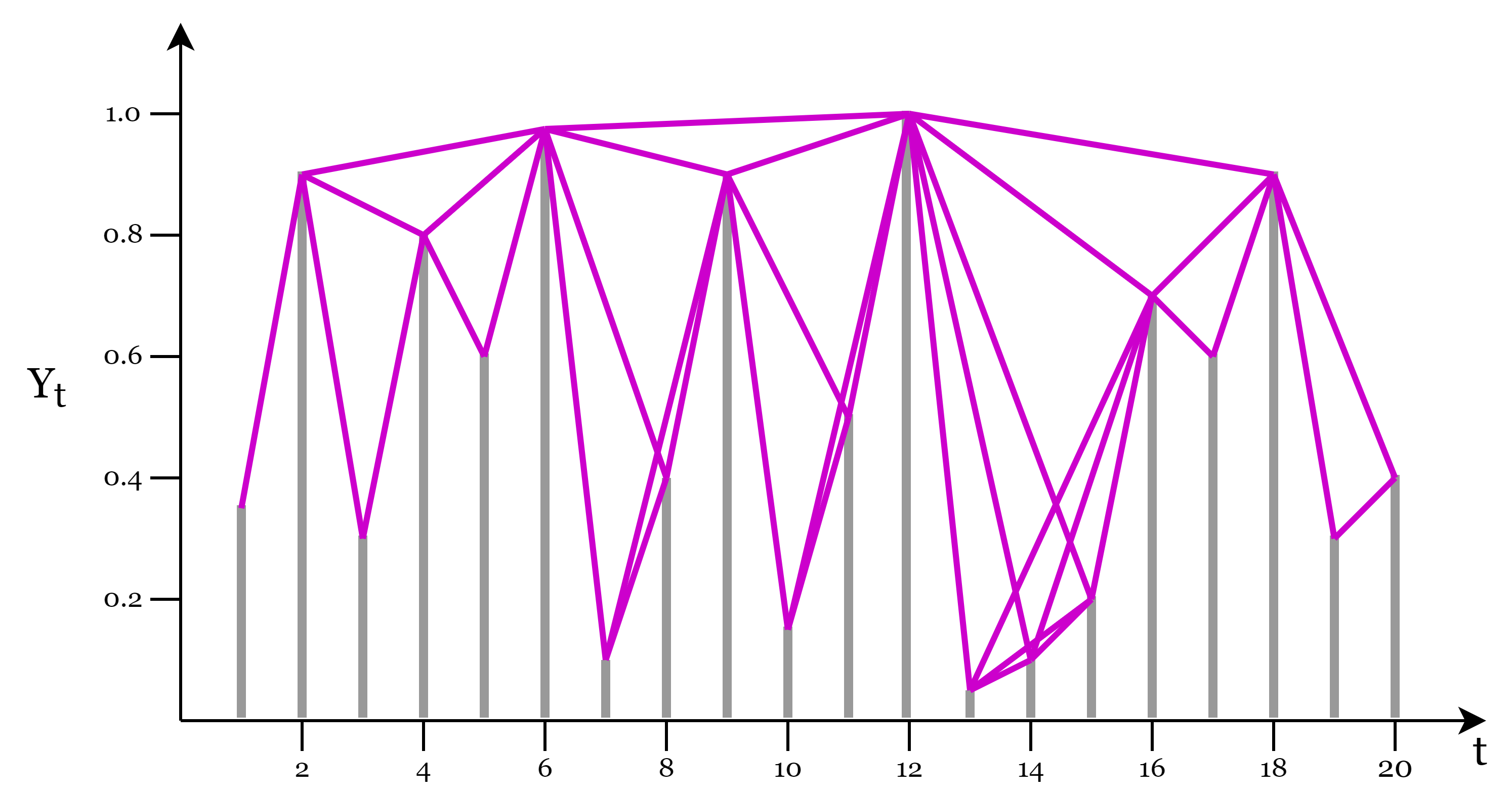}
		\label{figure:4a}
	}
	\quad
	\quad
	\subfloat[NVG]{
		\includegraphics[scale = 0.25,keepaspectratio,valign=m]{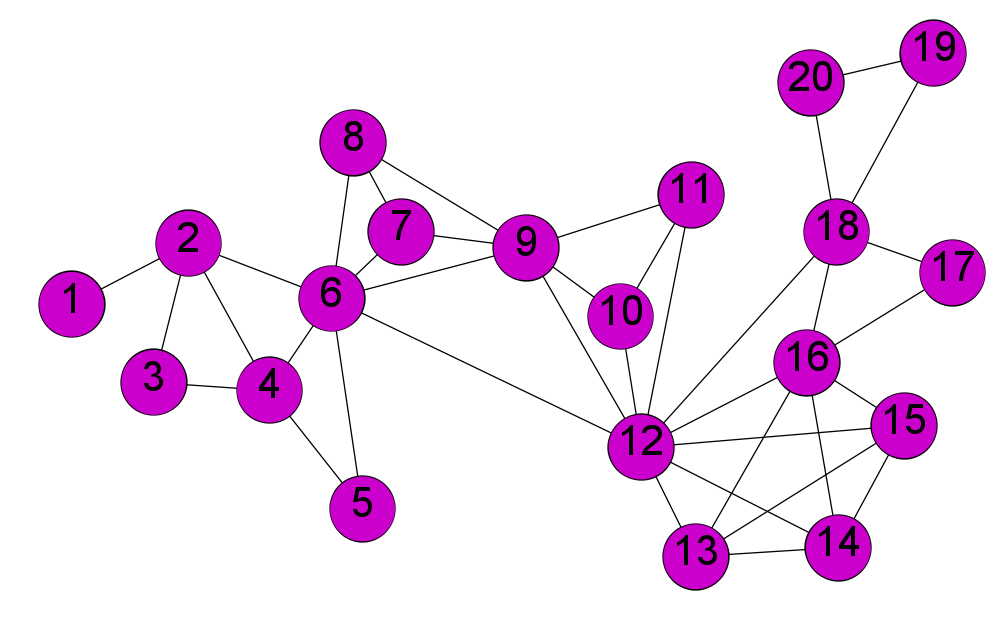}
	}
	\caption[Figure 4]{On the left side, we present the plot of a toy time series and, on the right side, the network generated by the natural visibility algorithm. The purple lines in the time series plot represent the lines of visibility (and hence the edges of the graph) between data points.}
	\label{figure:4}
\end{figure}

Lacasa and co-authors~\citep{Lacasa2008} proposed the first method based on the concept of visibility, the natural visibility graph (NVG), or simply visibility graph (VG). This method is based on the idea that each observation of the time series is seen as a vertical bar with height equal to the  numerical value of the observation and that  these vertical bars are laid in a landscape, the top of a bar is visible from the tops of other bars. Each node in the graph corresponds to a time stamp $t$ of the time series, so the nodes are serially ordered. Two nodes are connected if there is a line of visibility between the corresponding data bars that is not intercepted. 
This idea is illustrated in Figure~\ref{figure:4a}.
\vspace{-1.5mm}

Formally,  the set of nodes $\{v_i\}$ of a NVG are numbered sequentially in time and two nodes $v_i$ and $v_j$ are connected (have visibility) if any other observation $(t_k, Y_k)$ with $t_i<t_k<t_j$ satisfies:
\begin{equation}
	Y_k < Y_j+(Y_i-Y_j)\frac{(t_j-t_k)}{(t_j-t_i)}.
\end{equation}

The NVG algorithm is easy to implement and, in terms of computational complexity, 
has quadratic  complexity ($\mathcal{O}(T^2)$) thus becomes very slow for  very long time series. However, a more efficient algorithm based on the divide and conquer technique  proposed in~\cite{Lan2015}  has a  complexity of $\mathcal{O}(T\log T)$.

Visibility graphs are always connected because each node $v_i$ sees at least its neighbors, $v_{i-1}$ and $v_{i+1}$ and  are always undirected.
However, a direction can be defined considering the direction of the time axis.
The network is also invariant under affine transformations of the  data~\citep{Lacasa2008} because the visibility criterion is invariant under rescheduling of both the horizontal and vertical axis, as well as in vector translations, that is, each transformation $Y'= aY + b,$ for $a \in \mathbb{R}$ and $b \in \mathbb{R},$ leads to the same NVG.

A NVG has, typically, a distinct topology characterized by hubs representing maxima of the time series, since these have visibility contact with more nodes when compared to other points~\citep{donner2012visibility}. 
However, this is not  always necessarily true since  a process  with a concave behavior over a certain period of time, as for example the Conway series~\citep{Lacasa2008},  can lead to highly connected nodes that do not coincide with local maxima~\citep{donner2012visibility}.
The presence of  hubs that correspond to (local) maxima of time series typically give rise to a topological structure represented by communities. 
These communities usually reflect the temporal order of the observations. The communities may also be formed by  fragments of the time series, in which case do not reflect the temporal order of the observations~\citep{yao2017visibility}. 

\newpage
Lacasa and co-authors~\citep{Lacasa2008} have shown that NVGs inherit several structural properties of the time series. In fact, the method maps periodic series~\footnote{Periodic time series are series where the process has a regular seasonal pattern. However, there are many processes where the data appears highly periodic, but does not repeat itself exactly. 
These processes are called pseudo-periodic time series.} 
on regular graphs, random series on random graphs~\footnote{Random networks are graphs obtained by a random process or by a probability distribution~\citep{erdHos1960evolution}. 
There are many random graph models, but the \textit{Erdős Rényi model}~\citep{erdHos1960evolution} is the most fundamental and widely studied of them~\citep{newman2010networks}. This model is mathematically referred by $G(n,p)$ where $n$ is the number of nodes and $p$ is the probability with that the edges between each distinct pair of nodes are connected. 
In this model the probability that a node has $k$ edges (degree distribution) follows a Poisson distribution 
$P(k) = \frac{e^\lambda \lambda^k}{k!}$~\citep{barabasi1999emergence}.} 
and fractal series on scale-free graphs~\footnote{Scale-free networks have a degree distribution (the probability $P(k)$ that a given node has a degree, i.e., a number of edges that connects the node to the other nodes, equal to $k$) that follows the power law $P(k) \sim k^\gamma$. This means that most nodes have few edges in contrast to the existence of some nodes with a high degree (hubs)~\citep{barabasi1999emergence}.}. 
In particular, for periodic series, the graphs have a regular structure where the degree distribution presents a number of peaks related to the  period of the series. 
For white noise processes the resulting graph is completely random, producing a degree distribution that is an exponential function.
For fractal time series, the degree distribution is a power law related to the fractality of series.

In~\cite{Lacasa2009}, the authors used the NVGs to quantify long-range dependence and fractality in time series. 
These authors  concluded that fractal processes result in scale-free visibility graphs with a degree distribution that is a power law $P(k) \sim k^{\gamma}$ with exponent $\gamma$ which is a linear function of the Hurst coefficient. 
These results were applied to study the fractality of energy dissipation rates in turbulent flows~\citep{liu2010statistical}, to study turbulent heated jets behaviors~\citep{charakopoulos2014application} and to diagnose Alzheimer’s disease automatically based on the data electroencephalography (EEG)~\citep{ahmadlou2010new}, among others. 

Recently, a new definition of motif~\footnote{Network \textit{motifs}~\citep{milo2002network} are small subgraphs (typically with 3 to 5 nodes) originally defined as patterns that occur more often than expected, that is, whose frequency is higher than in randomized networks; nowadays the term motif is also used to refer to all such small patterns, regardless of its overrepresentation, and their frequencies can serve as a rich network fingerprint~\citep{milo2004superfamilies}.} 
in the context of NVG was presented in~\cite{iacovacci2016sequential}:  \textit{sequential NVG $n$-node motifs}. According to this definition, motifs are the set of NVG subgraphs formed by the sequence of nodes $\{g, g+1, \ldots, g+n-1\}$, where $n < T$ and $g \in [1, T-n+1]$). 
This  definition differs from traditional network motifs in that they require nodes to be labeled according to the  temporal order  of the nodes induced by the construction of NVG. One advantage of this definition is that the computation of these motifs is extremely efficient, linear time $\mathcal{O}(T)$, using the theory developed in~\cite{Iacovacci2016}. 
Similarly to  traditional motifs networks, we can compare the relative occurrence of each motif to distinguish different time series processes, see more details in~\cite{iacovacci2016sequential}. 

Donner and Donges~\citep{donner2012visibility} address the  problem of missing data in time series. 
They assessed the effect that missing data produces in  the topological properties of NVGs associated with Gaussian white noise processes, using two strategies: missing data is simply ignored in the generation of the NVG;  the NVG is fragmented into subgraphs corresponding to times before and after the missing observations.~\cite{donner2012visibility} concluded that ignoring the missing data  had a considerable effect mainly on closeness centrality but not on degree distribution and local clustering coefficient, as well. When the missing data is taken into account and the NVG becomes a set of subgraphs, path-based metrics are the most impacted. As expected the impacts become stronger as the missing data increases. 
Interestingly, those authors also found out that the effects of node-level metrics are slightly reduced for  missing data occurring in runs  rather than randomly. 

Given the potential demonstrated, the method was applied in the study of energy dissipation rates in three-dimensional turbulence~\citep{liu2010statistical},  financial time series~\citep{Yang2009,qian2010universal,long2013visibility,Zhuang2014},  heart rate variability~\citep{Shao2010,hou2016visibility},  classification of sleep stages~\citep{zhu2014analysis},  classification of  time series~\citep{Li2018},  hurricane occurrence in the United States~\citep{elsner2009visibility} and to perform  forecasting~\citep{zhang2017novel,zhang2018forecasting}.

\newpage
Note that NVGs are not able to  distinguish time series with certain traits. As an example, consider a  time series with  a deterministic increasing trend. Its NVG and the NVG of its symmetric time series which presents a deterministic decreasing trend are similar  graphs with the same properties. This disadvantage may be overcome defining NVGs with directed edges.

\subsubsection{Horizontal Visibility Graphs}

{\label{subsec:hvg}}

\begin{figure}[hbt!]
	\centering 
	\subfloat[Toy time series]{
		\centering
		\includegraphics[scale = 0.6,keepaspectratio,valign=m]{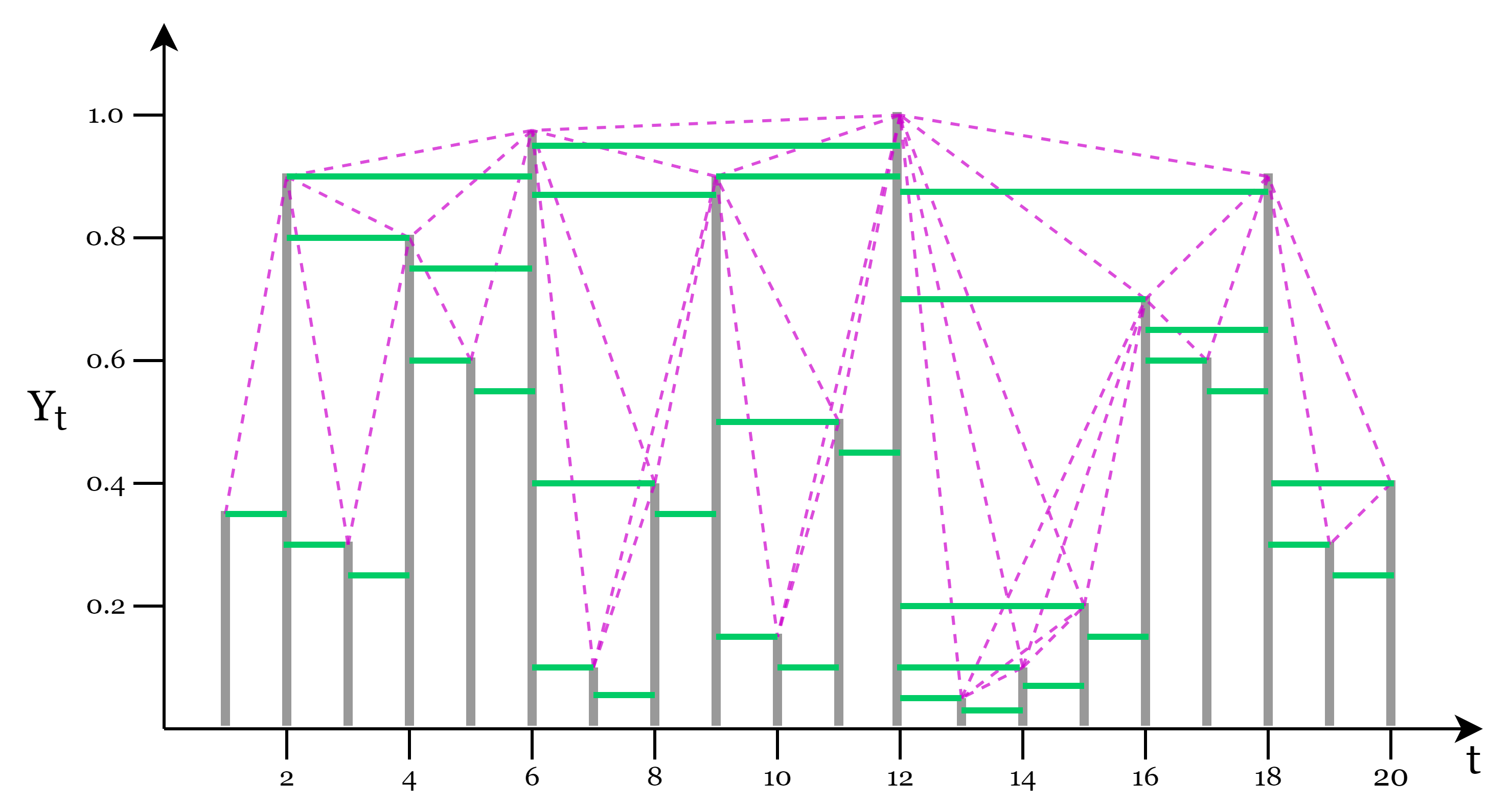}
	}
	\quad
	\quad
	\subfloat[HVG]{
		\centering
		\includegraphics[scale = 0.25,keepaspectratio,valign=m]{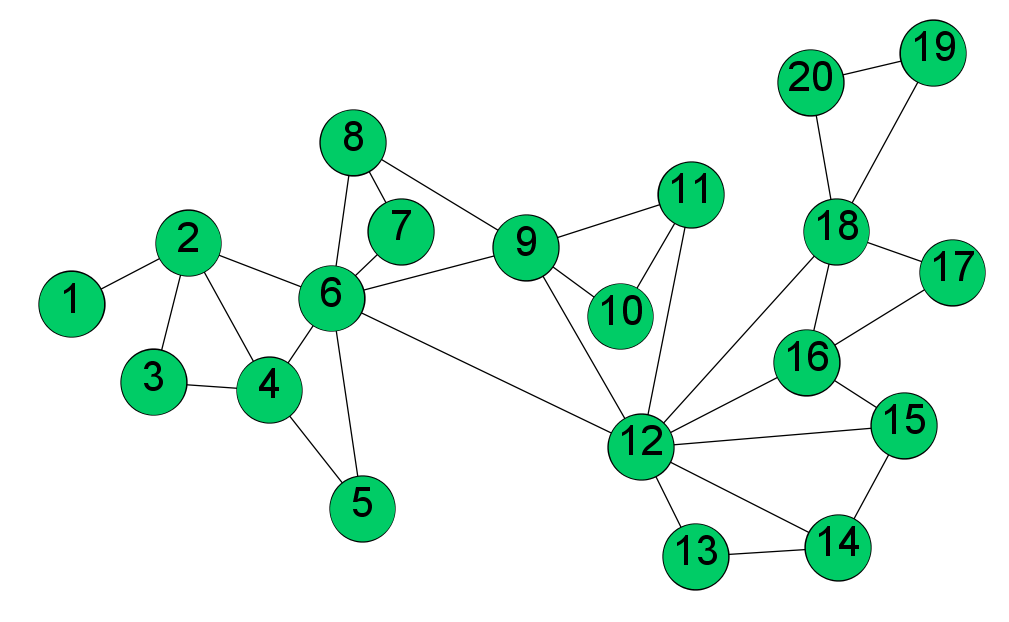}
	} 
	\caption[Figure 5]{On the left side, we present the plot of a toy time series and, on the right side, the network generated by the horizontal visibility algorithm. The green lines represent the horizontal lines of visibility between the data points and the purple lines the natural visibility, to comparison.}
	\label{figure:5}
\end{figure}

In order to reduce the computational complexity associated with NVGs, Luque and co-authors~\citep{Luque2009} proposed a simplified NVG method called the horizontal visibility graph, HVG, which inherits all NVG graph characteristics mentioned above. The construction of HVGs differs from that of NVGs in that the visibility lines are only horizontal. 
Figure~\ref{figure:5} gives a simple illustration of this algorithm, with a toy time series and the resulting network. 
\vspace{-1.5mm}

Formally, two nodes $v_i$ and $v_j$ are connected, if the following condition is met:
\begin{equation}
	Y_i, Y_j > Y_k, 
\end{equation}
for all ($t_k,Y_k)$ such that $t_i < t_k < t_j$. 

In terms of computational complexity, the generation of HVGs  has time complexity of the construction $\mathcal{O}(T \log T)$ \citep{yela2020online} and, in the case of noisy (stochastic or chaotic) processes the algorithm has an average-case time complexity $\mathcal{O}(T)$~\citep{Luque2009}.

The HVG is always a subgraph of the NVG for a particular time series. This is well illustrated in  Figures~\ref{figure:4} and~\ref{figure:5} where we can easily verify that all the edges present in the HVG are present in NVG, but  the converse is not true, e.g. edges $(6,8)$ and $(12,15).$ 
 HVG nodes will always have a degree less than or equal to that of  the corresponding NVG nodes. Therefore,  there is some loss of quantitative information in HVGs in comparison with NVGs~\citep{Luque2009} which may be crucial in the analysis of certain time series. 
However, it has no impact on the qualitative characteristics of the graphs, since the graphs preserve some part of the data information, namely the local information (the closest time stamps). 
Another characteristic of HVGs is that it is always an outerplanar graph~\footnote{An outerplanar graph is a planar graph, (i.e., it can be drawn on the plane so that no edges cross each other), where all nodes are incident to the infinite face, that is, no node is totally surrounded by edges.} and has a Hamilton path, that is, a path that passes through all the nodes of a graph only once. In fact,~\cite{gutin2011characterization} have shown that this is a necessary and sufficient condition for a graph to be  an HVG.

Luque and co-authors~\citep{Luque2009} have formally established several relationships between properties of HVGs and characteristics of the underlying time series.   

The first such property states that uncorrelated random series is mapped into an HVG with an exponential degree distribution given by:
\begin{equation}
	P(k) = \frac{1}{3}\left(\frac{2}{3}\right)^{k-2}, \quad \textrm{for } k \geq 2.
	\label{eq:deg_dist}
\end{equation}
This result holds for all  probability distributions. 
It follows that, for uncorrelated random series, the associated average degree is~\citep{nunez2012detecting}:
\begin{equation}
	\bar{k} = \sum kP(k) = \sum_{k=2}^\infty \frac{k}{3} \left( \frac{2}{3}\right)^{k-2} = 4.
	\label{eq:avg_deg}
\end{equation}
Additionally,~\cite{Lacasa2010} suggested that the degree distribution  follows the exponential law, $P(k) \sim \exp^{(- \lambda k)}$, where the value of $\lambda$ depends on the type of process generating the time series:  $\lambda < \ln(\frac{3}{2})$ for chaotic processes,  $\lambda > \ln(\frac{3}{2})$ for stochastic processes and  $\lambda = \ln(\frac{3}{2})$ for uncorrelated processes. 
However, note that~\cite{ravetti2014distinguishing}  found that for the R$\ddot{\text{o}}$ssler system~\footnote{The R$\ddot{\text{o}}$ssler system is a complex system of three ordinary differential equations~\citep{rossler1976equation}, which define a continuous-time process that exhibits chaotic behavior, where the predictability of the behavior decreases exponentially with lead time.} the rule is not valid, and~\cite{zhang2017visibility} found results in which negatively correlated processes lead to lower $\lambda$ values than the critical value. These results show that there are exceptions to the rule and that using $\lambda$ to distinguish chaotic from stochastic processes requires further investigation. 

A second property obtained by~\cite{Luque2009} refers to the relationship between hubs of the graph and the extreme values of the data. This equivalence was obtained numerically. 
Furthermore, based on geometric arguments,~\cite{Luque2009} obtained the following relation between the clustering coefficient, a measure to capture the degree to which the nodes of a graph tend to cluster, and the degree in an HVG: 
\begin{equation}
	C_i = \frac{k_i-1}{\binom{k_i}{2}} = \frac{2}{k_i},
\end{equation}
where $\binom{k_i}{2}$ denotes the number of possible triangles and $k_i-1$ is the number of visible nodes that are also visible from $v_i$. This relation allows also to deduce the local clustering coefficient distribution~\citep{nunez2012visibility}, substituting $k_i$ by $C_i$ in Equation~(\ref{eq:deg_dist}). 

The HVG associated to an infinite periodic series of period $P$ is a representation of a concatenation of a motif. Nuñez and co-authors~\citep{nunez2012detecting} proved that the average degree of the resulting HVG is given by:
\begin{equation}
	\bar{k} = 4 \left( 1 - \frac{1}{2P} \right).
	\label{eq:avg_deg2}
\end{equation} 
This result implies that  time series are mapped on HVGs with $2 \le \bar{k} \le 4$. The lower bound is reached for constant series (HVG is a chain graph and so each node is only connected to its two closest neighbors) and the upper bound for aperiodic series (random and chaotic process).

Nuñez and co-authors~\citep{nunez2012detecting} proposed a method for calculating the hidden periodicity in a periodic noisy signal, a very common problem in real world data analysis and difficult to solve  with traditional periodicity detection algorithms, such as spectral analysis. 
The method involves the construction of a modified HVG, called the filtered HVG (f-HVG), where two nodes $ v_i $ and $ v_j $ are connected if the following condition is meet:
\begin{equation}
	Y_i, Y_j > Y_k + f, 
\end{equation}
for all $t_i < t_k < t_j$ and $f \in [\min (Y_t), \max (Y_t)]$ is a real-valued scalar that acts as a filter. 
To filter the noise of a time series, its f-HVG is generated for increasing values of $f$, and at each step, the average degree is calculated. 
The result is a decrease of the average degree, with initial value $4$ (for $f=0$), an asymptotic value of $2$ (lower bound of $\bar{k}$) and the plateau obtained for the distribution allows to obtain the period through Equation~(\ref{eq:avg_deg2}). 
This method delivered very promising results including, specific cases where autocorrelation analysis yields misleading results.

The definition of sequential motif presented for NVGs in~\cite{iacovacci2016sequential} also apply similarly to HVGs. 
The definition of \textit{sequential HVG $n$-node motifs} is  the same way as that for an NVG. 
The frequency of sequential HVG motifs is also useful for discriminating different dynamic process. 

HVGs were used to analyze seismic signals~\citep{Telesca2012}, evaluate the complex dynamics of tourism systems~\citep{Baggio2016}, construct the so-called Feigenbaum graphs in order to study unimodal maps and their spectral properties~\citep{flanagan2019spectral}, study the volatility behavior of returns for financial time series~\citep{zhang2015volatility}, analyze heartbeat rates of healthy subjects, congestive heart failure subjects, and atrial fibrillation subjects~\citep{xie2019tetradic} and predicting catastrophes~\citep{zhang2018predicting}.

In a recent work,~\cite{Li2018} show that the topological properties of the NVGs and HVGs are useful features for constructing accurate classification models using generic classifiers for the classification of real time series, since combining both graphs allows capturing global (in the case of NVGs) and local (in the case of HVGs because they are more sensitive to local variations) features. This approach allows to increase the classification accuracy of a vast set of time series of different domains and it turns out to be better when compared to the traditional approaches, such as time series features obtained from common statistical measures.

Several variants of NVGs and HVGs have been proposed in the literature. The most relevant are presented in the following sections.

\subsubsection{Directed Visibility Graphs} 

{\label{dhvg_alg}}

\begin{figure}[hbt!]
	\centering 
	\subfloat[Toy time series]{
		\includegraphics[scale = 0.6,keepaspectratio,valign=m]{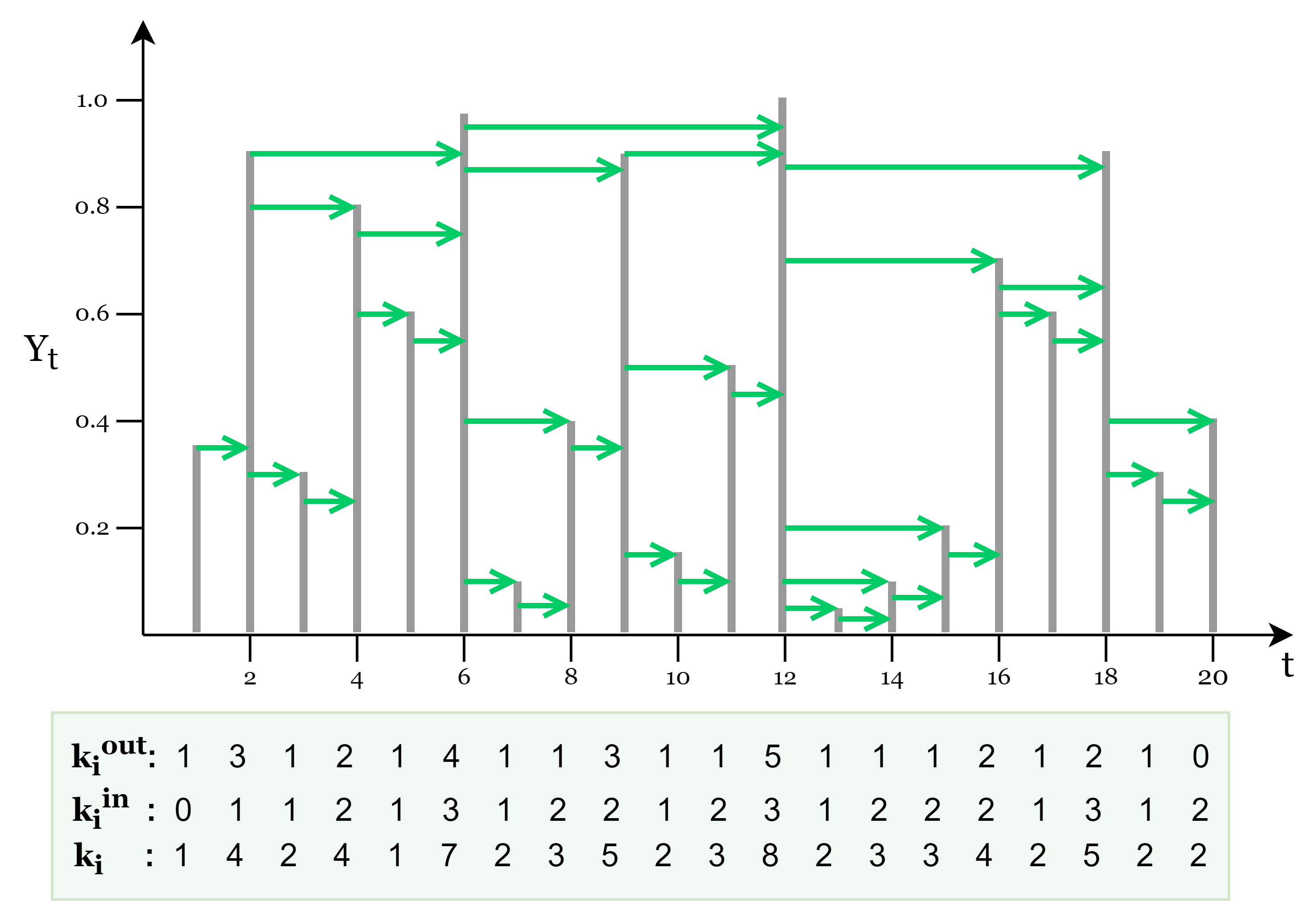}
	} 
	\quad
	\quad
	\subfloat[DHVG]{
		\centering
		\includegraphics[scale = 0.25,keepaspectratio,valign=m]{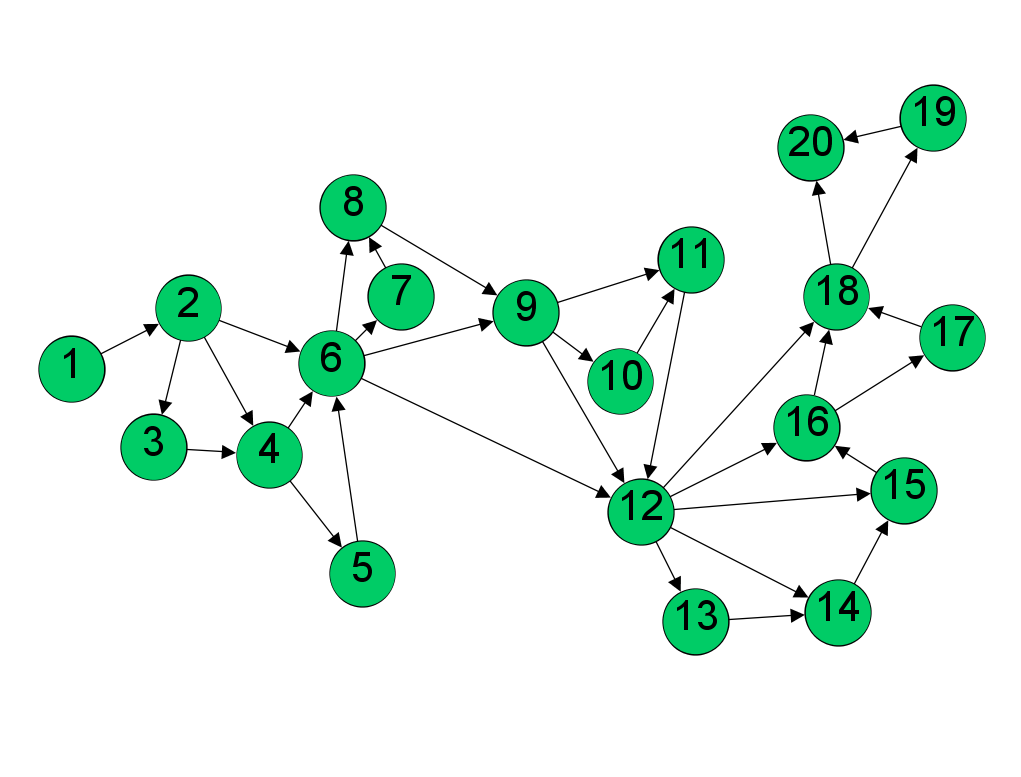}
	}
	\caption[Figure 6]{Illustrative example of directed horizontal visibility algorithm and corresponding \textit{out}-degree ($k_i^{out}$), \textit{in}-degree ($k_i^{in}$) and \textit{total}-degree ($k_i$). 
	On the left side, we present the plot of a toy time series and, on the right side, the network generated by the directed horizontal visibility algorithm. The green directed lines represent the directed horizontal lines of visibility between data points. }
	\label{figure:6}
\end{figure}

Given that time has a natural direction,~\cite{lacasa2012time} introduced directed horizontal visibility graphs, DHVG, by defining an HVG with edges $(v_i,v_j)$, $i < j.$ 
The adjacency matrix is no longer a symmetric matrix. 
An example of the representation of this algorithm is illustrated  in Figure~\ref{figure:6}. 
This variant is extended straightforwardly to the NVGs.
\vspace{-1.5mm}

\newpage
DHVG were proposed as a simple and well-defined tool for measuring time series irreversibility~\footnote{A stationary time series  is reversible if $\{ Y_1, \ldots, Y_T \}$ and $\{ Y_T, \ldots, Y_1 \}$ have the same joint probability distributions.} notoriously difficult to assess with traditional  algorithms, like algorithms that involve time series symbolization, which normally involve the choice of extra parameters and the results may depend on that choice. 
The degree of irreversibility of a time series is calculated as the Kullback-Leibler  distance between the \textit{in}- and \textit{out}-degree distributions, $P(k^{in})$ and $P(k^{out})$~\citep{lacasa2012time}. These authors proved that for a bi-infinite sequence of i.i.d. random variables, both the \textit{in} and \textit{out}-degree distributions of the corresponding DHVGs are equal and given by:
\begin{equation}
	P(k^{in}) = P(k^{out}) = \left( \frac{1}{2} \right)^k, \quad k = 1,2,3, \ldots .
\end{equation}
In the same line of research~\cite{donges2013testing} followed a similar approach. Donges and co-authors proposed a set of rigorous statistical tests for time series irreversibility based on both visibility algorithms (NVG and HVG).
The authors compare the degree and local clustering coefficient distributions taking into account the past ($P(k^r)$ and $ P(C^r)$) and the future ($P(k^a)$ and $ P(C^a)$) separately. It is conjectured that if a time series is reversible the distributions $P(k^r)$ and $P(k^a)$ (or $P(C^r)$ and $ P(C^a)$) should be similar, while for irreversible time series we should find statistically significant deviations between the distributions.

\subsubsection{Limited Penetrable Visibility Graphs} 

{\label{subsec:lpvg}}

\begin{figure}[hbt!]
	\centering 
	\subfloat[LPVG algorithm]{
		\includegraphics[scale = 0.6,keepaspectratio,valign=m]{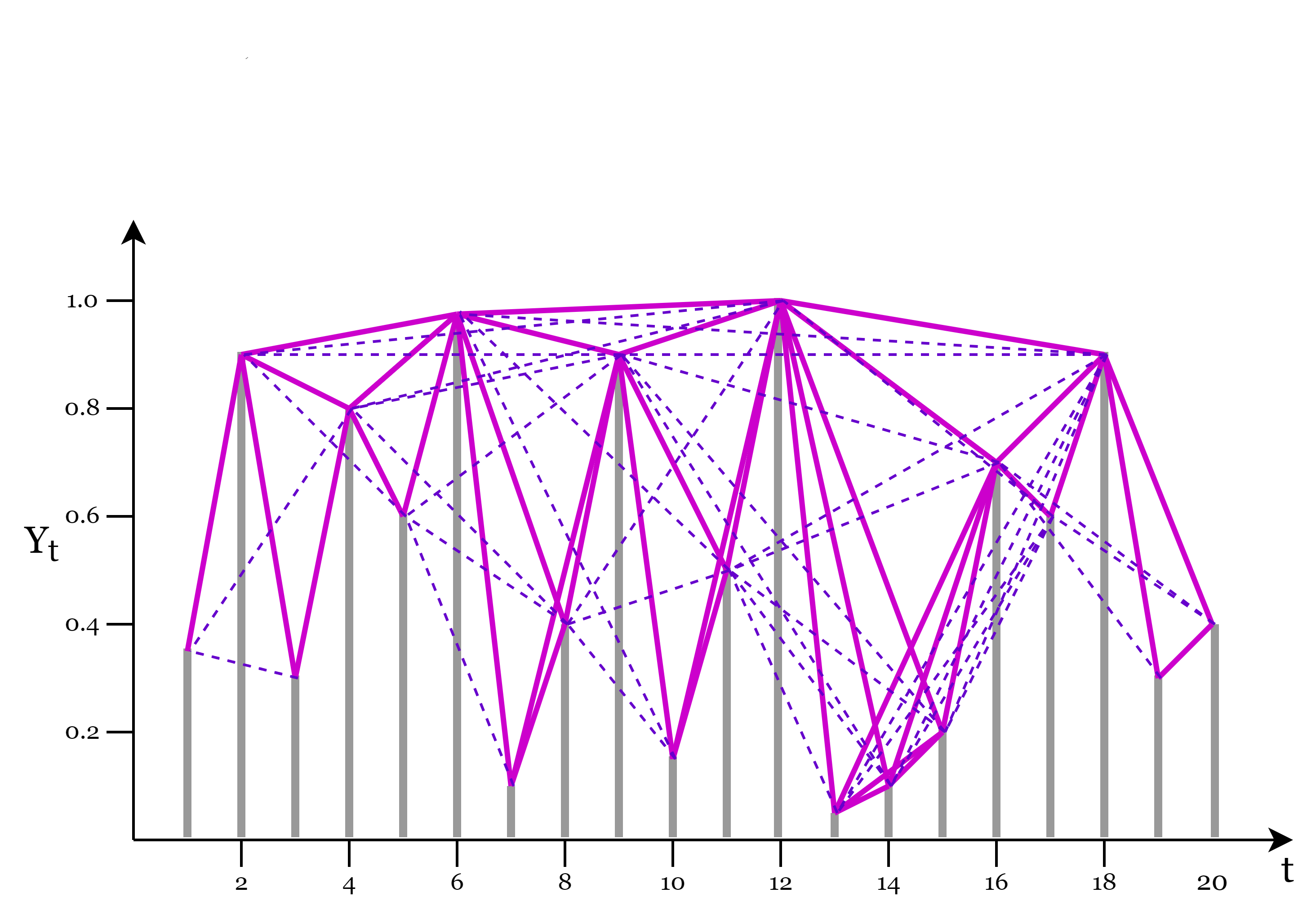}
		\label{figure:7a}
	} 
	\quad
	\quad
	\subfloat[PNVG vs. DNVG algorithm]{
		\centering
		\includegraphics[scale = 0.6,keepaspectratio,valign=m]{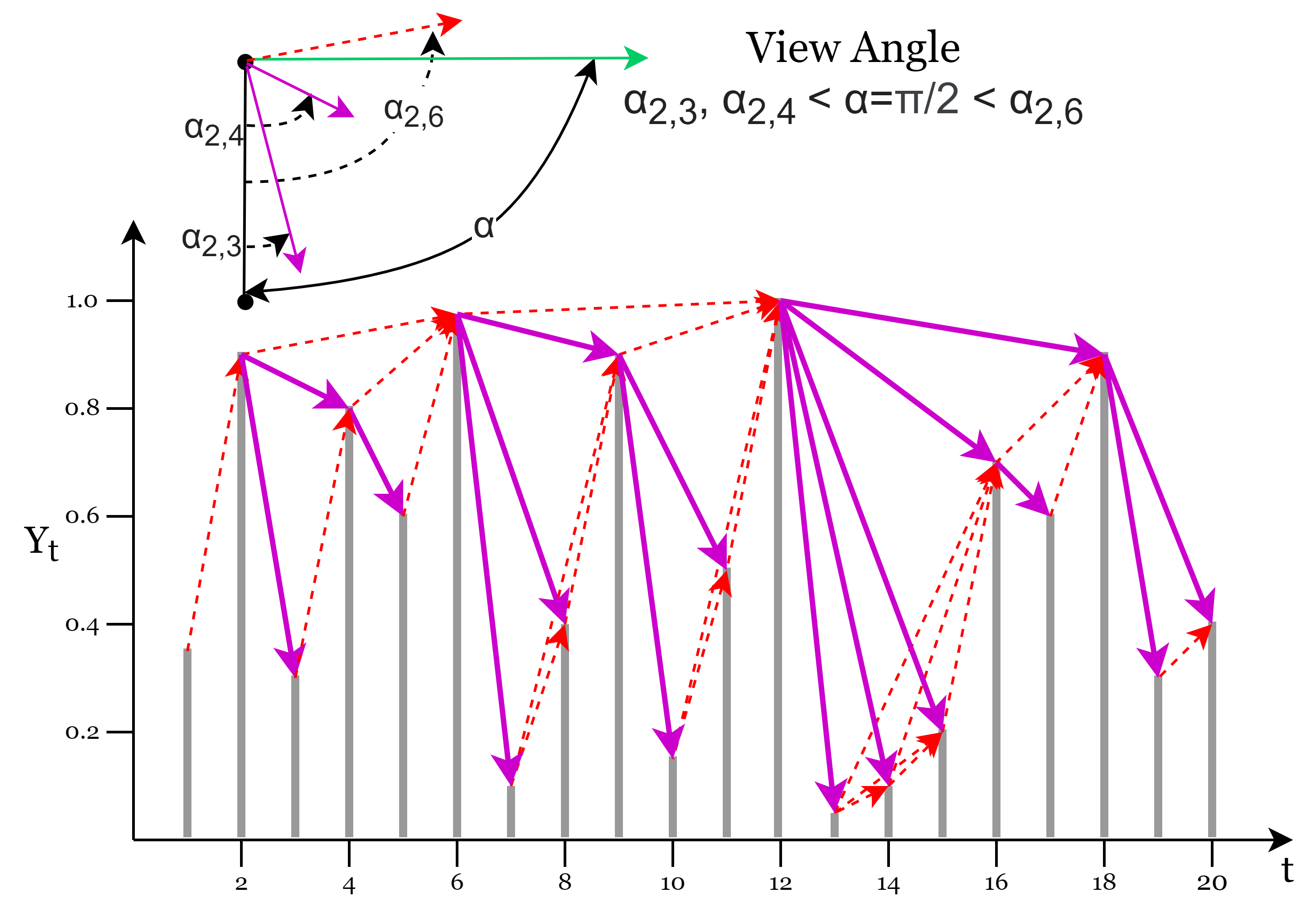}
		\label{figure:7b}
	}
	\caption[Figure 7]{(a) Illustrative example of limited penetrable visibility graph algorithm where the limit $l = 1$. The purple lines show the edges between points that have direct visibility (as in NVG, $l = 0$) and the blue dashed lines are the extra edges imposed by LPVG algorithm, where two points can be seen with only one higher intermediate point. (b) Illustrative example of parametric natural visibility graph algorithm with comparison with NVG algorithm. }
\end{figure}

The limited penetrable visibility graph, LPVG, was first proposed by~\cite{ning2012limited}  to improve the NVG and HVG mappings. 
Similarly to the NVG method, the nodes are numbered sequentially in time but now  have a limit $l$ of visibility and the nodes $v_i$ and $v_j$ are connected if:
\begin{equation}
	Y_{i+l} < Y_j+(Y_i-Y_j)\frac{t_j-(t_i+t_l)}{(t_j-t_i)}, \quad l < j-i,
\end{equation}
where $l$ is a limited penetrable distance established to reduce the effect of noise intrinsic to data. 
So, the nodes $v_i$ and $v_j$ have mutual visibility through $l$ intermediate bars (data) in the time series. 
The presence of noise in the data causes the connections on NVGs to break easily, while the nodes should, in principle, have more connections to other nodes, except for the maxima data~\citep{pei2014wlpvg}.
A small illustration of this method for $ l = 1 $ is represented Figure~\ref{figure:7a}.
\vspace{-1.5mm}

A limited penetrable version for HVG was developed by~\cite{gao2016multiscale},  and called limited penetrable horizontal visibility graph (LPHVG). 
The graphs are constructed using  the above algorithm, except that the visibility condition is horizontal visibility. 
Note that when $l=0$, the LPVG (LPHVG) is reduced to NVG (HVG). 
Other versions of this method exist, namely, a directed version, the directed limited penetrable horizontal visibility graph (DLPHVG), and an image version, the image limited penetrable horizontal visibility graph (ILPHVG)~\citep{wang2018degree}. 

LPVGs are very promising: as the NVGs they  inherit several properties of the time series and are able to detect differences between random and chaotic series, detect the location of inverse bifurcations in chaotic dynamical systems and, in addition, as an advantage over NVGs, they show good tolerance to noise interference~\citep{wang2016functional}. 
However, it is necessary to pay attention to the choice of parameter $l$ so that not too much information is incorporated in the graphs. 

The idea of limited natural visibility was applied to analyze  abnormalities of EEG signals from  Alzheimer’s disease~\citep{wang2016functional}, EEG signals under manual acupuncture~\citep{pei2014wlpvg}, and signals from an electromechanical system in the process industry~\citep{wang2016complex}. 

Ren and Jin~\citep{rensequential} applied the idea of sequential motifs (see Subsection~\ref{nvg_alg}) to LPVGs and introduced a measure of motif entropy to estimate the complexity of network structure, rather than looking at the motif occurrence frequency. They obtained a better robustness and the ability to distinguish different processes when compared to visibility-graph motif entropy. 

\cite{wang2018exact} presents some exact results (similar to those obtained for the HVGs presented in Subsection~\ref{subsec:hvg}) on the topological properties of the LPHVG associated with bi-infinite time series of i.i.d. random variables with a probability density $f(x)$.  
The degree distribution of the associated LPHVG with the limited penetrable distance $l$ is given by:
\begin{equation}
	P(k) = \frac{1}{2l+3} \left( \frac{2l+2}{2l+3} \right)^{k-2(l+1)},
	\label{eq:deg_dist_lp}
\end{equation}
where $l=0,1,2, \ldots,$ and $k$ is the degree of a node. From  Equation~(\ref{eq:deg_dist_lp}) we obtain the  average degree $\bar{k}$: 
\begin{equation}
	\bar{k} = \sum k P(k) = 4(l+1).
\end{equation}
And based on the Equation~(\ref{eq:deg_dist_lp}) we can deduce the minimum and maximum local clustering coefficient of the LPHVG associated to i.i.d. random series as (see~\citealt{wang2018exact} for more details):
\begin{eqnarray}
	C_{min}(k) &=& \frac{2}{k} + \frac{2l(k-2)}{k(k-1)}, \quad l = 0,1,2, \ldots ; \quad k \ge 2(l+1) \\
	C_{max}(k) &=& \frac{2}{k} + \frac{4l(k-3)}{k(k-1)}, \quad l = 0,1,2, \ldots ; \quad k \ge 2(2l+1).
\end{eqnarray}
For an infinite periodic series of period $P$ the average degree depends on the period $P$:
\begin{equation}
	\bar{k} = 4(l+1) \left( 1- \frac{2l+1}{2P} \right), \quad l \ll P.
\end{equation}

LPHVG was employed in the analysis of EEG signals and biphasic-flow signals where they characterize the behaviors underlying the systems~\citep{gao2016multiscale}, in the analysis of chaotic series and energy and oil price series~\citep{wang2018exact}, and to distinguish between random, periodic and chaotic signals using motifs~\citep{wang2018degree}.

\subsubsection{Weighted Visibility Graphs} 

{\label{subsec:wvg}}

The method proposed by Supriya and co-authors~\citep{Supriya2016} is a fairly simple modification to the traditional NVG algorithm, it considers the NVG edges  as  directed and weighted. 
For a given time series the corresponding weighted visibility graph (WVG) (or weighted directed visibility graph, WDVG) is constructed as follows: a directed NVG is constructed as described in Subsection~\ref{dhvg_alg}, and a weight $ w_{i,j} $ equal to the view angle between the observations $(t_i, Y_i)$ and $(t_j, Y_j)$ in time series is assigned to the edge that connect the corresponding nodes. The angle is given by: 
\begin{equation}
    \alpha_{i,j} = \tan^{-1} \left( \frac{Y_j - Y_i}{t_j - t_i}\right), \quad i < j.
    \label{eq:pnvgang}
\end{equation}

We should note that Equation~(\ref{eq:pnvgang}) allows the attribution of not only positive weights but also negative weights to the edges of the WVG. However, the analysis of networks with negative weights is more complex,   standard methods and techniques do not apply straightforwardly~\citep{kaur2016analyzing}, and therefore is less common. 
For this reason,~\cite{Supriya2016} consider the absolute value of the view angle ($|\alpha_{i,j}|$) between the observations of the time series. 
We can see that the sign of the angle between the observations is an indicator of an increase (positive angle) or decrease (negative angle) of the values of the underlying process, thus  providing information about changes in trends throughout the series.

We can consider a parametric version of WVGs,  called parametric natural visibility graph, PNVG or PNVG$(\alpha)$, proposed by Bezsudnov and Snarskii~\citep{Bezsudnov2014}. 
This algorithm adds a restriction to the edge assignment in WVG determined by a threshold parameter $ \alpha $ associated with the visibility angle. 
So, the PNVG($\alpha$) imposes that an weighted edge $(v_i, v_j, w_{i,j})$ is established from node $v_i$ to node $v_j$ only if the view angle between those nodes is less than threshold parameter, that is, $\{ (v_i,v_j,w_{i,j}) \in E(G_{PNVG}) \text{ } | \text{ } \alpha_{i,j} < \alpha, w_{i,j} = |\alpha_{i,j}| \}$, as shown in the Figure~\ref{figure:7b}.

The PNVG($\alpha$) is a directed acyclic graph, it is always a subgraph of the underlying NVG (and WVG) and, consequently, invariant under affine transformations of the time series; and it can  be a connected or  disconnected graph. 
 
PNVG and WVG were designed to try to overcome the loss of quantitative information of the NVG's binary matrix. 
Other than \textit{arc tangent} weights can be associated  to the edges of the graphs.
The arc tangent is a direct measure of the concept of visibility and a direction for future work may be to analyze its properties. 
However, the characteristics of the resulting network will always be dependent on this weight. 

Specific topological measures were defined for this mapping, namely, \textit{relative average degree}, \textit{relative average length of edge} and \textit{relative number of clusters}, which aim to compare the average degree, average length of edges and number of clusters with the corresponding measures in the underlying NVG (i.e., when $\alpha = \pi$). 
The authors in~\cite{Bezsudnov2014} showed that these measures are useful for distinguish, identify and describe various time series. They performed tests in different synthetic time series and in real heart rate data to distinguish different series associated with people with different health conditions, where they obtained good results. 
However, PNVG is not a parameter-free method and it always involves choosing the parameter $\alpha$ that conditions the final results. 

In~\cite{Supriya2016} the authors applied this approach to EEG reference data associated with epileptic activity and showed that the metrics such as modularity (that measures how good a division of the graph is in specific communities) and weighted average degree of WVG constructed for this data help to distinguish convulsion signals from normal signals, detecting the sudden fluctuations in signals. The accuracy of the results surpassed many other methods.

Note that other strategies for assigning weights to the edges of (natural or horizontal) visibility graphs, that are different from those of the visibility angles, can be considered. Examples include~\cite{bianchi2017multiplex} and~\cite{xu2018novel} leading to undirected and weighted horizontal and natural visibility graphs.

\subsubsection{Difference Visibility Graphs} 

A variant of the NVG and HVG consists of the creation of difference visibility graphs, DVG~\citep{zhu2014analysis}. 
This algorithm subtracts the HVG edge set from the NVG edge set. 
Thus, the degree $k_{i}$ of a node $v_i$ of the DVG is: $k_{i} = k_{i,NVG} - k_{i,HVG},$ 
since the HVG is always a subgraph of the NVG. For the average degree we have $\bar{k} = \bar{k}_{NVG} - \bar{k}_{HVG}$. 
DVGs will very possibly be disconnected and may have isolated nodes (which have no connections). The DVGs were used together with HVGs in~\cite{zhu2014analysis,an2019novel} to extract graph metrics, as average degree, degree distribution and degree sequence, to classify automatically real EEG data in order to detect different sleep stages. DVGs obtained intermediate precision values when compared to other strategies~\citep{an2019novel}.

\subsection{Transition Networks}

Transition networks are a type of networks that are constructed from time series based on the concept of  transition between symbols that are assigned to represent the series. 
In general, the construction of transition networks involves two fundamental steps: assigning a symbol encoding to the time series data;  mapping the symbols and the transition function between the symbols into the nodes and the edges, respectively. 
The first step  transforms the time series into a set of symbols or states by partitioning either the time range, reconstructing the time series in a new phase space~\footnote{\label{fnote:rec_phasespace}The reconstruction of phase space (or state space) is the basis of the analysis of non-linear processes, mostly adopted in areas such as physics. It allows reconstructing the complete dynamics of a system from a single time series~\citep{bradley2015nonlinear}. It consists of embedding $(Y_1, \ldots, Y_t)$ into a $w$-dimensional space with a time delay of embedding $\tau$ with states $\{\vec{z}_1, \vec{z}_2, \ldots, \vec{z}_{T-(w-1)\tau}\}$, where $\vec{z}_i = (Y_i, Y_{i+\tau}, \ldots, Y_{i+(w-1)\tau})$~\citep{takens1981detecting}.}, or the support of the time series. The first partition strategy leads to coarse-graining phase space graphs, to ordinal partition transition networks and to visibility graphlets networks, while the second leads to quantile graphs.
In the second step of the transition network construction process, the edges are established by  transition probabilities  between the symbols obtained in the first step, computed as the relative frequency of symbol sequences $s_i,s_j,$  leading to directed and weighted edges. 
The resulting transition network is a directed and weighted graph whose adjacency matrix is a Markov matrix. 
An unweighted representation can  be obtained by omitting the information on the probability/quantity of transitions between different states, or a less dense graph can be obtained considering a limit for the probability of transitions between states in order to exclude rare transitions (for example, due to noise in deterministic dynamical systems)~\citep{zou2018complex}.

\subsubsection{Quantile Graphs}

{\label{map:QG}}

\begin{figure}[hbt!]
	\centering 
	\subfloat[Toy time series]{
		\centering
		\includegraphics[scale = 0.6,keepaspectratio,valign=m]{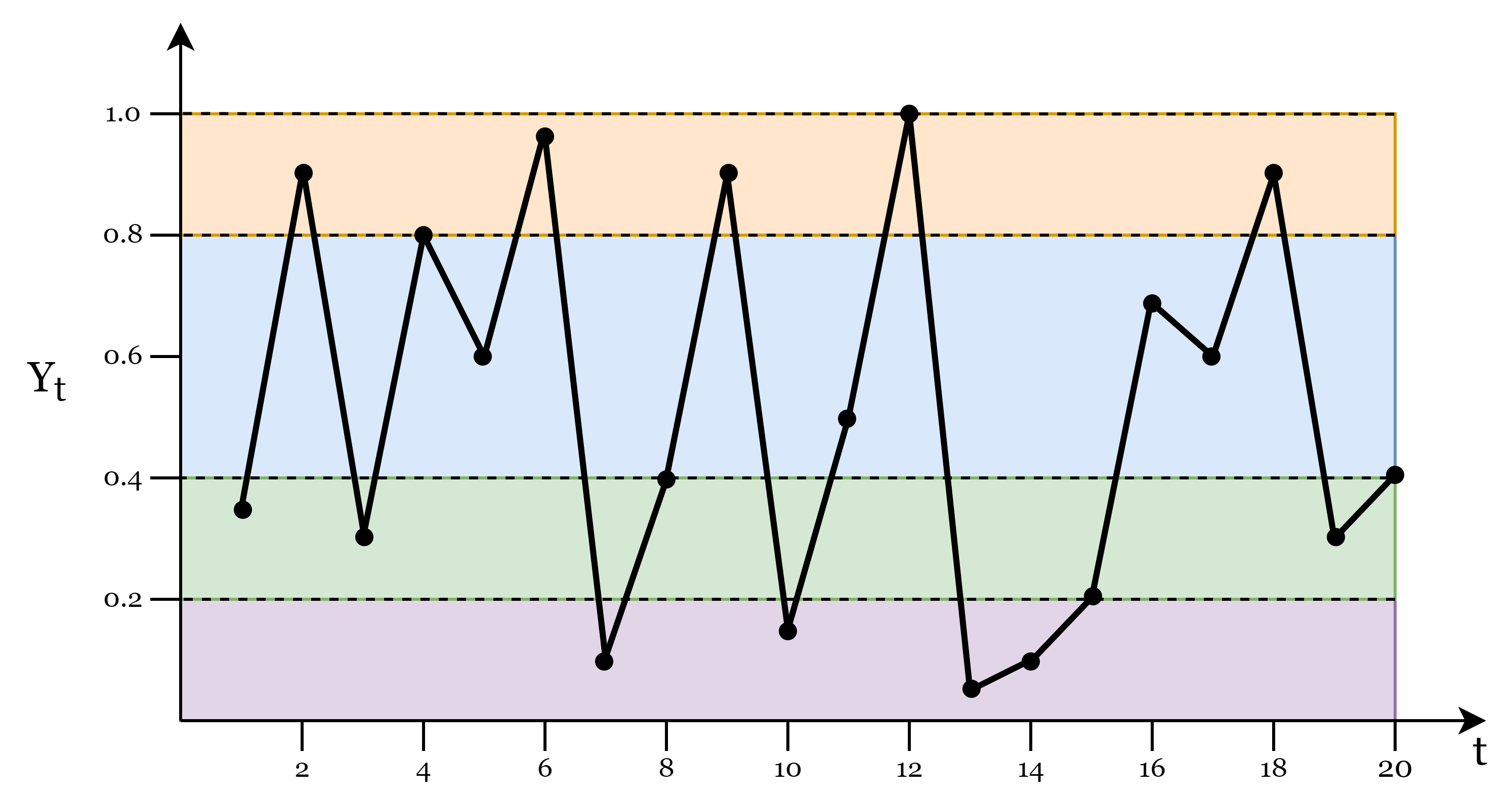}
	}
	\quad
	\quad
	\quad
	\subfloat[QG]{
		\centering
		\includegraphics[scale = 0.6,keepaspectratio,valign=m]{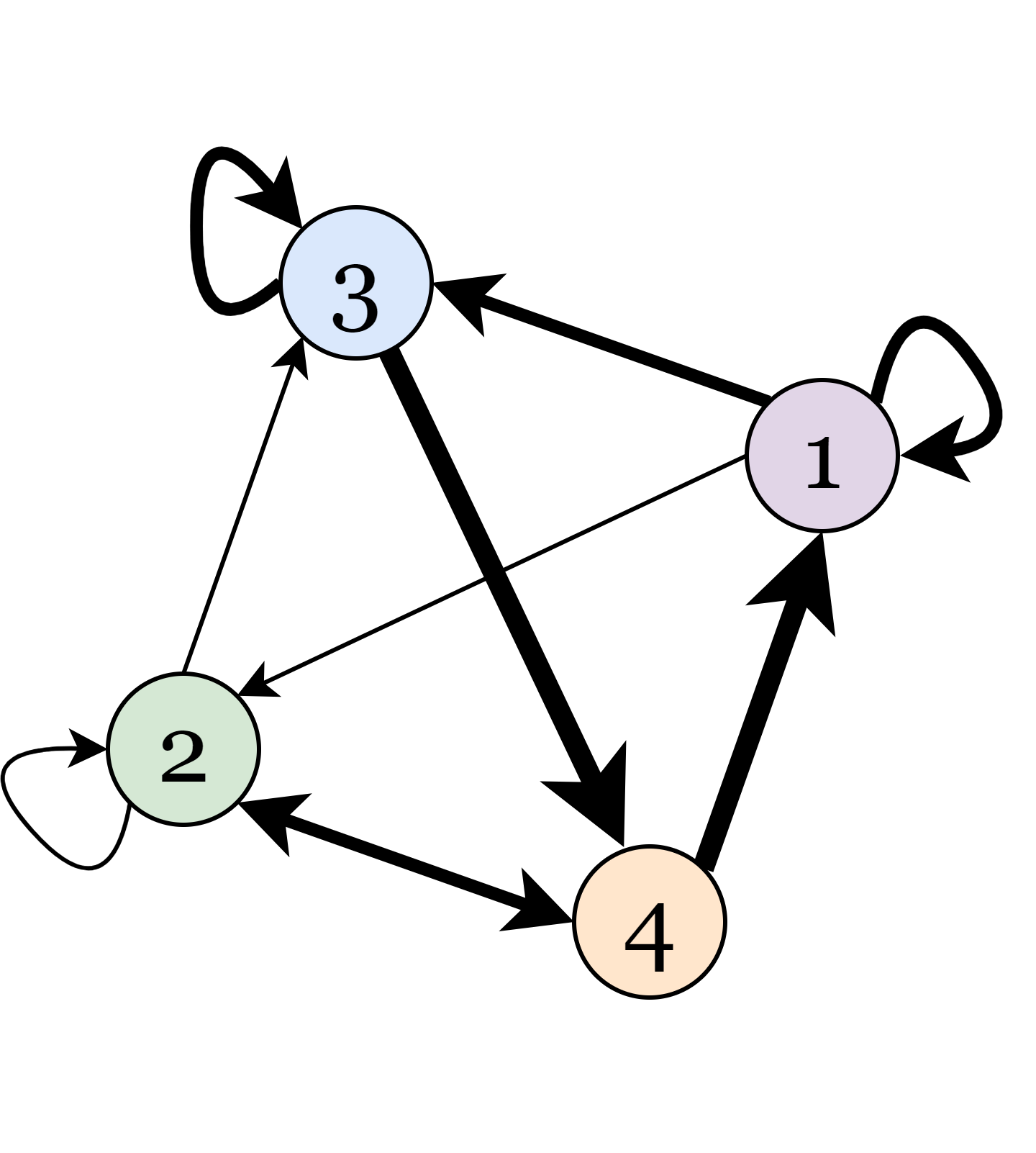}
	} 
	\caption[Figure 8]{Illustrative example of the quantile graph algorithm for $Q=4$. On the left panel we present the plot of a toy time series and on the right panel the network generated by the quantile graph algorithm. The different colors in the time series plot represent the regions corresponding to the different quantiles. In the network, edges with larger weights represented by thicker lines correspond to the repeated transitions between quantiles.}
	\label{figure:8}
\end{figure}

Quantile graphs (QG) were introduced by Campanharo and co-authors~\citep{Campanharo2011} based on the idea of assigning the time series observations to bins~\citep{Shirazi2009}.  In the construction of quantile graphs the bins  are defined by $Q$ quantiles, $q_1, q_2, ..., q_Q.$  Each quantile, $q_i,$ is associated to a node $v_i$ of the graph so the graph has as many nodes as the number of quantiles. Two nodes $v_i$ and $v_j$ are connected by a weighted directed edge $(v_i, v_j, w_{i,j})$, where the weight $w_{i,j}$ represents the transition probability between quantile ranges. The adjacency matrix becomes a Markov transition matrix, where $\sum_{j=1}^Q w_{i,j} = 1$, for each $i = 1, \ldots, Q$. This mapping is illustrated in Figure~\ref{figure:8}. 
\vspace{-1.5mm}

The number of quantiles is usually much less than the length of the time series ($Q \ll T$) and the resulting networks are weighted, directed and contain self-loops. 
If the number of quantiles is too large  the resulting graph may not be connected.  
On the other hand, QGs present a significant loss of information, represented by large weights assigned to self-loops, when $Q$ is  small. 
\cite{Shirazi2009} proposed a method based on the chi-squared statistic to obtain the optimal number of bins (quantiles). 
The connectivity of QGs represents the causal relationships contained in the dynamics of the process. 

Campanharo and co-authors~\citep{Campanharo2011} reconstructed a time series from a QG  using only the information contained in the adjacency matrix and without prior knowledge about the original series. 
These authors have shown that the resulting series presents statistical properties namely, autocorrelation, power spectrum and marginal distributions equivalent to those of the original series. 

\newpage
Recently,~\cite{campanharo2016hurst} developed a variation of the QGs method in order to estimate the Hurst exponent of fractional noise. This variation involves the introduction of a parameter $\phi$ such that two nodes $v_i$ and $v_j$ are connected by an edge with weight $w_{i,j}^\phi$ whenever two observations $Y_t$ and $Y_{t+\phi}$ belong  to the quantiles $q_i$ and $q_j,$ respectively, where $\phi = 1, \ldots \phi_{max} < T$. If $\phi = 1$, the method is reduced to QGs. The Hurst exponent, $H$, can be estimated as a function of the mean jump length: $\Delta(\phi) \sim \phi^H$, where 
\begin{equation}
	\Delta(\phi) = \frac{1}{L} \sum_{l=1}^L \delta_{l,\phi}(i,j),
\end{equation}
$\delta_{l,\phi}(i,j) = |i-j|$, with $i,j = 1, \ldots, Q$, is the jumps of length in a random walk on the graph and $L$ is the number of jumps. 

Shirazi and co-authors~\citep{Shirazi2009} were able to quantify the effect of long range correlations and of the marginal distributions, using distance and clustering measures. They tested the method on synthetic white noise data and on real data from turbulence and stock market index series. 
\cite{liu2018encoding} also used this approach  to  identify behavior patterns of the  processes underlying  synthetic and real world time series data from the visual analysis as well as the analysis of the  communities of the resulting networks.
Campanharo and co-authors~\citep{Campanharo2011} were able to show that time series with different properties are mapped into networks with different topological properties. In particular, as randomness increases in time series, the  corresponding QG networks become increasingly random as well, becoming similar to small-world models~\footnote{Small-world networks are networks characterized by a high clustering coefficient, like a regular graph, and  a low characteristic path length, like a random graph. This means that most nodes have few edges but can be reached from another node through a small number of edges~\citep{watts1998collective}.}. 
Later, \cite{Campanharo2015} showed that the QGs are capable of capturing and quantifying time series derived from long-range correlated processes and chaotic processes. They also verified that as the time series changes from periodic to chaotic, the corresponding networks, initially regular, become more and more random with larger values of the average path length~\footnote{The arithmetic mean of the shortest path lengths among all pairs of nodes in the graph, where the path length is the number of edges, or the sum of the edge weights if the graph is weighted, in the path.} and clustering coefficient.
The work developed by~\cite{campanharo2018application} and~\cite{pineda2020quantile}  takes advantage of the quantile graphs to classify EEG data.

\newpage
\subsubsection{Ordinal Partition Transition Networks}

{\label{map:OP}}

Ordinal patterns methods are based on the idea of a set of sequential patterns defined for a sequence of consecutive observations, where each node of the network represents one of the defined patterns and the edges are weighted according to the transition frequency between two consecutive patterns.

\begin{figure}[hbt!]
	\centering 
	\subfloat[Toy time series]{
		\centering
		\includegraphics[scale = 0.6,keepaspectratio,valign=m]{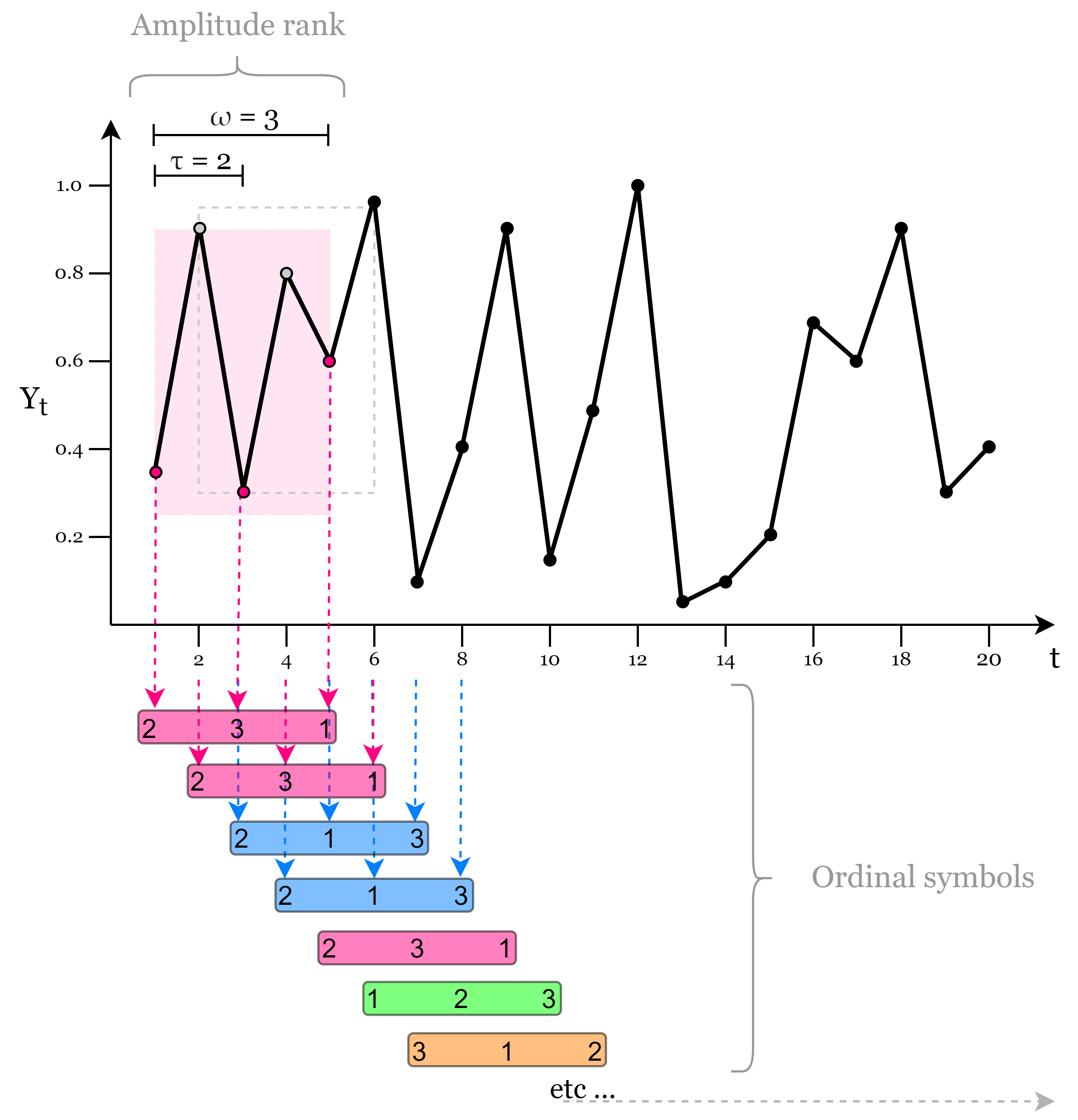}
		\label{figure:9a}
	}
	\quad
	\quad
	\subfloat[Ordinal Partition Transition Network]{
		\centering
		\includegraphics[scale = 0.6,keepaspectratio,valign=m]{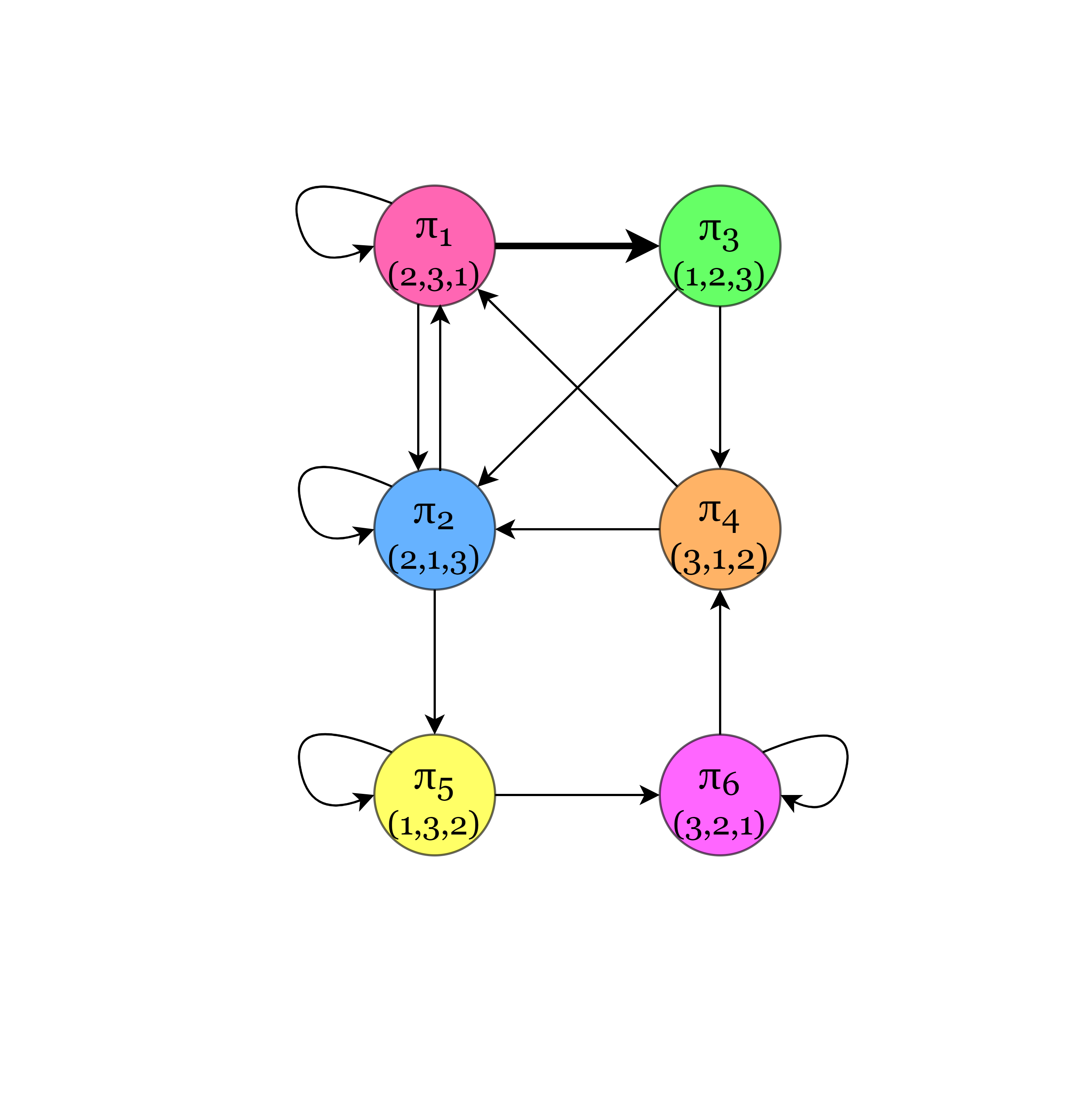}
	} 
	\caption[Figure 9]{Illustrative example of ordinal partition transition network algorithm. On the left side we illustrate the method of embedding with window size $w = 3$ and lag $\tau=2$ and the method of find its ordinal pattern, based in the amplitude rank of its elements. On the right side we show the resulting networks.}
\end{figure}

Formally, to construct a network of ordinal patterns (or ordinal partition transition network, OPTN) for a univariate time series 
 the time series is embedded in a $w$-dimensional space using a time delay $\tau.$ 
For each of these vectors $\vec{z}_i$, define the corresponding ordinal pattern by assigning  ranks to the data $Y_t$ in descending order, $\pi_i = (R_1, R_2, \ldots, R_w)$, where $R_j \in \{ 1,2, \ldots, w \},$ $R_j \neq R_k$ if $j \neq k$. If the observations $Y_j$ and $Y_k$ for time $j$ and $k$ have equal amplitudes we consider the order of time, we take the first to occur as smallest~\citep{small2013complex,mccullough2015time} (see Figure~\ref{figure:9a}). 
Finally, the set of patterns obtained is mapped to nodes in a network where the edges are allocated between nodes based on the transition sequence of the  symbols $\pi_i$. The weight ($w_{i,j}$) associated with the edge $(v_i,v_j)$ is the probability of transition between consecutive symbols. 
\vspace{-1.5mm}

McCullough and co-authors~\citep{mccullough2015time}, applied the ordinal partitions method to the R$\ddot{\text{o}}$ssler system and to other chaotic series. They found that both the visual topological structure of the networks, time series with periodic behavior are mapped in networks with a ring structure and chaotic time series in networks with band or tube-like structures, and the simple topological metrics as average \textit{out}-degree and variance of \textit{out}-degrees, convey the dynamics underlying the processes. 
In particular, the average \textit{out}-degree and the variance of \textit{out}-degrees allow to detect changes in the dynamic behavior in a similar way to the Lyapunov exponent~\footnote{The Lyapunov exponent, $\lambda$,~\citep{lyapunov1892general}  measures the speed with which trajectories in phase space approach (contraction) or move away from (expansion) each other~\citep{kiel1996chaos}. It can be interpreted as a qualitative measure of the sensitivity to the initial conditions of the chaotic systems and, consequently, instability of a process.},  distinguishing dynamics of different states and determining points of change. 
These networks also proved to be useful for the analysis of real world time series, such as: series from the externally driven diode resonator circuit~\citep{mccullough2015time},  and electrocardiograms (ECG) series for analysis of dynamical changes, as human cardiac dynamics~\citep{cao2004detecting,mccullough2017multiscale}.

A number of studies have begun to emerge around the so-called \textit{forbidden patterns}, which are patterns that do not occur in a time series. These studies have shown that measures around these patterns, namely the counting of forbidden patterns, are important to distinguish deterministic time series with very high levels of noise from random series~\citep{amigo2007true}, to detect of determinism in noisy data~\citep{mccullough2016counting,sakellariou2016counting}, and to distinguish between healthy patients and unhealthy patients with varying heart conditions based in ECG data~\citep{kulp2016using}.

Recently,~\cite{pessa2019characterizing} showed that properties of ordinal networks (namely, the average weighted shortest path lengths) can be used for estimating the Hurst exponent of time series with high precision, outperforming the quantile graphs~\citep{campanharo2016hurst} and the detrended fluctuations analysis~\citep{shao2012comparing}.

\subsubsection{Coarse-Grained Phase Space Graphs}

{\label{subsec:CGPSG}}

Networks based on coarse-graining of phase space consist of the idea of reconstructing (creating sets of symbol groups) the phase space of a dynamic process, where each of these symbol groups are mapped into a node of network and the edges are established between nodes representing successive partitions in time. A weight representing the amount or probability of transition between these partitions (nodes) is assigned to the edge.

The algorithm involves the following steps~\citep{Gao2009,wang2016time}: start by embedding a time series in a $w$-dimensional space with a time delay $\tau$, similarly to ordinal partition transition networks (Subsection~\ref{map:OP}); 
then the vectors (points in  the phase space) $\vec{z}_t = (Y_t, Y_{t+\tau}, \ldots, Y_{t+(w-1)\tau})$ are classified into $D = w$ different symbol groups ($\Psi_j = \{\psi_i\}_{i=1}^D$), according to intervals predefined by a set of values that delimit the phase space $\{ \alpha_1, \ldots, \alpha_{D-1} \}$~\citep{wang2016time,zou2018complex}:
\begin{equation}
	\left\{\begin{array}{l}
		\psi_1 \quad \textrm{if } Y_t < \alpha_1 , \\
		\psi_i \quad \textrm{if }  \alpha_{i-1} \le Y_t < \alpha_{i} , \ 1< i <D-1  \\
		\psi_D \quad \textrm{if } Y_t \geq \alpha_{D-1} .\\
	\end{array}\right.
\end{equation}

Note that this transformation process is similar to the transformation underlying the QG algorithm (Subsection~\ref{map:QG}), but applied separately to each component of the embedding vector. While in QGs, quantiles delimit the time series support, here the set of values $\{ \alpha_1, \ldots, \alpha_{D-1} \}$ delimit the phase space. 
Finally, each \textit{unique} vector of symbols $\Psi_j$ is mapped into a node of network. Directed and weighted edges $(v_i,v_j, w_{i,j})$ are defined by the transition probabilities between  symbol vectors $\Psi_i$ and $\Psi_j.$

\cite{wang2016time} have shown, based on the distribution of weighted degree of nodes, that increasing the dimension $w$ of the phase space leads to enhancing the heterogeneity of the networks and that the method allows distinguishing and characterizing different series, such as white noise, chaotic Lorenz systems\footnote{Lorenz's system, as R$\ddot{\text{o}}$ssler, is a complex system of three ordinary differential equations~\citep{lorenz1963deterministic}. It defines a chaotic continuous-time process.}, and gasoline price series. 

Weng and co-authors~\citep{weng2017memory}  proposed an extension of  this algorithm with the purpose of characterizing the memory (short or long) of the process.  Taking into account that the temporal information is lost after the coarse-graining transformation and that the transition probability matrix is an estimate of the process dynamics,  Weng and co-authors proposed to incorporate the temporal information into an additional topological dimension, such as an edge attribute. In this network, a node $v_i$ is connected to the node $v_j$ with an attribute $t,$ $(v_i, v_j; t)$, when  a transition from $\Psi_i$ to $\Psi_j$ occurs at time stamp $t$. This makes it possible to construct a memory network whose memory effect appears to be able to accurately differentiate various types of time series, namely, white noise, 1/f noise, AR model, periodic and chaotic time series. 

\newpage
\subsubsection{Visibility Graphlets Networks}

{\label{map:VGN}}

Given the ability to incorporate the topological characteristics (global and local) of the time series in the visibility graphs, visibility graphlets networks were proposed by~\cite{stephen2015visibility} in order to monitor and, consequently, understand the evolutionary behaviors of a time series.  
The method consists of five fundamental steps~\citep{stephen2015visibility, mutua2016visibility}. 
First, a time series is incorporated into a $w$-dimensional space without time delay ($\tau = 1$), similarly to sliding a window of length $w$ across the time series; 
each resulting vector  $\vec{z}_k = (Y_k, Y_{k+1}, \ldots, Y_{k+w-1}),$ with $k = 1, 2, \ldots, T-w+1,$ is mapped to a directed visibility graph (visibility graphlet) $G_k,$ following the algorithms described in Subsections~\ref{nvg_alg} and~\ref{dhvg_alg},  representing the $k$-th sate of the time series; 
the successive visibility graphs are then connected by a directed edge resulting in the following state chain network: $G_1 \to G_2 \to \ldots \to G_{T-w+1}.$ 
In order to obtain a network of distinguishable states, each graph $G_k$ is, iteratively,  compared to the others across the chain network and if any two graphs are identical, i.e., the adjacency matrices are the same, the latter graph is replaced by the former which is the reference state. For example, if $\boldsymbol{A}_1 = \boldsymbol{A}_5,$ the graph (state) $G_5$ is replaced by $G_1.$ 
Finally, the visibility graphlets network is defined so that the nodes represent the unique graphlets and the directed and weighted edges are established between pairs of nodes with weight equal to the number of edges between the successive corresponding graphlets in the chain network. 

Visibility graphlets network is a directed and weighted transition network where the time series states are represented by visibility graphlets and the edges reflect the relative frequency of states sequences, i.e., the transfer behaviors of the unique states. 
This network can also be seen as a temporal network~\citep{holme2012temporal} and a network of networks~\citep{kivela2014multilayer}, where the corresponding network science methodologies can be used~\citep{stephen2015visibility}. 
This mapping method is a "mixture" of two underlying concepts, visibility and transition probability, which can benefit from the characteristics mapped by both, namely, structural features that fully describe the states of the time series and the dynamic transfers of the underlying processes. 
However, like all methods that involve incorporation in the phase space, the selection of the parameter $ w $ is a problem. If $ w $ is too small, the number of different graphlets is very limited. On the other hand, the number of different graphlets increases geometrically with the increase of $ w $ implying that the time series is sufficiently large. 
See~\cite{stephen2015visibility} for a more detailed discussion. 

In~\cite{stephen2015visibility}, the authors studied some properties in a set of synthetic and real world time series, namely, degree of nodes and transmission probability related to edge weights for short-term prediction, and long-term persistence related to the occurring positions of some motifs on time series. 
\cite{mutua2016visibility} used the method in discrete and continuous chaotic time series and showed that the generated networks capture the dynamic properties of the systems, distinguishing chaotic zones that result in networks with a complex structure characterized by hubs and non-chaotic zones, such as periodic ones that result in regular networks.

\subsection{Proximity Networks}
Mappings based on the concept of proximity use measures of distance or similarity  to calculate the distance between the points of the time series incorporated in the multidimensional phase space. 
These methods map states of the time series into nodes of the network and create edges between those nodes based on some measure of distance or similarity. The ability to connect different nodes based on proximity measures allows capturing significant information about the topology of the dynamical systems, enabling to identify,  for example,  different regimes or patterns along the series via the similarity of successive states~\citep{donner2011recurrence}. 

Proximity networks include \textit{cycle networks}, \textit{correlation networks}, and \textit{recurrence networks}. 

\newpage
\subsubsection{Cycle Networks}

Cycle networks were proposed by Zhang and co-authors~\citep{zhang2006detecting,Zhang2006}  to represent pseudo-periodic time series. 

Formally, to construct a cycle network from a pseudo-periodic time series $\{Y_t\}_{t=1}^T,$ first  segment the series into $u$ consecutive cycles $\{ c_1, c_2, \ldots, c_u \}$, not necessarily of the same length, according to the local minima (or maxima). For each pair of cycles $c_i = \{ Y_i, \ldots,  Y_{l_i}\}$ and $c_j = \{Y_j, \ldots, Y_{l_j}\}$ ($i,j = 1,2,\ldots,u$ and $i \neq j$) 
 compute the correlation index $r_{i,j}$  defined as the maximum of the cross-correlation between the two cycles (assuming that $l_i \leq l_j$ without loss of generality)~\citep{zhang2006detecting}: 
\begin{equation}
	r_{i,j} = \underset{l=0,1,\ldots,l_j-l_i}{\max} \frac{\mbox{\rm cov}[(Y_i, \ldots, Y_{l_i}), (Y_{1+l}, \ldots, Y_{l_j+l})]}{\sqrt{\mbox{\rm var}(Y_i, \ldots, Y_{l_i})} \sqrt{\mbox{\rm var}(Y_{1+l}, \ldots, Y_{l_j+l})}},
\end{equation}
where $\mbox{\rm cov}$ is the covariance and $\mbox{\rm var}$ is the variance.  
This means that, if the cycles are not of the same length, the shortest, $c_i$ is shifted relative to the longest, $c_j$ by $l_j - l_i$ steps and the correlation coefficient between $c_i$ and the corresponding part of $c_j,$ in each step, is calculated. Finally, the highest value is chosen as the correlation coefficient between $c_i$ and $c_j$. 
The graph $G$ is constructed by assigning a node $v_i$ to each cycle $c_i$ and defining an undirected and unweighted edge between two nodes $v_i$ and $v_j$  if the correlation index is above a certain threshold $r_{i,j} > \alpha.$

The definition of the edges may be alternatively based on other distance measures, such as \begin{equation}
	d_{i,j} = \underset{l=0,1,\ldots , l_j - l_i}{\min} \frac{1}{l_i} \sum_{k=1}^{l_i} \| Z_k - X_{k+l} \|,
\end{equation}
for which  thresholds must also be set~\citep{Zhang2006}. 

Zhang and Small~\citep{Zhang2006} applied  cycle networks to the study of the R$\ddot{\text{o}}$ssler system and of  electrocardiogram (ECG) signals. The authors studied topological metrics of the resulting graphs, such as degree distribution, among others, and concluded that noisy periodic signals are mapped into random networks and chaotic time series into networks that exhibit small-world and scale-free features. 
In particular, they observed peaks in the degree distribution function of the graph that correspond to the unstable periodic orbits of the system, the nodes corresponding to these orbits form communities in the network.  
The metrics studied also allowed to distinguish between ECG of healthy volunteers, $P(k)$ varies smoothly, and those of unhealthy patients, $P(k)$ shows more prominent variations.

\subsubsection{Correlation Networks}

Correlation networks are essentially constructed based on a correlation measure, such as  the correlation matrix of the time series $Y_t$ proposed by~\cite{yang2008complex} based on the ideas of functional networks~\citep{eguiluz2005scale}, correlation of stock markets~\citep{bonanno2004networks} and the results obtained by cycle networks~\citep{Zhang2006}. 
The process consists of first considering individual state vectors $\vec{z}_i$ of time series which are extracted by embedding the series into a sufficiently large $w$-dimensional space. 
Each state vector is mapped into a node $v_i$ of an undirected and weighted network where the set of edges $\{(v_i, v_j, w_{i,j})\}$ are established in terms of the Pearson correlation coefficient calculated between the corresponding state vectors $(\langle \vec{z}_i,\vec{z}_j \rangle).$ Note that the Pearson correlation takes values between $-1$ and $1$, and so a negative weight may be assigned to the edges. However, as already mentioned, network analysis with negative weights is unusual and so the weights of correlation networks are usually established as the absolute value of the correlations, $w_{i,j} = |\rho_{i,j}|.$ 
The correlation matrix $[|\rho_{i,j}|]$ is the matrix of the correlation network, and note that other measures of correlation may be considered.

\newpage
As with cycle networks, a threshold $r_c$ also can be considered and a binary adjacency matrix can be produced. 
This threshold must be chosen properly because it determines the characteristics of the resulting network. If it is extremely small, pairs with weak correlations are also connected (noise). But if it is too large important information may be lost too~\citep{yang2008complex}.   

Correlation networks were applied to stock price series (return and amplitude series) resulting in a degree distribution function that follows a perfect Gaussian distribution. Additionally, the return series point to a random behavior, indicating that the global correlation characteristic can be modeled by an Erdős-Rényi network~\citep{yang2008complex}. 
The authors used the degree distribution to decide the best parameters for the algorithm. 
In contrast, Feng and He~\citep{feng2017construction}, used the cross-correlation as a measure of similarity between two points in phase space, and used the clustering coefficient and efficiency to decide the best parameters for the model. They analyzed the Lorenz system, white Gaussian noise and sea clutter time series. The results show that, for an unknown complex system, dynamic states, i.e., the behavior changes of the system, can be discovered by studying the community structures of complex networks. 
In particular, Feng and He identified changes in behavior from the contractive state to the open state in the sea clutter time series and for the white Gaussian noise did not identify clear community structures, the behavior changes of white Gaussian noise are completely random and therefore the resulting networks do not present a topology in communities.

\subsubsection{Recurrence Networks}

{\label{map_RN}}

Recurrence networks are quite popular in the research community. These networks are based on recurrence plots, a tool used in the study of  dynamical (non-linear or chaotic) systems~\citep{marwan2008historical}. 
For constructing a recurrence plot for a dynamical system from an observed time series, first embed the time series in a $w$-dimensional space (phase space reconstruction)  thus defining a set of vectors $\vec{z}_i.$ 
Then define a matrix $\boldsymbol{A}$, also called the recurrence matrix, where the elements $(i, j)$ are defined based on the proximity relation in phase space of the vectors $\vec{z}_i$ and $\vec{z}_j$~\citep{eckmann1995recurrence}. The proximity relation can be defined by different criteria, namely, fixed (probability) mass, where a fixed amount of nearest neighbor vectors is established, or fixed volume, where an $\varepsilon$-threshold of proximity is established. 
The matrix $\boldsymbol{A}$ is commonly defined with elements: 
\begin{equation}
	A_{i,j} = \left\{\begin{array}{l}
		1 \quad \textrm{if }  \|\vec{z}_i - \vec{z}_j\| \le \varepsilon \\
		0 \quad \textrm{otherwise}\\
	\end{array}\right.,
	\label{eq:RN}
\end{equation}
where $\varepsilon$ is a distance threshold and $\| \cdot \|$ can be any norm in phase space (for example, Euclidean, Manhattan, or maximum norm)~\citep{marwan2007recurrence}. 
The matrix $\boldsymbol{A}$ is a symmetric ($A_{i,j} = A_{j,i}$) matrix with constant values on the diagonal ($A_{i,i} = 1$), and it can be interpreted as the adjacency matrix of an undirected and unweighted network, the recurrence network, where the nodes represent $\{\vec{z}_i \}$ and the edges are defined by $A_{i,j}$, usually the self-loops are removed ($A_{i,i} = 0$). 
A direct generalization is considering the associated distance matrix between pairs of states, resulting in a weighted network. 
This type of networks, both weighted and unweighted, are spatial networks, where the nodes (states) are located in an $m$-dimensional space equipped with a norm in the phase space and which is reflected in the topological characteristics/features of the networks~\citep{Donner2010}. 
In the unweighted networks case, the probability of finding an edge between two states will decrease with the distance. 
For weighted networks, the direct use of the distance matrix is generally avoided because the more distant/dissimilar two states are, the stronger the connection between them.  
To avoid this other alternatives can be used, such as, the inverse of the distance $w_{i,j} = \frac{1}{\|\vec{z}_i - \vec{z}_j\|}$~\citep{strozzi2011recurrence}. 
Recently, a general alternative, that can be applied to any kind of network, was presented to recurrence networks where weights are given by: $w_{i,j} = \frac{\sqrt{k_i k_j}}{k_{max}},$ $k_{max}$ is the maximum degree in the unweighted recurrence network~\citep{jacob2019weighted}.

\newpage
The construction of these networks implies that they do not incorporate any temporal information on the underlying process and therefore do not explicitly depend on the presence of equally spaced observations, which is a major issue in the analysis of  time series observed in many phenomena~\citep{donner2011recurrence}. 
Recurrence plots are also one of the few time series analysis techniques that work well with non-stationary time series data.  However, in the case of more complex systems, the rich geometric structure of recurrence plots can become difficult to interpret, for example, for chaotic systems this happens due to their unstable periodic orbits~\citep{bradley2015nonlinear}.
The characteristics and topological properties of  recurrence networks depend essentially on their construction, namely on the following parameters: dimension of states $w$, time delay $\tau$ and threshold $\varepsilon$, the measure of  proximity between states.   
It is important to note $w=1$ meaning that the embedding in  the phase space is not required in which case the distance is computed between the points of time series~\citep{Marwan2009,zhao2020reciprocal}. 

The literature essentially highlights three types of recurrence networks, the \textit{$\kappa$-nearest neighbor networks}, the \textit{adaptive nearest neighbor network} and the \textit{$\varepsilon$-recurrence networks}, which we describe in more detail below.

\textbf{$\kappa$-Nearest Neighbor Networks, $\kappa$-NNN}

The $\kappa$-nearest neighbor version imposes a constraint on the quantity $\kappa$ of the nearest neighbor points (nodes $v_j$) of a given point in the  phase space (node $v_i$). This implies a direction in the edge $(v_i,v_j)$ since it is not necessarily true that, if the node $v_j$ is one of the $\kappa$ the closest neighbors of the node $v_i$ then $v_i$  is also one of the $\kappa$ closest neighbors to $v_j$, and so the adjacency matrix becomes an asymmetric matrix. 
This method was proposed by Small and co-authors~\citep{small2009transforming}, in order to overcome   the issue  of  cycle networks mappings that requires  time series with an oscillatory nature. 

$\kappa$-NNN imply that all nodes have the same amount $\kappa$ of \textit{out}-edges, $k_i^{out} = \kappa$, preserving a constant mass of the considered neighborhoods, i.e., the number of nodes is the same in all neighborhoods~\citep{donner2011recurrence}. However, the distribution of \textit{in}-edges $k_i^{in}$ is allowed to vary but the average degree is $\bar{k}^{in} = \kappa$. In particular, if $k_i^{in} \ll \kappa$, the node $v_i$ must be in a phase space region with decreased density compared to the remaining, and if $k_i^{in} \gg \kappa$, $v_i$ must be located in a densely populated region~\citep{donner2011recurrence}, thus giving information about the local geometry of the phase space.

\cite{xiang2012multiscale} studied the distribution of network motifs in $\kappa$-NNN and were able to classify different dynamic behaviors (such as maps and flows) using the frequency of motifs. 
These authors have also shown that  degree variations accompany the changes in the dynamics throughout the bifurcation process of the R$\ddot{\text{o}}$ssler system.

\textbf{Adaptive Nearest Neighbor Networks, ANNN}

In order to overcome the imposition   that all nodes have the same $k_i^{out} = \kappa$ in  $\kappa$-nearest neighbor networks, a new version, called adaptive nearest neighbor networks, was proposed~\citep{Xu2008,small2009transforming}. 
In ANNN if $v_j$ is one of the $\kappa$ closest neighbors of $v_i,$ then $v_i$ is also considered a neighbor of $v_j,$ even if it is not one of the $\kappa$ closest neighbors to $v_j.$ This leads to a symmetric adjacency matrix and the node degree to be variable, unlike $\kappa$-NNN, since there may be more than $\kappa$ neighbors to a given node.
\cite{Xu2008} and~\cite{liu2010superfamily} have shown that the motif distributions allow characterizing different processes, namely periodic, chaotic, noise and fractional, creating super-families. Unique fingerprints have also been found for specific dynamical systems within a family. Additionally,~\cite{liu2010superfamily}  analyzed three stock market indexes and concluded that the motif distributions are equivalent for the three series (since they have the same dynamic behavior characteristic of return stock series) and are also very similar to the distributions obtained for the fractional Brownian motion~\footnote{Fractional Brownian motion is a continuous-time Gaussian process where the increments need not be independent. The Hurst exponent, $H$, describes the raggedness of the resultant motion, $H \in [0,1]$, the higher the value is smoother (correlated) motion~\citep{zunino2007characterization}. The Hurst exponent is an important measure that quantifies the correlation of a time series, and it is used as a measure of long-term memory, the persistence of the process~\citep{zunino2007characterization}.}.

\textbf{$\varepsilon$-Recurrence Networks}

Equation~(\ref{eq:RN}) suggests that  the "neighborhood" of a single state vector $\vec{z}_i$ can be defined by a fixed distance in the phase space $\varepsilon$, considering fixed volumes (communities) in phase space~\citep{donner2011recurrence}. 
Thus,  an undirected and unweighted network can be built.

This version implies choosing the  distance threshold $\varepsilon$ which allows  to control the phase space resolution. If $\varepsilon$ is too small, the volume of the neighborhood will be small and therefore there will be almost no recurrence points and the information incorporated in the network will be insufficient. 
On the other hand, if $\varepsilon$ is too large we observe a general qualitative change in the network topology~\citep{jacob2016can}, namely, each node will behave like a hub, leading  to an excess of  recurring points and misleading information. 

Donner and co-authors~\citep{Donner2010,donner2011recurrence} studied properties of $\varepsilon$-recurrence networks at three different scales, namely local, intermediate and global on several paradigmatic systems: 
Hénon map, Bernoulli map, Lorenz system and R$\ddot{\text{o}}$ssler system. The authors studied graph properties as a function of the distance threshold $\varepsilon$ and proposed  specific measures like the local clustering coefficient 
to detect dynamically invariant objects, saddle points or unstable periodic orbits. Moreover, those authors suggested varying the embedding dimension  as a means to distinguish between chaotic and stochastic systems. 
\cite{zou2012power} have shown that the degree distribution of networks generated for 
one-dimensional maps with a local power-law in the invariant density is a scale-free distribution, resulting in scale-free networks. In a general case, the resulting exponent of the degree distribution does not need to coincide with the fractal dimension. In this direction,~\cite{donner2011geometry} demonstrated that the local and global transitivity of the networks are closely related with a generalized notion of fractal dimensionality (local and global, respectively). 

$\varepsilon$-recurrence networks were the most exploited and led to the well-established relationships between some topological metrics of networks and the properties and measurements of phase space.  
Donges and co-authors~\citep{donges2012analytical} propose an analytical framework for $\varepsilon$-recurrence network analysis  describing graph-theoretical recurrence network quantifiers. 
This framework shows that several standard measures of network analysis can represent discrete estimators of continuous measures of certain complex phase space properties. 
A theoretical relation is established between measures of the recurrence networks and phase space properties. 
$\varepsilon$-recurrence networks were also widely used to determine changes in the dynamics of theoretical~\citep{Marwan2009,iwayama2013change} and real world~\citep{donges2011identification,fukino2016coarse} systems.

\par\null

\section{Mapping Multivariate Time Series into  Complex Networks}

So far we have focused  on  approaches to mapping  univariate time series into  network structures. However, technological developments are producing a wealth of inter-linked multidimensional data, such as multivariate spatio-temporal data. Tools for the analysis of these high dimensional data sets are yet scarce, hinting at the possibility of using network science approaches. The literature on mapping multivariate time series to the network domain, summarized in Table~\ref{table:2}, is not as developed as for the univariate case. 
Even so, we can distinguish two  classes of methods, introduced in Figure~\ref{figure:1}: those that  map the multivariate time series into a single layer or monoplex network  and those that map the  multivariate time series  into a multiplex\footnote{Remember that a multiplex network is just a particular case of multilayer networks so the terms are not equivalent, see Subsection~\ref{susbsubsec:mnet}.} network. 
The mappings in the first class construct networks with nodes representing the (component) time series and edges representing the relationship between the node time series, computed as, e.g., causal relation. Mappings on the second class lead to multiplex networks where each layer is a network resulting from a univariate time series mapping. The layers  usually have the same nodes and  are connected via the edges which connect the same node across adjacent layers. 
Details on each approach and corresponding core results are given in the following sections. 

\begin{table}[hbt!]
    \centering
    \footnotesize
    \renewcommand{\arraystretch}{0.75}
    \begin{tabular}{|c c|l|c|c|c|c|c|c|}
        \cline{3-9}
         \multicolumn{1}{l}{} & \multicolumn{1}{l|}{} & \begin{tabular}{@{}l@{}}\textbf{Network Type} \\ \scriptsize(original reference)\end{tabular} & 
         \rotatebox[origin=c]{90}{\textbf{Node}} & \rotatebox[origin=c]{90}{\textbf{Edge}} & \rotatebox[origin=c]{90}{\textbf{Directed}} & \rotatebox[origin=c]{90}{\textbf{Weighted}} & \rotatebox[origin=c]{90}{\textbf{Param-free}} & \rotatebox[origin=c]{90}{\textbf{Pre pross}} \\
        \hline
        
        \parbox[c]{0.5mm}{\multirow{14}{*}{\rotatebox[origin=c]{90}{\textbf{Single}}}} & \parbox[c]{0.5mm}{\multirow{14}{*}{\rotatebox[origin=c]{90}{\textbf{Layer}}}}
            \rule{0pt}{12pt} & Correlation Network & \multirow{2}{*}{$Y_{\alpha,t}$} & \multirow{2}{*}{CM} & \multirow{2}{*}{\xmark} & \multirow{2}{*}{\cmark} & \multirow{2}{*}{\cmark} & \multirow{2}{*}{\xmark} \\
            & & \tiny~\citep{eguiluz2005scale} & & & & & &  \\
            
            \rule{0pt}{10pt} & & Long-Run Variance Decomposition Network & \multirow{2}{*}{$Y_{\alpha,t}$} & \multirow{2}{*}{VD} & \multirow{2}{*}{\cmark} & \multirow{2}{*}{\cmark} & \multirow{2}{*}{\xmark} & \multirow{2}{*}{VAR} \\
            & & \tiny~\citep{diebold2014network} & & & & & &  \\
            
            \rule{0pt}{10pt} & & Causal Effect Network & \multirow{2}{*}{$t$} & \multirow{2}{*}{LCR} & \multirow{2}{*}{\cmark} & \multirow{2}{*}{\cmark} & \multirow{2}{*}{\xmark} & \multirow{2}{*}{CDA} \\
            & & \tiny~\citep{runge2015identifying} & & & & & &  \\
            
            \rule{0pt}{10pt} & & Ordinal Partition Transition Network & \multirow{2}{*}{$\pi_{i}$} & \multirow{2}{*}{TP} & \multirow{2}{*}{\cmark} & \multirow{2}{*}{\cmark} & \multirow{2}{*}{\xmark} & \multirow{2}{*}{PS} \\
            & & \tiny~\citep{zhang2017constructing,ruan2019ordinal} & & & & & &  \\
            
            \rule{0pt}{10pt} & & Pattern Interdependent Network & \multirow{2}{*}{$G_{i}$} & \multirow{2}{*}{TP} & \multirow{2}{*}{\cmark} & \multirow{2}{*}{\cmark} & \multirow{2}{*}{\cmark} & \multirow{2}{*}{PS;VG} \\
            & & \tiny~\citep{ren2020pattern} & & & & & &  \\
            
            \rule{0pt}{10pt} & & Inter-system recurrence network & \multirow{2}{*}{$\vec{z}_i^{[\alpha]}$} & \multirow{2}{*}{DM} & \multirow{2}{*}{\xmark} & \multirow{2}{*}{\xmark} & \multirow{2}{*}{\xmark} & \multirow{2}{*}{PS} \\
            & & \tiny~\citep{feldhoff2012geometric} & & & & & &  \\
            
             \rule{0pt}{10pt} & & Joint Recurrence Network & \multirow{2}{*}{$\vec{z}_i^{[\alpha]}$} & \multirow{2}{*}{DM} & \multirow{2}{*}{\xmark} & \multirow{2}{*}{\xmark} & \multirow{2}{*}{\xmark} & \multirow{2}{*}{PS} \\
            & & \tiny~\citep{feldhoff2013geometric} & & & & & &  \\
         
        \hline
        
        \parbox[c]{0.5mm}{\multirow{4}{*}{\rotatebox[origin=c]{90}{\textbf{Multiple}}}} & \parbox[c]{0.5mm}{\multirow{4}{*}{\rotatebox[origin=c]{90}{\textbf{Layer}}}}
            \rule{0pt}{12pt} & Multiplex Visibility Network & \multirow{2}{*}{$t$} & \multirow{2}{*}{V} & \multirow{2}{*}{\xmark} &  \multirow{2}{*}{\xmark} & \multirow{2}{*}{\cmark} & \multirow{2}{*}{\xmark} \\
            & & \tiny~\citep{lacasa2015network} & & & & & &  \\
            
            \rule{0pt}{10pt} & & Multiplex Recurrence Network & \multirow{2}{*}{$t$} & \multirow{2}{*}{DM} & \multirow{2}{*}{\xmark} &  \multirow{2}{*}{\xmark} & \multirow{2}{*}{\xmark} & \multirow{2}{*}{PS} \\
            & & \tiny~\citep{eroglu2018multiplex} & & & & & &  \\
         
        \hline
        
    \end{tabular}
    \caption{Comparison of (multivariate) time series mappings based on the properties of the corresponding algorithms and of the resulting networks. Notation: CM - correlation measures, VD - variance decomposition, LCR - lagged causal regression, TP - transition probability, DM - distance measures, V - natural and/or horizontal visibility, VAR - vector autoregression model, CDA - causal discovery algorithm, PS - phase space, VG - directed visibility graph.}
    \label{table:2}
\end{table}

\newpage
\subsection{Correlation Networks}

Formally, a correlation network (CN) is defined as an undirected and weighed graph $G=(V(G),E(G))$ where $V(G) = \{Y_{i,t}\}_{i=1}^m$ and $E(G) = \{(v_i,v_j, w_{i,j}) | (v_i,v_j) \in V(G) \wedge w_{i,j} = \rho_{i,j}(0) \wedge i \neq j\}$ (see in~\ref{sec:ts}). Thus, inter-dependencies between the $m$ time series  are represented by the (contemporaneous) correlation.

Other approaches to represent the inter-dependencies and establish the edges of the network rely on suitable correlation measures such as cross-correlations~\citep{nakamura2016constructing} or  partial correlations~\citep{epskamp2018tutorial}, eventually subject to thresholding. The thresholding (edges are established only if the correlation exceeds a predefined  value) helps to remove spurious edges.
Alternatively, correlations can be replaced by any  similarity measure, including distance measures. The resulting similarity/distance matrix is then used to construct  a monoplex network~\citep{mori2016distance}. 

\begin{figure}[hbt!]
	\centering 
	\subfloat[Toy time series]{
		\centering
		\includegraphics[scale = 0.6,keepaspectratio,valign=m]{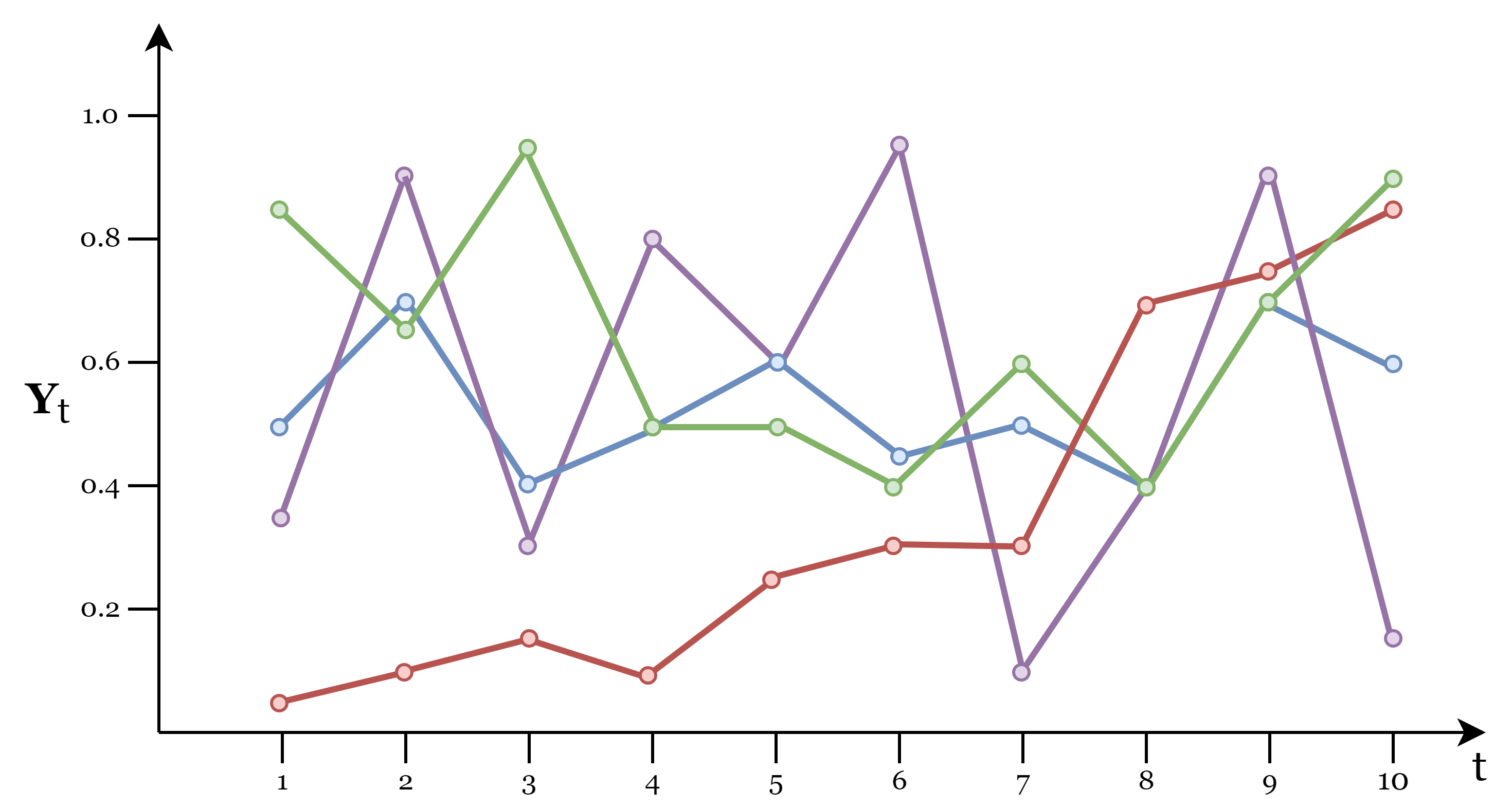}
	}
	\quad
	\quad
	\quad
	\subfloat[CN]{
		\centering
		\includegraphics[scale = 0.6,keepaspectratio,valign=m]{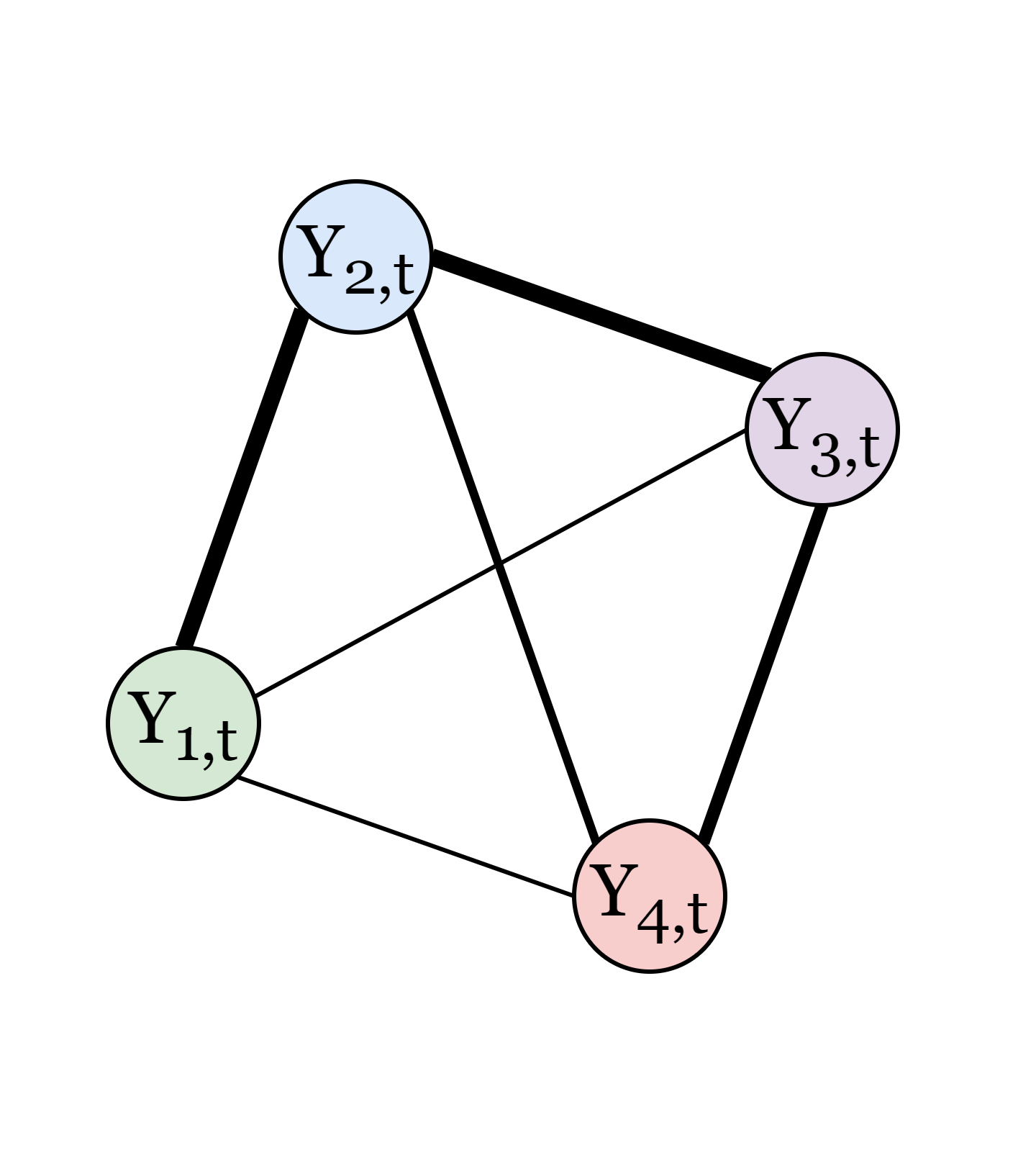}
	} 
	\caption[Figure 10]{Illustrative example of the correlation network algorithm. On the left side we present the plot of a toy multivariate time series and on the right side the network generated by the correlation algorithm (using contemporaneous cross-correlation). The different colors represent the  time series. Higher correlation values result in edges in the network with larger weights represented by thicker lines.}
	\label{figure:10}
\end{figure}

Figure~\ref{figure:10} presents  an example of a correlation network representation of  a toy multivariate time series.  Nodes that represent similar time series  are linked by thicker edges.

\vspace{-1.5mm}

Correlation networks (also known as functional networks)  have been widely used in neuroscience~\citep{eguiluz2005scale,bullmore2009complex}, financial~\citep{tumminello2010correlation,cai2010hierarchical,gao2015influence} and climate science~\citep{tsonis2012origins,dijkstra2019networks} areas. In particular, Eguiluz and co-authors~\citep{eguiluz2005scale} applied this approach to  functional fMRI data in order to connect brain zones based on their similarities and their relationships. They have shown that the degree distribution obeys a power law, indicating scale-free networks which implies that there are always a few zones of the brain (hubs) with relations to most other regions of the brain, that is, predominant zones.

Gao and co-authors~\citep{gao2015influence} used correlation networks to analyze the interactions between  companies of the different sectors using clustering metrics. The objective is also to study the influence that some companies and/or sectors have on others. The results show important information about fluctuations of the series and show that correlation networks can be very useful for financial risk analysis.

\subsection{Long-Run Variance Decomposition Networks}

The mapping of multivariate time series into correlation networks is based on the contemporary dependence of time series, but in multivariate time series, cross-sectional dependencies may arise in different leads/lags~\citep{lanne2016generalized}. We  introduce now a class of networks that reflect, in the directed and weighted edges, the presence of contemporaneous as well as lagged partial correlations between  time series $Y_{i,t}$ and $Y_{j,t}$. The edges of the network are established via Vector AutoRegression (VAR) models and their analysis, namely  Forecast Error Variance Decomposition (FEVD).  
The process $\boldsymbol{Y}_t$ follows a VAR($p$) model, where $p$ is the order of the VAR, if it satisfies the following equation:
\begin{equation}
	\boldsymbol{Y}_t = c_0 + \sum_{i=1}^{p} \boldsymbol{\Phi}_i \boldsymbol{Y}_{t-i} + \boldsymbol{a}_t, 
\end{equation}
where $c_0$ is an $m$-vector of constants, 
$\boldsymbol{\Phi}_i$ $(i = 1,2, \ldots, p)$ is an $m \times m$-matrix of coefficients to be estimated 
and $\boldsymbol{a}_t$ is an $m$-vector of uncorrelated errors with zero mean and covariance matrix $\boldsymbol{\Sigma},$ with elements $\sigma_{i,j}$. Under stationarity conditions $\boldsymbol{Y}_t$ may be represented as an infinite moving-average representation (MA model)
\begin{equation}
    \boldsymbol{Y}_t = \sum_{j=0}^{\infty} \boldsymbol{\Theta}_j \boldsymbol{a}_{t-j}.
\end{equation}

The FEVD, carried out typically  on the infinite moving-average representation of the VAR process,   calculates the  proportion of the prediction error variance of the $i$-th variable which is assignable to its own (lagged) shocks and to the (lagged) shocks of the other variables. A problem with the FEVD is that it is dependent of the ordering of the variables. To overcome this issue,~\cite{diebold2014network} and~\cite{lanne2016generalized} propose to use the generalized forecast error variance decomposition (GFEVD)\footnote{See~\cite{pesaran1998generalized} for more details.} which is independent of the ordering of the variables. 
Thus, a long-run variance decomposition network (LVDN) is a network where the nodes $v_i$ and $v_j$ represent the time series $Y_{i,t}$ and $Y_{j,t},$ respectively, and the edges $(v_i,v_j, w_{i,j})$ are given by $w_{i,j} = \frac{d_{i,j}^h}{\sum_{j=1}^T d_{i,j}^h} $, where $d_{i,j}^h,$ denotes the $(i,j)$-th $h$-step ahead variance decomposition component, that is, the fraction of variable~$i$ $h$-step ahead forecast error variance due to shocks in variable $j$~\citep{diebold2014network}, calculated as:
\begin{equation}
	d_{i,j}^h = \frac{\sigma_{j,j}^{-1} \sum_{k=0}^{h-1} (\boldsymbol{\epsilon}_{i}^{'} \boldsymbol{\Theta}_{k} \boldsymbol{\Sigma}\boldsymbol{\epsilon}_{j})^2}{\sum_{k=0}^{h-1} (\boldsymbol{\epsilon}_{i}^{'} \boldsymbol{\Theta}_{k} \boldsymbol{\Sigma} \boldsymbol{\Theta}_{k}^{'} \boldsymbol{\epsilon}_{i})},
\end{equation}
where $\boldsymbol{\epsilon}_j$ is a vector of orthogonalized shocks with the $j$-th element unity and zeros elsewhere. 
Note that the adjacency matrix $D^h = [d_{i,j}^h]$ is then normalized. 
 
Diebold and Yilmaz~\citep{diebold2014network} applied the method to 13 years of major US financial institutions' stock return volatilities, and they found qualitative relations between the measures of \textit{in} and \textit{out}-degree with traditional economic risk measures.
Given the interpretability and qualities of the results obtained, Barigozzi and Hallin~\citep{barigozzi2017network} proposed an improvement of this approach to make it applicable in the analysis of large sets of time series. Note that the LVDN involves the estimation of VAR models, which becomes unbearable for large sets of time series. This approach involves the use of a generalized dynamic factor model~\citep{barigozzi2015generalized}.

An \texttt{R} package "Nets: Network Estimation for Time Series"~\citep{barigozzi2019nets} has been developed based on~\cite{diebold2014network} and~\cite{barigozzi2017network} and on the concept of long run partial correlation. It allows the estimation of sparse long run partial correlation networks, based on the estimated VAR parameters and the concentration matrix of the VAR residuals using LASSO~\citep{meinshausen2006high} regressions.

In finance, these methods have been widely applied to study interconnections between financial institutions to identify possible contagion channels.
The main disadvantage of these methods is that they are heavily parameterized, as they are based on time series models that involve the estimation of several parameters and restrictive assumptions, and this is one of the major problems of the analysis of multivariate series.

\subsection{Causal Effect Networks} 

Causal effect networks~\citep{runge2015identifying}, like LVDNs, allow  to encode the inter-relations between components of multivariate time series in different time lags and to distinguish the directionality of these relations. 
The corresponding graphs are directed and weighed, the nodes represent the individual observations $Y_{i,t}$ with $i=1,\ldots,m$ at each time $t$ and the edges are established based on a causal discovery algorithm\footnote{There are several methods of causal discovery based on different approaches, examples include approaches based on Granger causality, structural causal models, among others. For an overview of methodological frameworks and challenges of these approaches see~\cite{runge2019inferring}.}~\citep{runge2019inferring}. 
More specifically, the causal discovery algorithm is only used as a variable selection for a subsequent lagged causal regression and the construction of the network consists of the following three main steps. 
First, we select causal parents, $\mathcal{P}(Y_{j,t}),$ for each component, $Y_{j,t},$ using the causal discovery algorithm~\citep{runge2014quantifying} that iteratively tests the conditional correlation between $Y_{j,t}$ and the remaining components at a range of time lags $0 < h \le h_{max},$ $h_{max}$ is a maximum time lag. Then we estimate the lagged causal regression matrix $\boldsymbol{C}(h)$ of shape $(m,m,h_{max})$ using the selected parents:
\begin{equation}
	C_{i \to j}(h) = b_{j,i \cdot \mathcal{P}(Y_{j,t})}(h) \quad \textrm{for } h = 1,\ldots,h_{max} \textrm{ and } i,j = 1, \ldots, m,
\end{equation}
where $b$ is  the standardized regression coefficient of $Y_{i,t-h}$ in the multiple regression model of $Y_{j,t}$ on 
$\{ Y_{i,t-h},$ $\mathcal{P}(Y_{j,t}) \}$ using ordinary least squares regression~\citep{runge2015identifying}. 
Finally, we construct the causal effect graph where the edges are established from the threshold of the causal regression matrix $|C_{i \to j}(h)| \ge \theta$ chosen to obtain a given link density. Self-loops are not counted and multiple edges are only counted once. 
Note that the advantage of causal effect networks over LVDNs is that the later can exploit sophisticated versions of conditional mutual information and therefore do not need to resort to parametric time series models. 

Causal effects networks also arise from the need to eliminate possible spurious edges  added by pairwise association measures, such as cross-correlation. These edges result from transitivity effects (leading to indirect paths) or other processes  making it difficult to analyze the causal interactions among multiple nodes~\citep{runge2015identifying}. 
However, the construction of the causal effect networks implies several assumptions given the underlying statistical methods: causal sufficiency (common drivers of all variables are taken into account), causal Markov condition (all error terms of the nodes in the graph are independent) and stationarity. 
Note that different causal discovery methods can lead to different assumptions~\citep{runge2019inferring}. 

This network mapping approach can be complemented by multiple information at the edges, such as weights indicating causal edge strengths and additional attributes with the associated time lags. This also allows an extension from a single-layer to a multilayer network, where each layer would correspond to a different time lag~$h$.

Causal effect networks have been extensively explored  for the analysis of climate data, especially in complex spatio-temporal systems~\citep{runge2015identifying,kretschmer2016using,runge2019detecting,runge2019inferring}. 
In particular, Runge and co-authors~\citep{runge2015identifying}, introduce novel network measures based on causal effect theory, which differ  from the standard complex network tools by distinguishing direct from indirect paths, in order to identify components with high cumulative causal effect either as sources or as intermediate nodes on path of causal network.

\subsection{Ordinal Partition Transition Networks}

The success of OPTN presented in Subsection~\ref{map:OP} led to the extension of these type of networks to the multivariate context. 
Zhang and co-authors~\citep{zhang2017constructing}, built the first ordinal partition transition networks for multivariate time series $\{Y_{i,t}\}_{i=1}^m$, with an algorithm that differs from the univariate case. 
 Start by constructing the set of order patterns $\Pi = \{\pi_j\}$ with $j=1, \ldots, 2^m,$ where each $\pi_j$ represents a pattern set $(\pi_{Y_{1,t}}^\theta, \ldots, \pi_{Y_{m,t}}^\theta),$ $\theta \in [0,1],$ with $\pi_{Y_{i,t}}^1$ capturing the increasing trend and $\pi_{Y_{i,t}}^0$ the decreasing trend of the time series $Y_{i,t}.$ In short,  $2^m$ different combinations of order patterns are constructed depending on the signs of first-order differences of the $m$ time series, as exemplified in Table~\ref{tab:op}  for $m=3.$ 
 
\begin{table}[hbt!]
\centering
\renewcommand{\arraystretch}{1.25}
\begin{tabular}{|l||c|c|c|c|c|c|c|c|} 
\hline
\textbf{$\Pi$} & \textbf{$\pi_1$} & \textbf{$\pi_2$} & \textbf{$\pi_3$} & \textbf{$\pi_4$} & \textbf{$\pi_5$} & \textbf{$\pi_6$} & \textbf{$\pi_7$} & \textbf{$\pi_8$} \\
\hline
\hline
$\nabla Y_1$ & $\pi_{Y_1}^1, +$ & $\pi_{Y_1}^1, +$ & $\pi_{Y_1}^1, +$ & $\pi_{Y_1}^1, +$  & $\pi_{Y_1}^0, -$ & $\pi_{Y_1}^0, -$ & $\pi_{Y_1}^0, -$ & $\pi_{Y_1}^0, -$ \\   
\hline
$\nabla Y_2$ & $\pi_{Y_2}^1, +$ & $\pi_{Y_2}^1, +$ & $\pi_{Y_2}^0, -$ & $\pi_{Y_2}^0, -$ & $\pi_{Y_2}^1, +$ & $\pi_{Y_2}^1, +$ & $\pi_{Y_2}^0, -$  & $\pi_{Y_2}^0, -$ \\  
\hline
$\nabla Y_3$ & $\pi_{Y_3}^1, +$ & $\pi_{Y_3}^0, -$ & $\pi_{Y_3}^1, +$ & $\pi_{Y_3}^0, -$ & $\pi_{Y_3}^1, +$ & $\pi_{Y_3}^0, -$ & $\pi_{Y_3}^1, +$ & $\pi_{Y_3}^0, -$ \\ 
\hline
\end{tabular}
\caption{Order patterns in multivariate time series with three variables $(Y_{1,t}, Y_{2,t}, Y_{3,t})$. (Adapted from~\citealp{zhang2017constructing})}
\label{tab:op}
\end{table}

\begin{figure}[hbt!]
	\centering 
	\subfloat[Toy multivariate time series]{
		\includegraphics[scale = 0.6,keepaspectratio,valign=m]{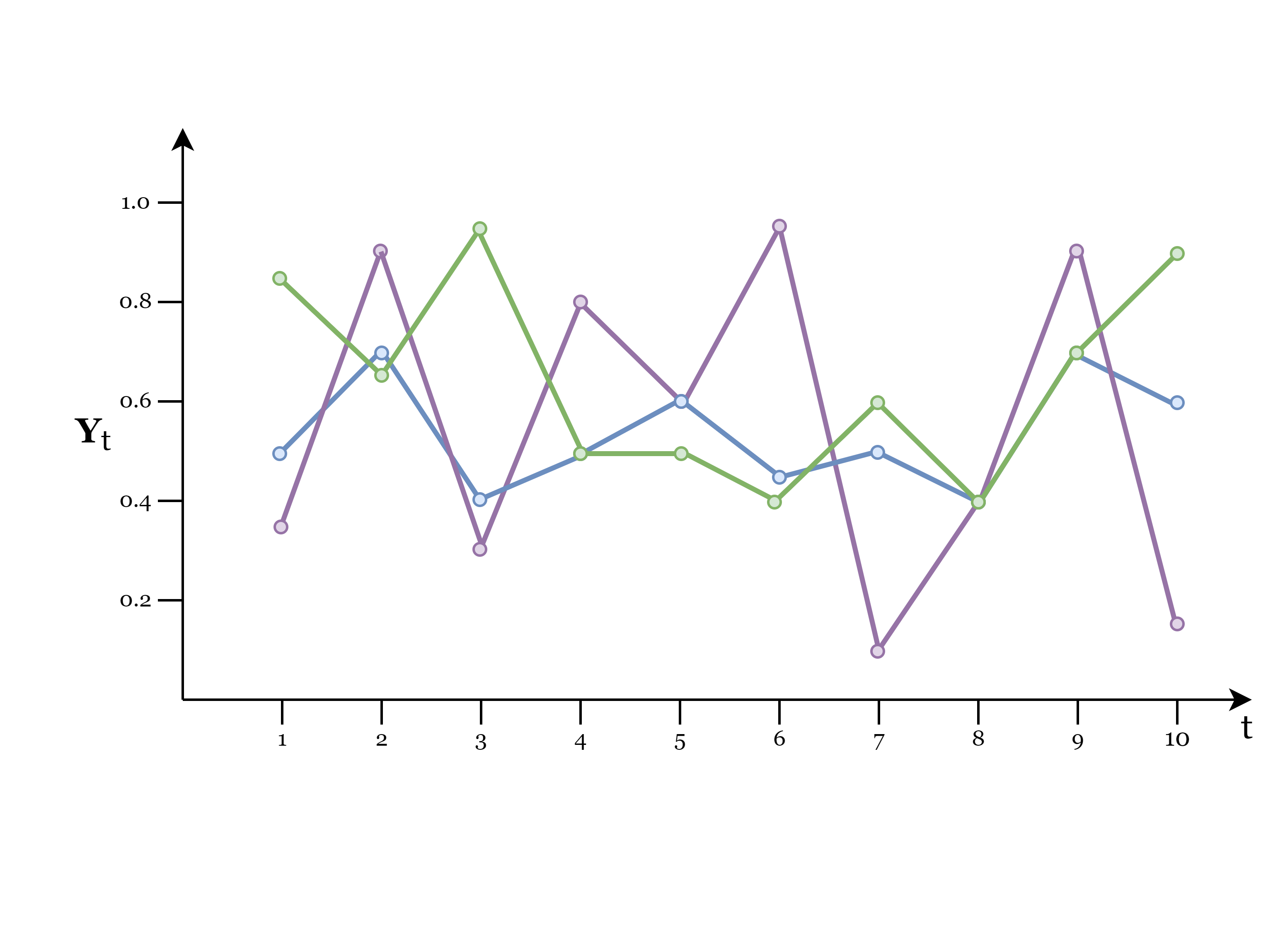}
	} 
	\quad
	\quad
	\subfloat[Ordinal pattern association]{
		\centering
		\includegraphics[scale = 0.6,keepaspectratio,valign=m]{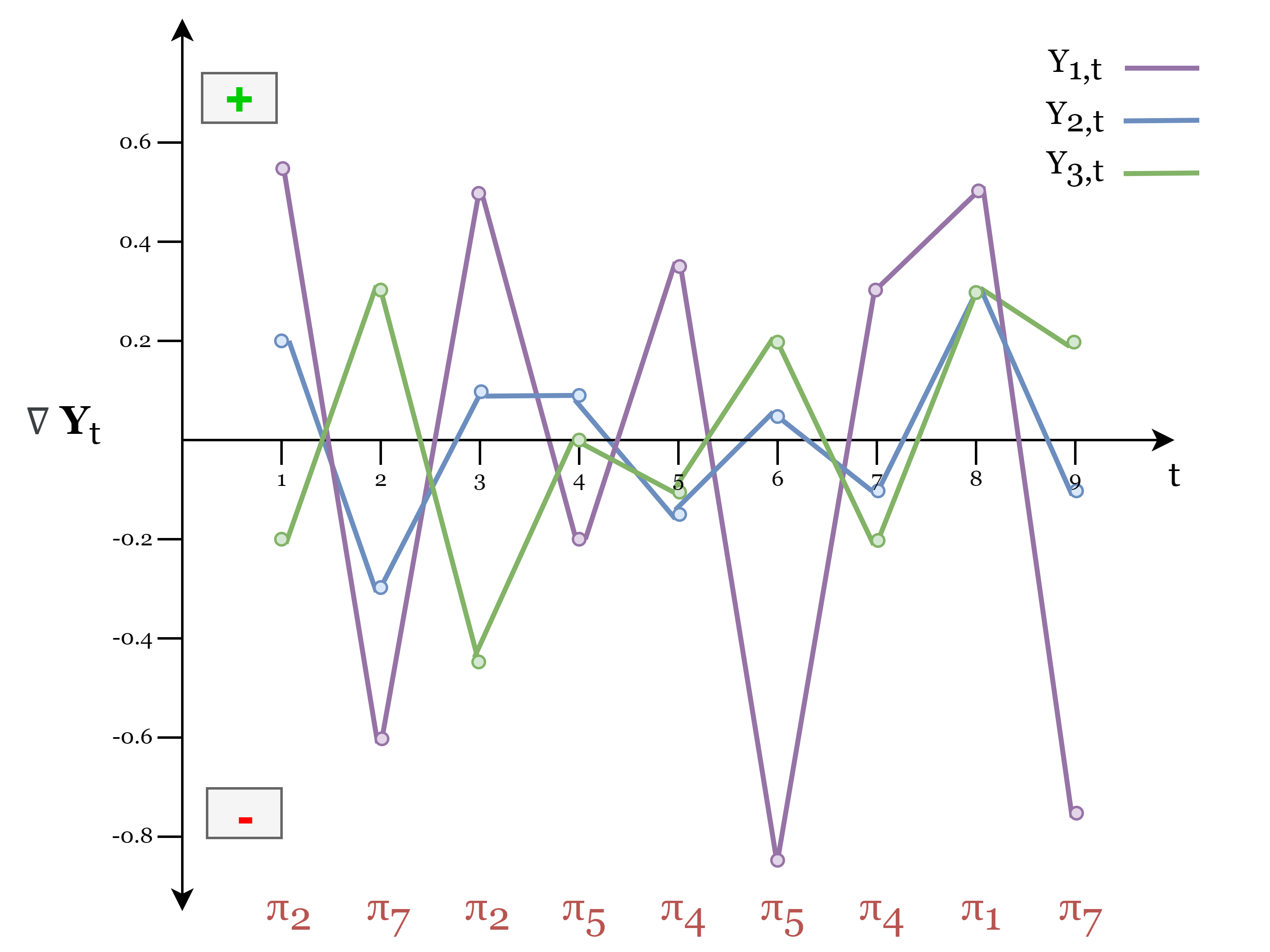}
	}
	\caption[Figure 11]{(a) Illustrative example of a toy three-dimensional series ($Y_{1,t}, Y_{2,t}, Y_{3,t}$). (b)  First-order differences  and corresponding  ordinal patterns.}
	\label{figure:11}
\end{figure}

Then apply the first-order differences $\nabla Y_{i,t} = Y_{i,t} - Y_{i,t-1},$ to each of the $m$ time series. Note that this corresponds to considering the time series of changes and that   many non-stationary time series become stationary with just this procedure. At each time $t$,  associate the  order pattern of the increasing trend  with the positive sign of the difference $\nabla Y_{i,t}$ or the order pattern of the decreasing trend with the negative sign of the difference $\nabla Y_{i,t}$ thus constructing the order pattern $\pi_{j_1},\ldots,\pi_{j_{T-1}},$ $j_1, \ldots, j_{T-1} \in \{ 1, \ldots, 2^m\}$, see Figure~\ref{figure:11} for an example. 
\vspace{-1.5mm}

Finally,  the ordinal partition transition network is the network with $2^m$ nodes representing the patterns $\pi_j$ and with directed weighted edges $(v_i, v_j, w_{i,j})$  established based on frequency of the transitions  from pattern $\pi_i$ to pattern $\pi_j.$ 
In~\cite{zhang2017constructing}, the authors also propose a measure of entropy to characterize ordinal partition transition dynamics, useful for capturing  local geometric changes of trajectories. 
Those authors have shown that for periodic multivariate time series, with different combinations of periods, the resulting network presents   clear evidence of  possible forbidden patterns and so the network has several disconnected nodes. This shows determinism of the time series,  while random uniform noise multivariate time series are mapped into a complete connected graph.

Guo and co-authors~\citep{guo2018cross} propose the \textit{cross ordinal pattern transition network} and the \textit{joint ordinal pattern transition networks} in order to compare the relative rate of change between the two processes by the signs of $(\nabla Y_{i_1,t} - \nabla Y_{i_2,t})$ or $(\nabla Y_{i_1,t} \cdot \nabla Y_{i_2,t})$, respectively. 
Note that  the last sign is related to the signs of changes in the component series while the former is related to amplitude of the changes.

\begin{figure}[hbt!]
	\centering 
	\subfloat[Toy bivariate time series and sequence of ordinal patterns]{
		\includegraphics[scale = 0.6,keepaspectratio,valign=m]{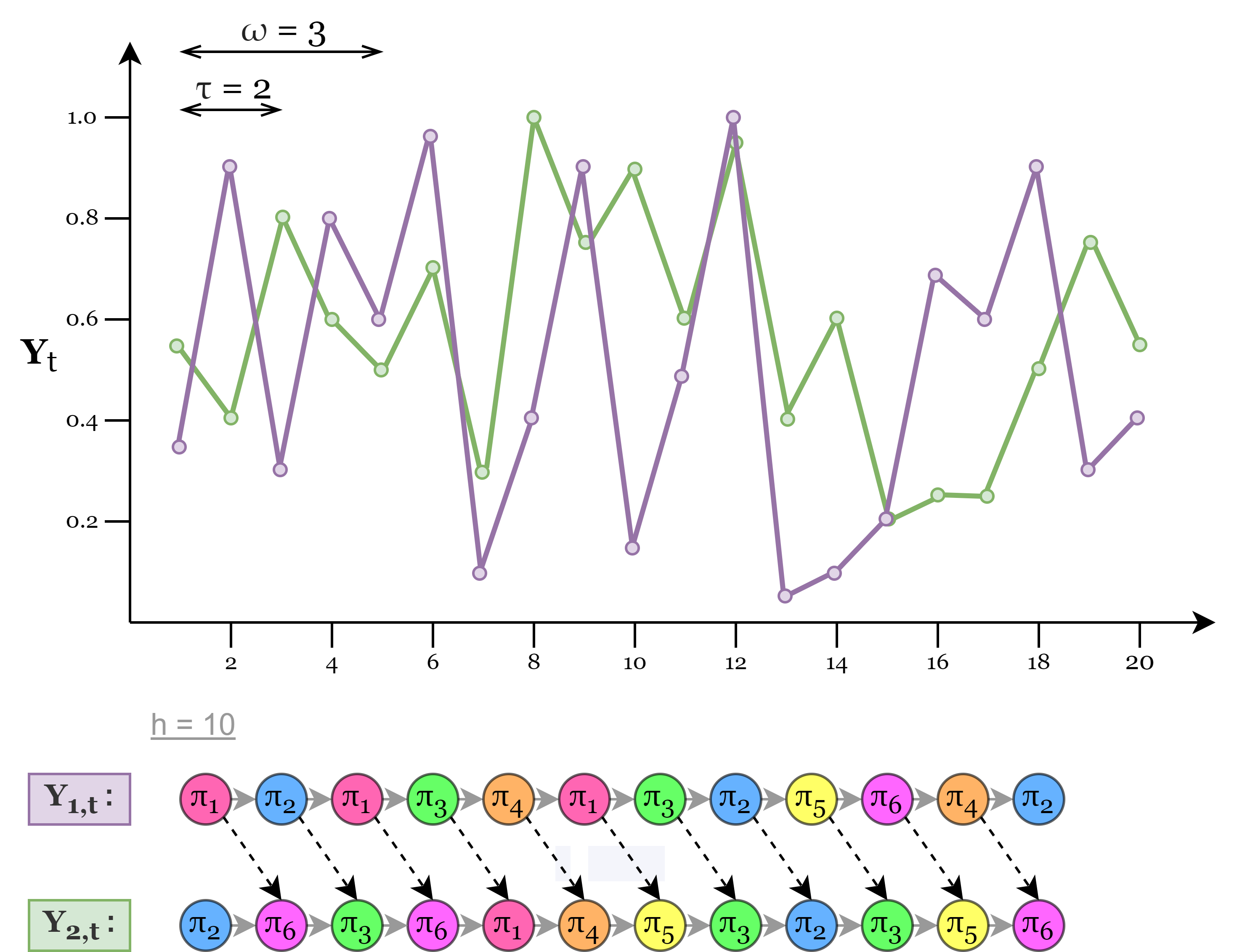}
		\label{figure:12a}
	} 
	\quad
	\quad
	\quad
	\subfloat[Bipartite OPTN]{
		\centering
		\includegraphics[scale = 0.325,keepaspectratio,valign=m]{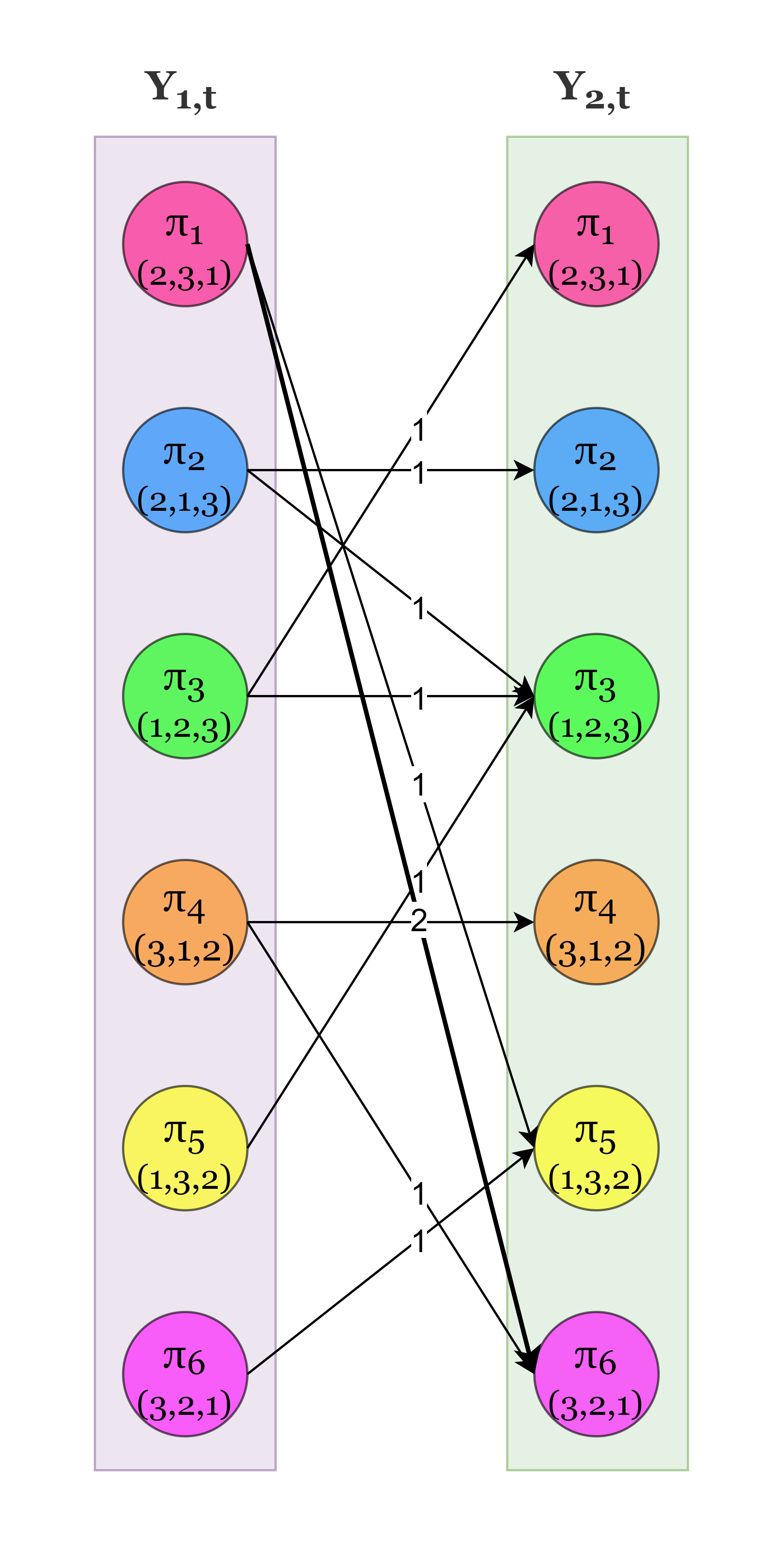}
		\label{figure:12b}
	}
	\caption[Figure 12]{(a) Illustrative example of a toy bivariate time series ($Y_{1,t}, Y_{2,t}$) and its ordinal pattern definitions and evolution. We show an embedding in a 3-dimensional space using a time delay $\tau = 2$ and  the schematic illustration of the OPTN analysis of the two series of ordinal patterns with a unidirectional coupling $Y_{1,t} \to Y_{2,t}$ with a coupling delay of $h = 1$. The time-lagged conditional co-occurrences of the patterns $\pi_i^{Y_{1,t}}$ and $\pi_j^{Y_{2,t+h}}$ are indicated by dashed arrows. (b) The network generated by the ordinal partition transition algorithm proposed in~\cite{ruan2019ordinal}.}
\end{figure}

Recently, Ruan and co-authors~\citep{ruan2019ordinal} proposed a method of ordinal partition transition networks to analyze bivariate time series, $\{Y_{i,t}\}_{i=1}^2,$ based on the initial method proposed for the analysis of univariate series (see Subsection~\ref{map:OP}). 
The method consists in generating the ordinal partition transition network (following the algorithm presented in Subsection~\ref{map:OP}) for each time series $Y_{1,t}$ and $Y_{2,t}$ and extracting the successive sequence of ordinal patterns of each series, removing self-transitions (self-loops in the network). 
Having the two sequences of ordinal patterns $\pi_i^{Y_{1,t}}$ and $\pi_j^{Y_{2,t}},$ as shown in Figure~\ref{figure:12a}, the next step is to calculate at each time $t$  the frequencies of co-occurrence of the ordinal patterns of $Y_{1,t}$ with the ordinal patterns of $Y_{2,t+h}$  i.e., $p(\pi_j^{Y_{2,t+h}} | \pi_i^{Y_{1,t}})$. 
In particular, when $h=0$ we analyze the simultaneous co-occurrence of ordinal patterns and when $h \neq 0$ we analyze the possible indications of causal relationships between the two systems. 
The resulting network have two different types of nodes corresponding to the ordinal patterns $\pi_i^{Y_{1,t}}$ of $Y_{1,t}$ and $\pi_j^{Y_{2,t}}$ of $Y_{2,t},$ respectively, and have directed and weighted edges $(\pi_i^{Y_{1,t}}, \pi_j^{Y_{2,t}}, w_{i,j})$ with $w_{i,j} = p(\pi_j^{Y_{2,t+h}} | \pi_i^{Y_{1,t}}).$ This results in a \textit{bipartite}~\footnote{A bipartite graph (or bigraph) is a graph $G$ whose nodes $V(G)$ can be divided into two disjoint and independent sets $X$ and $Y$ ($V(G) = X \cup Y$ and $X \cap Y = \emptyset$) and every edge in $E(G)$ connects a node in $X$ to a node $Y$.} OPTN (Figure~\ref{figure:12b}).
\vspace{-1.5mm}

Compared to the univariate case, the bipartite OPTN measures co-occurrence probabilities between symbols in two series rather than succession probabilities of symbols in a single time series.  
Ruan and co-authors~\citep{ruan2019ordinal} have also introduced a set of OPTN based complexity measures to infer the coupling direction and infer causality between two systems. Those authors used these measures  in several coupled stochastic processes and in climate time series.

\subsection{Pattern Interdependent Networks}

Based on the visibility graphlets networks (Subsection~\ref{map:VGN}) recently,~\cite{ren2020pattern} proposed the pattern interdependent networks to represent the cross-correlation patterns in a stationary bivariate time series, $\{Y_{i,t}\}_{i=1}^2$. 
The method consists in generating the state chain network (following the algorithm presented in~\ref{map:VGN}) for each time series component, resulting in a bi-graphlet series:  
\begin{equation}
	\left( \begin{smallmatrix} G_1^{[1]} \to G_2^{[1]} \to \ldots \to G_{T-w+1}^{[1]} \\ G_1^{[2]} \to G_2^{[2]} \to \ldots \to G_{T-w+1}^{[2]} \end{smallmatrix} \right)\nonumber.
\end{equation}
Similar to visibility graphlets networks, all unique graphlets $G_i$ are identified. From all the co-occurrent pairs $G_k^{[1]}$ and $G_k^{[2]},$ with $k = 1, 2, \ldots, T-w+1,$ the graphlet co-occurrent frequency matrix, $\boldsymbol{A},$ is generated by calculating the co-occurrent frequencies of all the possible pairs of the unique graphlets, where the element $A_{i,j}$ represents the occurring frequency of $\left( \begin{smallmatrix} G_i  \\ G_j  \end{smallmatrix} \right)$ in $\left( \begin{smallmatrix} G_k^{[1]}  \\ G_k^{[2]}  \end{smallmatrix} \right).$ 
$\boldsymbol{A}$ represents a bipartite network where the rows correspond to the total of unique graphlets (nodes) across the time series $Y_{1,t}$ and the columns to the unique graphlets (nodes) across the $Y_{2,t},$ and the edges are only established between different node types (similar to the bipartite OPTN illustrated in Figure~\ref{figure:12b}). 
Since the sets of nodes, unique graphlets $ G_i $ of each time series, are identical, they are merged and the bipartite network is converted into a directed and weighted network that represents the co-occurrent relationships between the unique visibility graphlets. 

Ren and co-authors~\citep{ren2020pattern} applied this method to synthetic bivariate time series and showed that a set of unique graphlets and the topological structure of the resulting networks is determined and dependent on the cross-correlation, and that the differences in features, such as Hurst exponent, of the time series components determine the symmetry of the edges of the network.

\subsection{Inter-System Recurrence Networks}

Two extensions of recurrence networks (Subsection~\ref{map_RN}) have been proposed in the literature, one directed to bivariate time series and the other to multivariate time series~\citep{feldhoff2012geometric,feldhoff2013geometric}. 
As in recurrence-based methods for univariate time series, we embed each time series $Y_{\alpha,t}$ with $\alpha = 1, \ldots, m$ in a $w$-dimensional, defining a set of vectors $\vec{z}_i^{[\alpha]}$ ($i = 1, \ldots, N_{\alpha}$).  

Cross-recurrence networks are based on cross-recurrence plots that aim to compare the dynamics of two time series $Y_{\alpha,t}$ and $Y_{\beta,t}$ in the same phase space~\citep{zbilut1998detecting,marwan2002nonlinear}. 
So, both time series are simultaneously embedded in the same phase space and the cross-recurrence plot is defined by the matrix $\boldsymbol{CR}^{[\alpha \beta]}$ with elements~\citep{marwan2002nonlinear}:
\begin{equation}
	CR^{[\alpha \beta]} = \left\{\begin{array}{l}
		1 \quad \textrm{if } \|\vec{z}_i^{[\alpha]} - \vec{z}_j^{[\beta]}\| \le \varepsilon \\
		0 \quad \textrm{otherwise}\\
	\end{array}\right.,
	\label{eq:CRN}
\end{equation}
where $i = 1, \ldots, N_{\alpha},$ $j = 1, \ldots, N_{\beta},$ and $\varepsilon$ is a distance threshold in the joint phase space of both processes. 
$\boldsymbol{CR}^{[\alpha \beta]}$ is asymmetric, since  $\|\vec{z}_i^{[\alpha]} - \vec{z}_j^{[\beta]}\| = \|\vec{z}_i^{[\beta]} - \vec{z}_j^{[\alpha]}\|$ 
does not hold for all $i,j,\alpha,\beta, $ 
and it can be non-square if we consider time series of different lengths,  $N_{\alpha} \neq N_{\beta}.$ 
This matrix can represent the adjacency matrix of a bipartite graph, corresponding to the cross-recurrence network, where the nodes belong to two distinct groups, namely, the state vectors $\{\vec{z}_i^{[\alpha]}\}$ and $\{\vec{z}_j^{[\beta]}\}$ and the edges are established only between nodes of different groups.

Inter-system recurrence networks $\boldsymbol{IR}$ arise from the combination of the recurrence matrices $\boldsymbol{R}^{[\alpha]}$ (for univariate time series $Y_{\alpha, t}$), with the cross-recurrence matrix  $\boldsymbol{CR}^{[\alpha \beta]}$~\citep{feldhoff2012geometric}:
\begin{equation}
	\boldsymbol{IR} = \left(
	    \begin{array}{cccc}
	        \boldsymbol{R}^{[1]} & \boldsymbol{CR}^{[12]} & \ldots & \boldsymbol{CR}^{[1m]}  \\
	        \boldsymbol{CR}^{[21]} & \boldsymbol{R}^{[2]} & \ldots & \boldsymbol{CR}^{[2m]} \\
	        \vdots & \vdots & \ddots & \vdots \\
	        \boldsymbol{CR}^{[m1]} & \boldsymbol{CR}^{[m2]} & \ldots & \boldsymbol{R}^{[m]} 
	    \end{array}
	\right).
	\label{eq:ISRN1}
\end{equation}
The adjacency matrix of an inter-system recurrence network is defined as
\begin{equation}
   \boldsymbol{A} = \boldsymbol{IR} - \boldsymbol{I}_N,
	\label{eq:ISRN2}
\end{equation}
where $\boldsymbol{I}$ is an identity matrix of size $N = \sum_{\alpha=1}^m N_{\alpha}$.  
This results in undirected and unweighted simple graph, where the nodes and edges obey a natural partition: subgraphs $G_{\alpha}$ correspond to $\boldsymbol{R}^{[\alpha]}$ (intra-system connectivity) and subgraphs $G_{\alpha \beta}$ to $\boldsymbol{CR}^{[\alpha \beta]}$ that includes only edges between nodes from different systems (inter-system connectivity). 
Note that the definition of closeness can vary between different pairs of systems~\citep{zou2018complex} if we consider different thresholds $\varepsilon_{\alpha \beta}$  for all $\alpha, \beta = 1, \ldots, m.$

Inter-system recurrence networks can be analyzed as a network of networks where the nodes represent subgraphs $G_{\alpha}.$ 
As such and in line with previous work for  the univariate case, Donges and co-authors~\citep{donges2011investigating} used specific graph measures  as discrete approximations of more general geometric properties to study the interconnections between systems. 
These measures proved to be useful to coupling detection from two paleoclimate records~\citep{feldhoff2012geometric}, distinguish between the dynamics of focal and non-focal EEG signals~\citep{subramaniyam2015dynamics} and characterize different oil–water flow patterns from  multi-channel measurements~\citep{gao2013multivariate,gao2016multivariate}.

\subsection{Joint Recurrence Networks}

Joint recurrence network is another extension of the recurrence networks. 
The difference for the networks introduced in the previous subsection is that joint recurrence networks  study the recurrence of different time series in their individual phase spaces, contrary  the same phase space. The joint recurrence matrix $\boldsymbol{JR}$ is defined as follows~\citep{romano2004multivariate}:
\begin{equation}
	JR = \left\{\begin{array}{l}
		1 \quad \textrm{if } \|\vec{z}_i^{[\alpha]} - \vec{z}_j^{[\alpha]}\| \le \varepsilon_{\alpha}, \ \alpha = 1, \ldots, m \\
		0 \quad \textrm{otherwise}\\
	\end{array}\right.,
	\label{eq:JRN}
\end{equation}
where $\varepsilon_{\alpha}$ is the selected distance threshold  for the individual time series $Y_{\alpha,t}.$
This matrix can be seen as the adjacency matrix $\boldsymbol{A}$ of a joint recurrence network:
\begin{equation}
   \boldsymbol{A} = \boldsymbol{JR} - \boldsymbol{I}_N,
	\label{eq:JRN2}
\end{equation}
where $\boldsymbol{I}$ is an identity matrix of size $N = \sum_{\alpha=1}^m N_{\alpha}.$ 
$\boldsymbol{A}$ represents the joint probability of $m$ simultaneous recurrences ($\| \vec{z}_i^{[\alpha]} - \vec{z}_j^{[\alpha]} \|,$  $\alpha = 1, \ldots, m$)  in the phase spaces~\citep{marwan2007recurrence}. 
Joint recurrence networks are undirected and unweighted simple graphs.
They can be constructed for time series with different phase spaces and require simultaneous observations, i.e., time series of the same length. 
So, unlike the recurrence networks and inter-system recurrence networks, the time information is taken into account. 
This type of network can be analyzed from the same point of view of the  recurrence networks for univariate time series, however, need to be reinterpreted in terms of the underlying joint recurrence structure~\citep{feldhoff2013geometric}. 
Note that joint recurrence networks are similar in construction  to multiplex recurrence networks, as we will see  in Subsection~\ref{subsec:mrn}, except that the former do not establish inter-layer edges~\citep{zou2018complex}.

This method has been successfully used to detect generalized synchronization in time series~\citep{feldhoff2013geometric}. 
It is expected that  joint recurrence will be increasingly unlikely with an increase in the number $m$ of processes. 
Therefore, Donner and co-authors~\citep{donner2015complex} proposed a version called $f$-joint recurrence networks that reduces the requirement of occurrence of simultaneous recurrences in all subsystems.

\subsection{Multiplex Visibility Networks}

{\label{subsec:mnvg}}

\begin{figure}[hbt!]
	\centering 
		\includegraphics[scale=0.47]{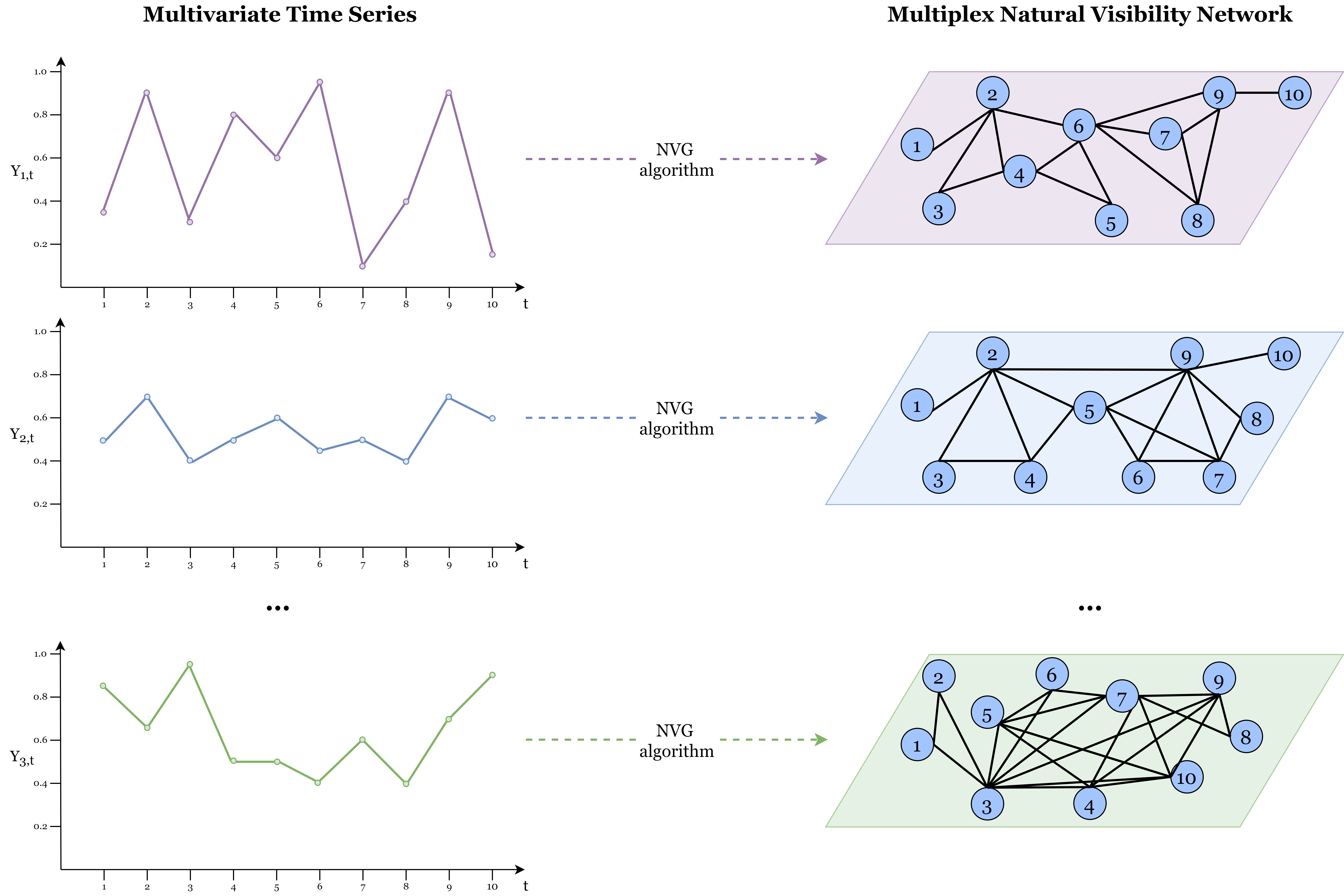}
	\caption[Figure 13]{Illustration of multiplex natural visibility graph algorithm for a toy multivariate time series.}
	\label{figure:13}
\end{figure}

Based on the  visibility methods of Section~\ref{section.visnet} and the definition of multilayer networks, Lacasa and co-authors~\citep{lacasa2015network} proposed an extension of the visibility mapping method for multivariate series analysis.
The networks resulting from this approach have been called \textit{multiplex visibility graphs}.
A multiplex visibility graph $M$ of $m$ layers is constructed so that each layer $\alpha$ corresponds to the NVG (Section~\ref{section.visnet}) associated with the series $Y_{\alpha,t},$ as illustrated in Figure~\ref{figure:13}.
\vspace{-0.3cm}

\vspace{-1.5mm}
The resulting graph $M$ is now represented by the adjacency matrix vector, $\mathcal{A},$ whose elements are the adjacency matrices of the layers, $\mathcal{A} = \{\boldsymbol{A}^{[1]}, \boldsymbol{A}^{[2]}, \ldots, \boldsymbol{A}^{[m]}\}$ with $A^{[\alpha]}_{i,j} = 1$ if and only if the node $v_i$ and node $v_j$ are connected by an edge in layer  $\alpha.$

A \textit{graph of layers} (or \textit{network of networks}) can be built  by projecting  the original multiplex visibility network into a (single layer) weighted graph of $m$ nodes, where each node represents one layer. The edge weights denote a relation measure between layers, e.g., the magnitude of mutual information~\citep{lacasa2015network} calculated by \textit{interlayer mutual information}~\footnote{Interlayer mutual information, $I_{\alpha,\beta}$, quantifies the common information by every two different layers $\alpha$ and $\beta$ based on the similarity of the degree distributions. Higher values of $I_{\alpha,\beta}$ indicating that the corresponding layers are associated/correlated and, consequently, the series they represent~\citep{lacasa2015network}.}.
We should note that the construction of multiplex visibility networks can be based on any of the visibility algorithms  for univarite time series mentioned  in Section~\ref{section.visnet}.

The multiplex structure was used for the study of cellular networks of diffusively coupled maps lattices (CML)~\citep{lacasa2015network}, to model spatio-temporal complex dynamics. They show that multiplex visibility graphs allow to quantify the amount of information flow between different series through the interlayer mutual information metric, and also to distinguish between different signal behaviors, between chaotic, periodic and multiband patterns. The analysis of this metric allowed locating data points of change between different patterns of behavior of a system.

Bianchi and co-authors~\citep{bianchi2017multiplex} used multiplex visibility networks method based on the weighted HVG algorithm (Subsection~\ref{subsec:wvg}) where the edge weights are given by: $w_{i,j} = 1 / \sqrt{(j-i)^2 + (Y_i - Y_j)^2}$, incorporating temporal and amplitude information of the data. 
They studied topological measures of the networks to characterize neuron activations. For example, in certain types of neuronal activities the local clustering coefficient of the corresponding nodes at the moment of the activity have high values which are replicated by the same nodes in the remaining layers. This type of metrics allows to identify neuronal activities and the propagation of this activity by different neurons. 
Another set of real data studied using these networks were the fMRI series~\citep{sannino2017visibility} that are of extreme importance for the diagnosis of mental and neurological diseases. The results show differences in brain activities connected to psychiatric disorders.

In~\cite{gao2019characterization} the authors studied multiplex limited penetrable horizontal visibility graphs to try to explain the difference in the accuracy of the results obtained in the task of classifying time series from individuals in normal and fatigue states. They found that some network metrics decrease in states of fatigue and that, consequently, there is less efficiency in the global information transfer. They concluded, therefore, that this transfer loss leads to a reduction in the accuracy of the classification between individuals in normal state and in fatigue.

\subsection{Multiplex Recurrence Networks}

{\label{subsec:mrn}}
 
Recently, Eroglu and co-authors~\citep{eroglu2018multiplex}, motivated by the idea that recurrences are a fingerprint of the characteristic properties of dynamic systems, the results from recurrence networks, the current and ever greater need for analysis and exploration of large amounts of data, and the previous method, multiplex visibility networks, proposed by Lacasa and co-authors~\citep{lacasa2015network},  proposed recurrence networks for the domain of multivariate time series, the so-called multiplex recurrence networks. The algorithm follows exactly the same process as the multiplex visibility graphs presented in Section~\ref{subsec:mnvg}, but using recurrence networks presented in ~\ref{map_RN} instead of the visibility graphs. 
Remembering, the quantities of interlayer mutual information allows us to deduce a weighted single layer network from the multiplex recurrence network.

Eroglu  and co-authors~\citep{eroglu2018multiplex}, used metrics as the interlayer mutual information and the average edge overlap~\footnote{The average edge overlap, $\omega$, calculates the expected number of layers in which a given edge is present.} in the multiplex network and also weighted network metrics, clustering coefficient and average path length, in the weighted single layer network, in order to detect the transitions between different dynamical regimes in coupled map lattices and real world paleoclimate time series.
They show that all the metrics capture similarities in the topological structures of the $m$ recurrence networks. In particular, high values of the first three metrics have been obtained for periodic behaviors, while lower values have been obtained to the chaotic behaviors. Additionally, average path length showed the opposite behavior. 

In~\cite{gao2019recurrence}, the authors propose to combine the multiplex recurrence networks and deep learning techniques to detect driver fatigue in the EEG signals of subjects under alert  and fatigue states. They construct a multiplex recurrence network from the EEG signal, used convolutional neural networks to extract and learn the features of the multiplex networks, and they perform a classification task. Those authors showed that this combination can achieve a high accuracy, and better results than traditional methods.

\par\null

\section{Conclusions}

In this review we present several  algorithms proposed in the literature to map univariate and multivariate time series into the complex network domain with the aim of producing new insights and overcome open issues in time series analysis, such as high-dimensionality, finding periodicities, classifying different dynamic processes, among others. 
For univariate time series there is a large body of literature and the  mappings may be classified according to three main underlying concepts: visibility, transition and proximity. 

Visibility concepts map  times series into  graphs with  nodes representing time stamps and edges defined by a geometric relationship between data values. The graph reflects both local and global properties of the time series, especially via local maxima. 
Visibility-based mapping methods do not require preprocessing of the time series data. They are also completely parameter-free, except for the limited and parametric versions. These networks have become very popular in the literature due to the fact that the geometric criteria associated to visibility mappings are intuitive and easy to understand.

Intuitively, transition networks represent the transition pattern of a time series based on different types of symbolic encoding of the data. With exception to time series with clear trending behavior, these networks are not explicit about the underlying time order since time series with sharp trend behavior are mapped into quantile graphs that are chain graphs allowing us to perceive the time order of the data. 
The construction of these networks involves the definition of symbols representing quantiles, order patterns or different dynamic states, requiring the choice of parameters that must  balance the loss of information induced by the partitioning.

The proximity networks represent the similarity of sliding windows of data over time, thus, reflecting how local properties of the time series evolve over time. The construction of proximity networks involves the definition of states of the time series as vectors, such as cycles or vectors in a phase space, requiring the selection of parameters. The resulting networks have nodes that represent these states of the time series and edges that are established using measures of similarity or distance between the states, with or without thresholding. The construction of transition and of proximity networks thus requires choosing several parameters that may influence the resulting graph and consequently the analysis of the time series. These networks are quite popular in the study of dynamical systems.

Several approaches to map multivariate time series into complex networks have been developed. Most focus on the idea of constructing a single layer network from multivariate time series, capable of capturing dependencies between the component processes, both contemporary and lagged, as well as the serial dependencies.  However, with technological advances and the growing need to analyze complex and high-dimensional data, new techniques have emerged, namely, methods that give rise to high-level structures: multiple layers networks. Although preliminary, this type of multivariate time series mapping has an important characteristic: all the methods developed for the univariate context can be extended to the context of multilayer networks, allowing the reuse of knowledge already acquired in the univariate case. 

A network science approach to time series analysis has been used in different application domains and allowed to characterize system dynamics, to distinguish different dynamics, to identify regime shifts and dynamical transitions,  to test for reversibility and  to forecast.  This survey hints at the conclusion  that this approach provides complementary information to traditional time series analysis. 
This approach can be leveraged to address  existing fundamental open issues. Among them we highlight the following: missing values, unequally spaced data, visualization of high-dimensional data, time series classification and clustering.
Data recorded using electronic devices such as sensors are prone to missing values and conventional imputation methods may cause bias in the data. The work of~\cite{donner2012visibility} addresses this problem of missing values via a network perspective  and, although the results are preliminary, they highlight a path to the solution of this issue. 
Unequally spaced data collected over time occurs frequently for several reasons, hindering the application of time series analysis methodology. The most common approach to this issue is data aggregation which leads to the loss of information. As far as we know, this problem has not yet been addressed from the point of view of networks.  
Visualization of high dimensional temporal data is a practical issue to whose solution may contribute strategies to reduce the dimensionality of the data by using suitable mapping methods (e.g. quantile graphs). 
Mining time large collections of time series data is increasingly accomplished via time series features, e.g. trend, length of time series, strength of trend, autocorrelations~\citep{kang2017visualising}. The features extracted from the time series networks may be added to the classical set of time series features to enhance the data characterization. In particular,~\cite{vanessa2018time} has shown that network metrics  can be used to accurately cluster  time series from different models, a relevant important contribution to time series mining. 

There is still much potential to be explored and this survey aims precisely to equip the reader with the knowledge capable of unveiling this potential.

\par\null

\section*{Acknowledgments}

This work was partially supported by the Portuguese funding agency, FCT - Fundação para a Ciência e a Tecnologia, through national funds, and co-funded by the FEDER, where applicable, and within the Center for Research and Development in Mathematics and Applications (CIDMA), references SFRH/BD/139630/2018 and UIDB/04106/2020.

\selectlanguage{english}
\FloatBarrier

\setlength{\parskip}{3.5pt}
\renewcommand{\baselinestretch}{0.85}

\bibliographystyle{apalike}
\bibliography{refs}

\end{document}